\documentclass[%
 reprint,
 amsmath,amssymb,
 aps,
]{revtex4-2}

\usepackage{graphicx}
\usepackage{dcolumn}
\usepackage{bm}
\usepackage{soul}
\usepackage{color}
\usepackage{hyperref}
\usepackage{natbib}

\begin{document}

\title{Accuracy of metaGGA functionals in describing transition metal fluorides \\ \vspace{0.3cm}}

\author{Dereje Bekele Tekliye$^{1}$ and Gopalakrishnan Sai Gautam$^{1,}$}
\email{saigautamg@iisc.ac.in}
\affiliation{$^{1}$Department of Materials Engineering, Indian Institute of Science, Bengaluru, 560012, Karnataka, India}

\begin{abstract}
Accurate predictions of material properties within the chemical space of transition metal fluorides (TMFs), using computational frameworks such as density functional theory (DFT), is important for advancing several technological applications. The state-of-the-art semi-local exchange-correlation (XC) functionals within DFT include the strongly constrained and appropriately normed (SCAN) and the restored regularized SCAN (r\textsuperscript{2}SCAN), both of which are meta generalized gradient approximation (metaGGA) functionals. Given their semi-local nature, both SCAN and r\textsuperscript{2}SCAN are susceptible to self-interaction errors (SIEs) while modelling highly correlated $d$ electrons of transition metals (TMs). Hence, in this work, we evaluate the accuracy of both SCAN and r\textsuperscript{2}SCAN functionals in estimating several properties of TMFs, including redox enthalpies, lattice geometries, on-site magnetic moments, and band gaps. Specifically, we consider binary fluorides of Ti, V, Cr, Mn, Fe, Co, Ni, and Cu. We observe both SCAN and r\textsuperscript{2}SCAN exhibit poor accuracy in estimating fluorination enthalpies among TMFs, which can be primarily attributed to SIEs among the $d$ electrons, given both functionals bind F$_2$ accurately. Thus, we derive optimal Hubbard \textit{U} corrections for both functionals based on experimental fluorination (or oxidation) enthalpies within binary TMFs. Note that our attempts at using the linear response theory to derive \textit{U} corrections yielded unphysical values for V, Fe, and Ni fluorides. While adding the fluorination-enthalpy-derived \textit{U} corrections to the metaGGA functionals does not significantly affect the lattice volumes and on-site magnetic moments (and in turn, the accuracy of these property estimations versus experiments), it does cause a significant increase in calculated band gaps. Also, we calculated the average Na intercalation voltage in Mn, Fe, Co, and Ni fluorides as a transferability check of our optimal \textit{U} values. Overall, we do recommend the incorporation of the Hubbard \textit{U} correction to improve predictions of redox enthalpies in other TMFs. With respect to band gap predictions, given the lack of experimental data, we suggest the use of the non-\textit{U}-corrected metaGGA functionals, given their observed physically-precise underestimation of empirical gaps. Finally, our study should advance the accuracy of DFT-based screening studies to unearth novel TMFs, which can be used in various applications, including energy storage, catalysis, and magnetic devices. 

\end{abstract}

\maketitle


\section{\label{sec:level1}Introduction}

Redox-active 3\textit{d} transition metal fluorides (TMFs) are widely studied and have been used in the fields of magnetism \cite{dubrovin2020incipient, dubrovin2018unveiling}, catalysis \cite{zheng2021improved, wang2023fluorine, ali2022utilization, han2019polarized}, electroceramics \cite{garcia2016strain, borisov2016multiferroic, reisinger2011phase}, and electrochemical energy storage \cite{wang2015ternary, zhang2019transition, park20212, foley2023polymorphism, lu2023weberite, dey2019topochemical, hwang2017naf, kim2010fabrication, hua2021revisiting}. For instance, fluorides exhibit considerable promise as prospective positive electrode (cathode) materials for batteries owing to the advantageous inductive effect associated with fluorine \cite{manthiram1989lithium}. Characterized by fluorine's high electronegativity, fluorine addition to a cathode material typically yields higher voltages and higher energy densities. Several fluoride-based cathode materials, such as LiMnF\textsubscript{3}, NaFeF\textsubscript{3}, NaMnF\textsubscript{3}, NaCoF\textsubscript{3}, and KVO\textsubscript{4}F have been investigated as potential candidates for intercalation-type and conversion-type rechargeable lithium-ion, sodium-ion, and potassium-ion batteries \cite{fan2018high, martin2018reversible, cao2017cubic, kitajou2017cathode, byeon2023effects}. Other notable conversion-based electrodes in the lithium-ion chemical space include the LiF+FeF\textsubscript{2}-LiFeF\textsubscript{3} and LiF-FeO systems, which are fluoride-based as well \cite{li2010reversible, kim2018conversion, fan2016situ, jung2018new, hwang2017naf}.    

Investigating materials for energy storage and other applications, including identifying new materials and understanding the underlying properties of existing materials, has greatly benefited from the utilization of quantum mechanical methods \cite{tekliye2022exploration, lu2021searching, cheng2015accelerating, sai2020exploring, sun2019map, chakraborty2017rational}, such as density functional theory (DFT \cite{hohenberg1964inhomogeneous, kohn1965self}). Notably, the choice of the exchange-correlation (XC) functional in DFT plays a crucial role in accurately describing the electronic interactions within a given material, and hence its resultant properties, such as redox enthalpies and ground state atomic configurations. Prominent XC functionals include the local density approximation (LDA  \cite{kohn1965self}), the generalized gradient approximation (GGA \cite{perdew1996generalized}), and the meta-generalized gradient approximation (metaGGA \cite{sun2015strongly, bartok2019regularized, furness2020accurate}), which represent increasing levels of accuracy, as illustrated using the Jacob’s ladder analogy \cite{car2016fixing}.

Among metaGGA functionals, the strongly constrained and appropriately normed (SCAN \cite{sun2015strongly}) functional satisfies all the 17 known constraints of an XC functional. SCAN incorporates the orbital kinetic energy density as a parameter in addition to the local electron density and its gradient, resulting in higher accuracy \cite{yang2019rationalizing, yao2017plane, sassnick2021electronic, fu2019density}. However, SCAN suffers from numerical instability and associated computational convergence difficulties \cite{bartok2019regularized}. To address SCAN’s shortcomings, the restored regularized SCAN (r\textsuperscript{2}SCAN \cite{furness2020accurate}) functional has been developed. r\textsuperscript{2}SCAN combines the numerical accuracy of SCAN with improved numerical stability \cite{kingsbury2022performance, swathilakshmi2023performance, dellostritto2023predicting}, while satisfying 16 out of the 17 known constraints for an XC functional. Thus, SCAN and r\textsuperscript{2}SCAN represent the state-of-the-art metaGGA functionals and have been widely used for both materials discovery and improving fundamental understanding \cite{sun2017thermodynamic, deng2022fundamental, ning2022reliable, kothakonda2023high, devi2022effect, jha2023evaluation, kumar2023study}.

Employing metaGGA functionals to predict and investigate correlated electron systems, such as TMFs, can still provide an erroneous description of the underlying electronic structure. For example, both SCAN and r\textsuperscript{2}SCAN exhibit residual self-interaction errors (SIEs \cite{perdew1981self}) while modelling transition metal oxides (TMOs), which are also highly correlated systems like TMFs. The residual SIEs often result in incorrect redox enthalpies, ground state polymorphs, lattice parameters, on-site magnetic moments, and electronic properties \cite{swathilakshmi2023performance, gautam2018evaluating, long2020evaluating, long2021assessing}. Typically, functionals suffer from SIEs owing to an overestimation of electronic delocalization within contracted and localized \textit{d} and/or \textit{f} orbitals. Thus any residual SIE is likely to influence the accuracy of the metaGGA functionals in modelling TMFs as well. One potential way of mitigating SIEs in correlated systems is adding a Hubbard \textit{U} correction \cite{anisimov1991band} (on the \textit{d} and/or \textit{f} orbitals) to the semi-local functionals. Adding the \textit{U} correction has been shown to remove several of the spurious predictions by SCAN and r\textsuperscript{2}SCAN in the TMO chemical space \cite{swathilakshmi2023performance, gautam2018evaluating, long2020evaluating}. Thus, it is useful to explore the accuracy of such Hubbard \textit{U} corrected metaGGA frameworks, specifically SCAN+\textit{U} and r\textsuperscript{2}SCAN+\textit{U}, on property predictions within TMFs. However, the magnitude of the \textit{U} correction is not known \textit{a priori}. 

Several techniques, with their own sets of pros and cons, have been used to identify an ‘optimal’ \textit{U} correction for a given transition metal (TM) system, including \textit{i}) theory-based approaches, such as linear response theory \cite{timrov2018hubbard, zhou2004first, moore2022high, lan2018linear, shishkin2016self}, and embedded Hartree-Fock calculations \cite{ mosey2007ab, mosey2008rotationally}, \textit{ii}) statistics-based approaches, such as machine learning-based Bayesian optimization \cite{yu2020machine}, and clustering-validation techniques \cite{Artrith@2022data}, and \textit{iii}) using experimental data, such as oxidation enthalpies \cite{gautam2018evaluating, long2020evaluating, wang2006oxidation, jain2011formation, lutfalla2011calibration}, and band gaps \cite{loschen2007first}. Statistics-based and experimental-data-based approaches typically provide an averaged \textit{U} correction for a given TM, i.e., a \textit{U} value that can be used across multiple oxidation states of the TM, while theoretical approaches provide an oxidation-state-specific (and often structure-specific) \textit{U} correction. Notably, previous studies have utilised experimental oxidation enthalpies to estimate optimal \textit{U} corrections for different functionals (e.g., GGA \cite{wang2006oxidation}, SCAN \cite{gautam2018evaluating, long2020evaluating}, and r\textsuperscript{2}SCAN \cite{swathilakshmi2023performance}) and for several TMs in oxide coordination. In turn, such \textit{U} corrections have been used successfully in several materials screening\cite{sai2020exploring, curtarolo2013high} and optimization\cite{lun2021cation, bartel2019role, zhong2023modeling} studies. However, such a widespread identification of optimal \textit{U} corrections that can be used with SCAN or r\textsuperscript{2}SCAN in the TMF chemical space has not been done, so far. 

In this work, we present a comprehensive investigation assessing the accuracy of the SCAN and the r\textsuperscript{2}SCAN metaGGA functionals, and their Hubbard \textit{U} corrected frameworks, in predicting the redox thermodynamics, lattice parameters, magnetic properties, and electronic structures of several TMFs. We verify that both SCAN and r\textsuperscript{2}SCAN estimate the binding energy of F\textsubscript{2} accurately. We identify an optimal \textit{U} correction, if necessary, by considering experimental oxidation (i.e., fluorination) enthalpies of binary 3\textit{d} TMFs, where the TMs are Ti, V, Cr, Mn, Fe, Co, Ni, or Cu. For a few compounds, we also use the linear response theory to demonstrate that the framework does provide unphysical \textit{U} corrections compared to the experimental-data-derived values. Overall, both SCAN and r\textsuperscript{2}SCAN exhibit similar errors in calculating properties across several TMFs, with their \textit{U}-corrected versions exhibiting similar improvements in accuracies compared to the corresponding `bare' functionals. Wherever possible, we verified the transferability of our \textit{U} values by calculating properties of compounds not used in estimating the \textit{U} values. Our work should provide a framework for computational screening approaches to more accurately predict properties within the correlated TMF space for energy storage and other applications.

\section{Methods}
\subsection{Computational methods}
All spin-polarized SCAN(+\textit{U}), and r\textsuperscript{2}SCAN(+\textit{U}) calculations were performed using the Vienna ab initio simulation package (VASP \cite{kresse1993abinitio, kresse1996efficient}), using the frozen-core projector augmented wave (PAW \cite{kresse1999ultrasoft, blochl1994projector}) potentials, as listed in \textbf{Table S1} of the supporting information (SI). We used the rotationally invariant Hubbard \textit{U} approach developed by Dudarev \textit{et al.} \cite{dudarev1998electron} in our calculations. The electronic kinetic energy was expanded using plane waves up to an energy of 520 eV. We employed a Gaussian smearing of width 0.05 eV to integrate the Fermi surface. We sampled the irreducible Brillouin zone with $\Gamma$\textendash centered Monkhorst-Pack \cite{monkhorst1976special} \textit{k}-point grids of a minimum density of 48 \textit{k}-points per Å (i.e., a real space lattice vector of 1 Å was sampled using 48 sub-divisions in the reciprocal space). The total energy and atomic force convergence criteria were set to 10\textsuperscript{-5} eV and \( |0.03| \) eV/Å, respectively, with no symmetry being preserved during relaxations of the cell volume, cell shape, and ionic positions within each structure. To calculate the binding energy of F\textsubscript{2} gas, we performed two calculations, one with an isolated F\textsubscript{2} molecule and another with an isolated F atom in $18 \, \text{\AA} \times 19 \, \text{\AA} \times 20 \, \text{\AA}$ asymmetric cells, and allowed the atomic positions to change in F\textsubscript{2}.

The initial structures of all compounds were obtained from the inorganic crystal structure database (ICSD \cite{hellenbrandt2004inorganic}). We utilized the conventional cell for all systems, except CuF and MnF\textsubscript{3}. In CuF, we used a $2\times2\times2$ supercell due to convergence difficulties associated with our electronic density of states (DOS) calculations with the conventional cell. In the case of MnF\textsubscript{3}, we used a $2\times2\times2$ supercell to account for the A-type (↑↑↓↓) antiferromagnetic (AFM) ordering \cite{wollan1958antiferromagnetic}. Unless specified, we initialised the on-site magnetic moment of a TM to its corresponding high spin configuration.  

For band gap calculation of TMFs, we used the generalized Kohn-Sham technique \cite{perdew2017understanding} and calculated the DOS for all systems considered. We used the optimized structure and the initial charge density from a prior structure relaxation for all DOS calculations. Subsequently, we introduced a set of zero-weighted $k$-points, corresponding to a density of 96 $k$-points per Å, with the $k$-points that were used for the structure relaxation retained their original weights. Finally, we performed a single self-consistent-field (SCF) calculation for each structure, sampling the electronic DOS over an energy range of $-$20 to 20 eV at intervals of 0.005 eV. The total energy convergence criterion was set to 10\textsuperscript{-6} eV in all DOS calculations.   

\subsection{Reaction energies and optimal \textit{U}}
We used experimental oxidation (i.e., fluorination) enthalpy per F\textsubscript{2} (\(\Delta H_\text{O}^\text{expt}\)) among binary TMFs to identify the optimal Hubbard \textit{U} correction for each TM. Specifically, we considered reactions of the type \( \text{MF}_x + \frac{{(y-x)}}{2}\text{F}_2 \rightleftharpoons \text{MF}_y \), where M = Ti, V, Cr, Mn, Fe, Co, Ni, or Cu. We collected the experimental thermochemical data (at 298 K, 1 atm) from the tables of the National Institute of Standards and Technology (NIST)-JANAF \cite{allison1998nist}, Kubaschewski \cite{kubaschewski}, Barin \cite{barin1995thermochemical}, and Wagman \cite{wagman1982nbs}. The experimental enthalpy of formation ($\Delta H_f^\text{expt}$ at 298~K) of each compound considered is compiled in \textbf{Table {\ref{tab:energies}}}. For FeF\textsubscript{3}, and NiF\textsubscript{3}, we obtained $\Delta H_f^\text{expt}$ from other literature sources \cite{johnson1981enthalpy, solov2012standard}. Note that we determined the thermochemical data obtained through bomb calorimetry to be more reliable for FeF\textsubscript{3} compared to the values reported in the NIST-JANAF tables, as detailed in Section~{\ref{sec:results}}. In the absence of an extrapolated value (from experimental data) of $\Delta H_f^\text{expt}$ at 0 K, we considered the measured value of $\Delta H_f^\text{expt}$ at 298 K to be equivalent to 0 K \cite{aykol2014local}. Wherever possible, we compared the accuracy of property predictions with optimal \textit{U} values derived using both 298 K and 0 K experimental data.    

We approximated the theoretical oxidation enthalpy (\(\Delta H_\text{O}^\text{theo}\)), based on calculated 0 K DFT total energies (i.e., \(H \approx E\), thus ignoring \(p-V\) contributions), as \(\Delta H_\text{O}^\text{theo} = E_{\text{MF}_y}^{\text{DFT}+U} - E_{\text{MF}_x}^{\text{DFT}+U} - \frac{{(y-x)}}{2} E_{\text{F}_2}^{\text{DFT}}\), where DFT indicates either SCAN or r\textsuperscript{2}SCAN. \textit{U} = 0 simply indicates a SCAN or r\textsuperscript{2}SCAN calculation without any Hubbard \textit{U} correction. To find the appropriate \textit{U} for a given fluorination reaction, we varied the \textit{U} value till the \(\Delta H_\text{O}^\text{theo}\) matched the \(\Delta H_\text{O}^\text{expt}\). In case of TMs with multiple oxidation states and multiple fluorination reactions, such as Mn and Cr, we determined the optimal \textit{U} by averaging the appropriate \textit{U} values obtained for the individual fluorination reactions (see \textbf{Figure S2} of SI). 

For TMFs involving V, Fe, and Ni, we employed the linear response theory \cite{cococcioni2005linear} to estimate \textit{U} corrections as well, to provide a point-of-comparison to the experimental-data-based \textit{U} values. In linear response theory, the effective interaction parameter, \textit{U}, associated with a specific site, I, in a structure is given as, 



\begin{equation}
U = (\chi_0^{-1} - \chi^{-1})_{II}
\end{equation}

where \(\chi\) and \(\chi_0\) are the interacting (or SCF) and non-interacting (or non-self-consistent-field $-$NSCF) response functions, respectively, to an on-site applied perturbation potential (\(\alpha_I\)) \cite{cococcioni2005linear}. For our linear response calculations, we used a \(2\times2\times2\)  supercell for V-fluorides, and the corresponding conventional cells for Fe- and Ni-fluorides. We varied \(\alpha_I\) across -0.2 eV, -0.1 eV, +0.1 eV, and +0.2 eV for calculating the response functions.    

\subsection{Crystal structures and magnetic configurations}

\begin{table}[b]
\caption{\label{tab:energies}
Experimental enthalpy of formation ($\Delta H_f^\text{expt}$ at 298 K), space group, ICSD collection code, and experimental on-site magnetic moments of TMFs. The ± sign in the magnetic moment column indicates an antiferromagnetic configuration. In the absence of quantitative experimental magnetic moments, only the configuration-type such as AFM, NM, and FM are given. No experimental information on the magnetic configuration is available for CrF\textsubscript{4}. }
\begin{ruledtabular}
\begin{tabular}{ccccc}
  &\(\Delta H_f^\text{expt}\) & Space  & ICSD  & On-site  \\
 Compound &(eV/atom) &  group & collection  &  magnetic  \\
  & &   & code &  moment (\(\mu_{B}\)) \\
\hline
TiF\textsubscript{3} & -3.713 &  \textit{R$\bar{3}$cR} & 16649 & AFM \cite{sheets2023frustrated} \\
TiF\textsubscript{4} & -3.414 &  \textit{Pnma} & 78737 & NM \\
VF\textsubscript{3} & -3.272 &  \textit{R$\bar{3}$cR}& 30624 & ±2.00 \cite{wollan1958antiferromagnetic} \\
VF\textsubscript{4} & -2.904 &  \textit{P121/n1} & 65785 & ±1.00 \cite{gossard1974magnetic} \\
CrF\textsubscript{2} & -2.687 &  \textit{P121/n1} & 31827 & ±3.60 \cite{chatterji2011magnCrF2_CuF2} \\
CrF\textsubscript{3} & -2.997 &  \textit{R$\bar{3}$cR} & 25828 & ±3.00 \cite{wollan1958antiferromagnetic}] \\
CrF\textsubscript{4} & -2.581 &  \textit{P42/mnm} & 78778 & -- \\
MnF\textsubscript{2} & -2.954 &  \textit{P42/mnm} & 14142 & ±5.00 \cite{strempfer2004magnetic} \\
MnF\textsubscript{3} & -2.770 &  \textit{C12/c1} & 19080 & ±4.00 \cite{wollan1958antiferromagnetic} \\
MnF\textsubscript{4} & -2.239 &  \textit{I41/a} & 62068 & ±3.85 \cite{lutar1988krf2} \\
FeF\textsubscript{2} & -2.479 &  \textit{P42/mnm} & 9166 & ±3.75 \cite{yang2012structural} \\
FeF\textsubscript{3} & -2.628 \cite{johnson1981enthalpy} &  \textit{R$\bar{3}$cR} & 41120 & ±5.00 \cite{wollan1958antiferromagnetic}\\
CoF\textsubscript{2} & -2.323 &  \textit{P42/mnm} & 9167 & ±2.57 \cite{chatterji2010CoF2} \\
CoF\textsubscript{3} & -2.043 &  \textit{R$\bar{3}$cR} & 16672 & ±4.40 \cite{wollan1958antiferromagnetic} \\
NiF\textsubscript{2} & -2.268 &  \textit{P42/mnm}  & 9168 & ±1.99 \cite{chatterji2010magnetoelastic} \\
NiF\textsubscript{3} & -2.114 \cite{solov2012standard} & \textit{R$\bar{3}$R} & 87944 & FM \cite{shen1999structure} \\
CuF & -0.346 & \textit{F4$\bar{3}$m} & 52273 & NM \\
CuF\textsubscript{2} & -1.834 & \textit{P121/n1} & 71833 & ±0.73 \cite{fischer1974magnetic} \\
\end{tabular}
\end{ruledtabular}
\end{table}

\label{sec:structures}
\begin{figure*}
    \centering
    \includegraphics[width=1\textwidth]{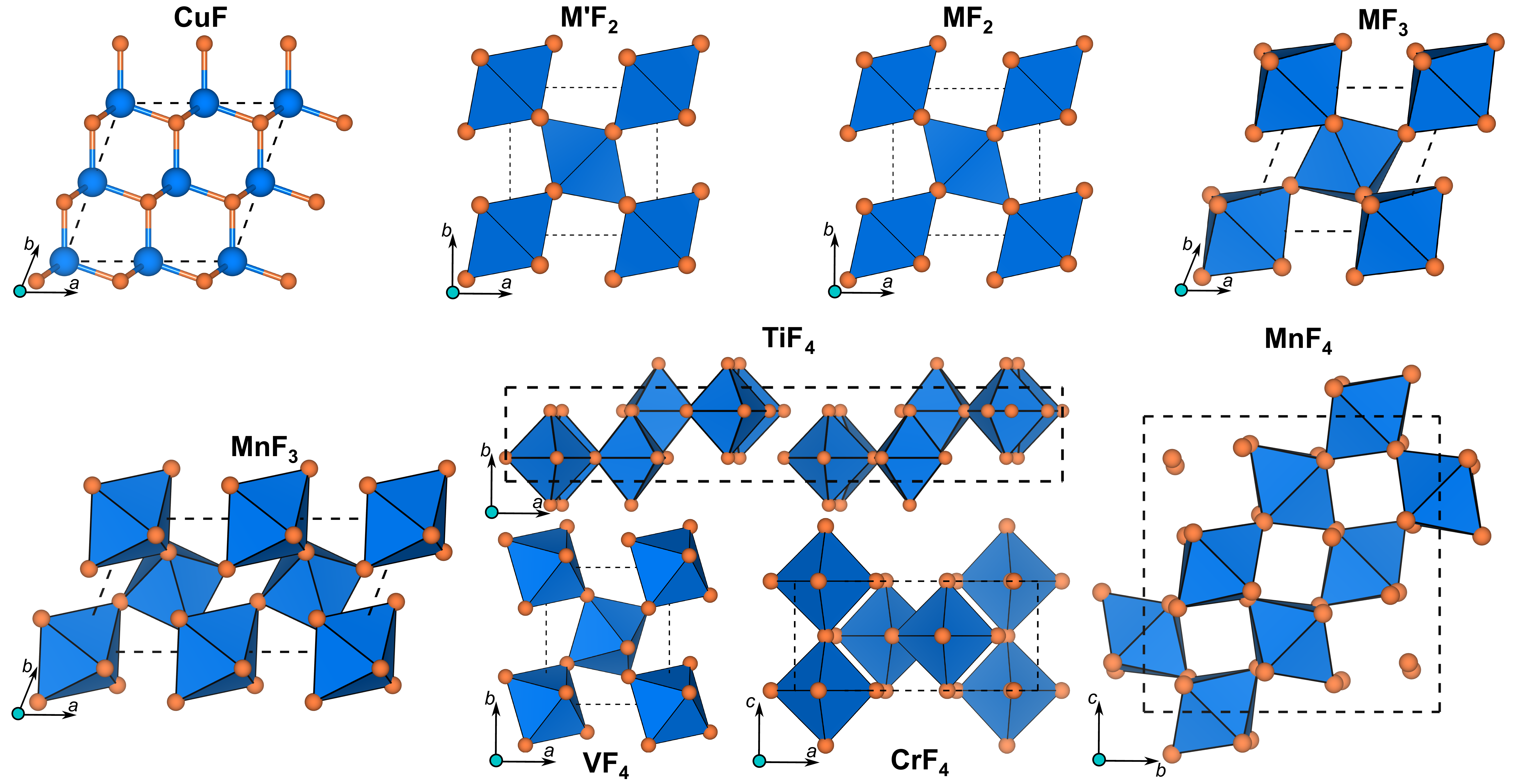}
    \caption{The initial crystal structures of 3\textit{d} TMFs. M$^\prime$ in M$^\prime$F\textsubscript{2} stands for Cr and Cu, while M in MF\textsubscript{2} signifies Mn, Fe, Co, or Ni. M in MF\textsubscript{3} indicates Ti, V, Cr, Fe, Co, or Ni. The 3\textit{d} TM and F atoms are represented by blue spheres/polyhedra and orange spheres, respectively. Dashed black lines signify the cell boundaries. }
    \label{fig:struct}
\end{figure*}

\textbf{Figure {\ref{fig:struct}}} displays the initial crystal structures of all 3\textit{d} TMFs that we examined in this study, as obtained from the ICSD. The space groups and ICSD collection codes of all structures are compiled in \textbf{Table {\ref{tab:energies}}}. Blue spheres/polyhedra in \textbf{Figure {\ref{fig:struct}}} indicate the TM/MF\textsubscript{6} octahedra while orange spheres indicate F atoms. The antiferromagnetic (AFM) and ferromagnetic (FM) configurations of TMFs that were considered in this work are compiled in \textbf{Figures~S2} and \textbf{S3} of the SI. We did not consider any specific magnetic ordering for CuF and TiF$_4$ owing to their non-magnetic (NM) $d^{\rm 10}$ and $d^{\rm 0}$ electronic configurations, respectively. 

CuF is known to crystallize in a cubic symmetry with the \(F\bar{4}3m\) space group and we show the \(2\times2\times2\) supercell of the primitive that we used in our calculations in \textbf{Figure {\ref{fig:struct}}} \cite{ebert1933kristallstrukturen}. The MF\textsubscript{2} (M = Mn, Fe, Co, or Ni) compositions crystallize in the well-known tetragonal rutile-type structure in the \textit{P42/mnm} space group \cite{baur1971rutile, uber1958rutile}, while CrF\textsubscript{2} and CuF\textsubscript{2} both exhibit a distorted rutile-type arrangement, characterized by their monoclinic \textit{P121/n1} space group \cite{proce1957, burns1991rietveld}. Most of the MF\textsubscript{3} (M = Ti, V, Cr, Fe, and Co) compositions crystallize in a rhombohedral structure within the \(R\bar{3}cR\) space group \cite{Siegel:a01775, jack1951crystal, knox1960structures, hepworth1957crystal, leblanc1985single}, with NiF\textsubscript{3} adopting a similar rhombohedral structure within the \(R\bar{3}R\) space group \cite{shen1999structure}. MnF\textsubscript{3} exhibits a monoclinic crystal structure within the \textit{C12/c1} space group, with \textbf{Figure {\ref{fig:struct}}} showing a \(2\times1\times1\) supercell to accommodate the A-type AFM ordering \cite{Hepworth:a01996}. MF\textsubscript{4} compositions exhibit diverse crystal structures, such as  orthorhombic (\textit{Pnma} space group) TiF\textsubscript{4} \cite{bialowons1995titantetrafluorid}, monoclinic (\textit{P121/n1}) VF\textsubscript{4} \cite{becker1990vanadiumtetrafluorid}, and tetragonal (\textit{P42/mnm} and \textit{I41/a}) CrF\textsubscript{4} and MnF\textsubscript{4}  \cite{kramer1995struktur, muller1987kristallstruktur}.  

We initialized the magnetic configuration of each structure according to its known experimental ground state configuration, wherever possible \cite{sheets2023frustrated, wollan1958antiferromagnetic, gossard1974magnetic, chatterji2011magnCrF2_CuF2, strempfer2004magnetic, lutar1988krf2, yang2012structural, johnson1981enthalpy, chatterji2010CoF2, chatterji2010magnetoelastic, solov2012standard, shen1999structure}, as listed in \textbf{Table {\ref{tab:energies}}} and depicted in \textbf{Figure~S2}. The presence of the $\pm$ sign or the AFM notation in \textbf{Table {\ref{tab:energies}}} indicate antiferromagnetic configurations. In general, all AFM TMFs examined in this work have similar (G-type or ↑↓↑↓) ordering, with the exception of MnF\textsubscript{3}, which exhibits an A-type (↑↑↓↓) ordering \cite{wollan1958antiferromagnetic}. Unlike other MF\textsubscript{3}, NiF\textsubscript{3} is usually represented as a Ni[NiF\textsubscript{6}] chemical formula to indicate the presence of Ni in the II and IV oxidation states. Notably, upon initialising the structure with both FM and AFM configurations, we obtained the FM configuration to be the ground state in NiF\textsubscript{3} with the SCAN functional, where the Ni atoms exhibited both II and IV oxidation states, in agreement with experiment \cite{shen1999structure}. Subsequently, we initialised NiF\textsubscript{3} in the FM configuration for all SCAN+\textit{U} and r\textsuperscript{2}SCAN(+\textit{U}) calculations.

The ground state magnetic configurations are not experimentally known for TiF\textsubscript{3}, CrF\textsubscript{4}, and MnF\textsubscript{4}. While TiF\textsubscript{3} is known to be AFM \cite{sheets2023frustrated}, the specific configuration of moments is not known. Therefore, we considered the G-type AFM ordering for TiF\textsubscript{3} as it is the only possible AFM arrangement within the two-Ti-atom conventional cell. Considering other complex AFM arrangements in TiF\textsubscript{3} will require larger supercells and significant computational expense. In the cases of CrF\textsubscript{4} and MnF\textsubscript{4}, we considered the FM and all possible AFM configurations within the conventional cell and determined the ground state using the SCAN functional. Specifically, we considered FM + two AFM configurations and FM + five AFM orderings for CrF\textsubscript{4} and MnF\textsubscript{4}, respectively (see \textbf{Figure~S3}). Subsequently, we initialised CrF\textsubscript{4} and MnF\textsubscript{4} in the SCAN-calculated ground state magnetic configuration for all SCAN+\textit{U} and r\textsuperscript{2}SCAN(+\textit{U}) calculations. 

\section{Results}
\label{sec:results}
\subsection{Fluorine binding}

To examine whether F\textsubscript{2} is under or overbound by SCAN and/or r\textsuperscript{2}SCAN, we evaluated the binding energy of the F\textsubscript{2} molecule with both functionals. In the case of SCAN, we calculate a binding energy of -1.57~eV, which is slightly lower but in close alignment with the experimental value of -1.64~eV \cite{barin1995thermochemical}, indicating a marginal underestimation (by 3.9\%). On the other hand, the r\textsuperscript{2}SCAN-calculated binding energy (-1.63~eV) is in excellent agreement with the experimental value, indicating no errors arising from the electronic description of the F\textsubscript{2} molecule. Thus, our binding energy calculations suggest that both functionals describe the F\textsubscript{2} molecule with sufficient accuracy, and any errors in calculated fluorination enthalpies within TMFs should predominantly arise from SIEs caused by delocalized \textit{d} electrons on the TM centers. 

\subsection{Fluorination energetics of TMFs}

\begin{figure*}
    \centering
    \includegraphics[width=1\textwidth]{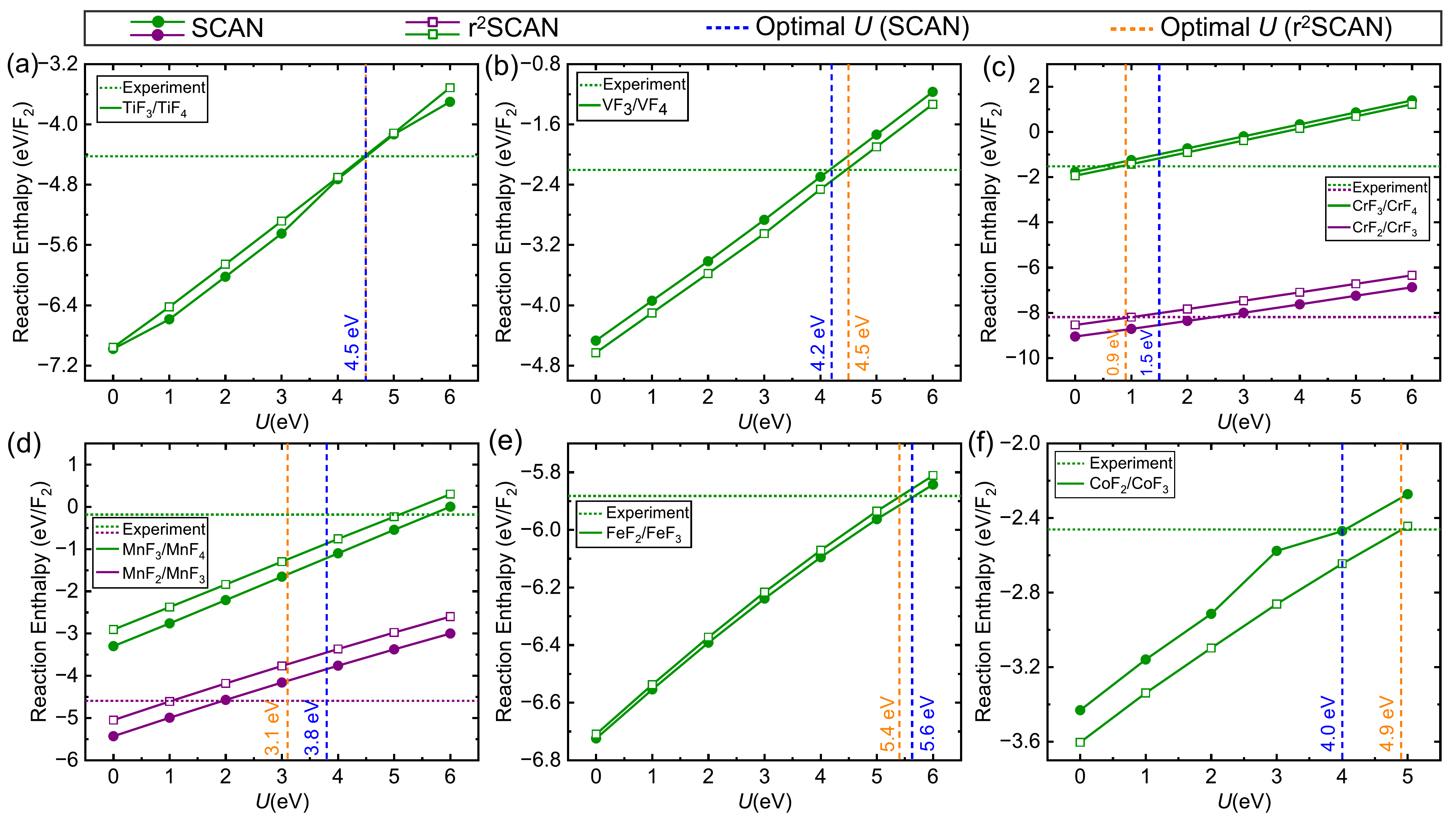}
    \caption{Oxidation (or fluorination) reaction enthalpy (solid line) variation with increasing \textit{U} within the SCAN+\textit{U} and r\textsuperscript{2}SCAN+\textit{U} functionals for (a) Ti, (b) V, (c) Cr, (d) Mn, (e) Fe, and (f) Co-fluorides. For systems with multiple possible oxidation states (e.g., Cr and Mn), each reaction is represented by a different colour. Horizontal dotted line of a given colour in each panel reflects the experimental oxidation enthalpy for the reaction considered (at 298~K). Vertical blue and orange dashed lines indicate optimal \textit{U} (magnitude is annotated as blue and orange text) for SCAN+\textit{U} and r\textsuperscript{2}SCAN+\textit{U} functional, respectively.}
    \label{fig:enthalpyu}
\end{figure*}

\textbf{Figure {\ref{fig:enthalpyu}}} displays the calculated fluorination (or oxidation) enthalpies of Ti\textsuperscript{+3}/Ti\textsuperscript{+4} (panel a), V\textsuperscript{+3}/V\textsuperscript{+4} (b), Cr\textsuperscript{+2}/Cr\textsuperscript{+3}/Cr\textsuperscript{+4} (c), Mn\textsuperscript{+2}/Mn\textsuperscript{+3}/Mn\textsuperscript{+4} (d), Fe\textsuperscript{+2}/Fe\textsuperscript{+3} (e) and Co\textsuperscript{+2}/Co\textsuperscript{+3} (f) as a function of \textit{U} in SCAN+\textit{U} (solid circles) and r\textsuperscript{2}SCAN+\textit{U} (open squares) frameworks. Except for Mn\textsuperscript{+2}/Mn\textsuperscript{+3} and Cr\textsuperscript{+2}/Cr\textsuperscript{+3} (represented by purple lines and symbols) all fluorination reactions in \textbf{Figure~{\ref{fig:enthalpyu}}} are indicated by green lines and symbols. The horizontal green and purple dotted lines indicate the experimental fluorination enthalpies (at 298 K) for the corresponding fluorination reactions. The optimal \textit{U} in each panel, which minimizes the error between the calculated and experimental fluorination enthalpies, is represented by a vertical dashed blue line for SCAN+\textit{U} and vertical dashed orange line for r\textsuperscript{2}SCAN+\textit{U}, respectively. 

Noticeably, the calculated fluorination enthalpies for most TMs evaluated using SCAN and r\textsuperscript{2}SCAN (\textit{U }= 0 eV) show significant discrepancies versus the corresponding experimental values. For example, the experimental fluorination enthalpy of Ti\textsuperscript{+3}/Ti\textsuperscript{+4} (-4.432 eV/F\textsubscript{2}) is estimated to be more negative (i.e., magnitude is overestimated) both by SCAN (-6.98 eV/F\textsubscript{2}) and r\textsuperscript{2}SCAN (-6.96 eV/F\textsubscript{2}). Similarly, the experimental fluorination enthalpies (in eV/F\textsubscript{2}) of V\textsuperscript{+3}/V\textsuperscript{+4} (-2.20), Cr\textsuperscript{+2}/Cr\textsuperscript{+3} (-8.18), Cr\textsuperscript{+3}/Cr\textsuperscript{+4} (-1.52), Mn\textsuperscript{+2}/Mn\textsuperscript{+3} (-4.60), Mn\textsuperscript{+3}/Mn\textsuperscript{+4} (-0.18), Fe\textsuperscript{+2}/Fe\textsuperscript{+3} (-5.88), and Co\textsuperscript{+2}/Co\textsuperscript{+3} (-2.46) are consistently overestimated by SCAN (-4.46, -9.04, -1.76, -5.44, -3.3, -6.72, -3.5 eV/F\textsubscript{2}, respectively) and r\textsuperscript{2}SCAN (-4.62, -8.54, -1.94, -5.04, -2.90, -6.72, -3.60 eV/F\textsubscript{2}, respectively). Thus, both SCAN and r\textsuperscript{2}SCAN overestimate experimental fluorination enthalpies for all TMs considered, which can be attributed to residual SIEs and erroneous description of the ground state electronic structures. Therefore, it is necessary to use the Hubbard \textit{U} corrected SCAN (SCAN+\textit{U)} or r\textsuperscript{2}SCAN (r\textsuperscript{2}SCAN+\textit{U}) frameworks to obtain more accurate estimations of TMF properties. 

Given that the magnitude of \textit{U} is not known \textit{a priori}, we determine the optimal \textit{U} for all 3\textit{d} TMs based on the corresponding fluorination energies among the binary fluorides. For example, in Ti, we estimate an optimal \textit{U} of 4.5 eV for both SCAN and r\textsuperscript{2}SCAN (\textbf{Figure~{\ref{fig:enthalpyu}}a}), which minimizes the error between experimental and calculated oxidation enthalpies for the Ti\textsuperscript{+3}/Ti\textsuperscript{+4} reaction. Similarly, we obtain an optimal \textit{U} of 4.2, 1.5, 3.8, 5.6, and 4.0 eV for V, Cr, Mn, Fe, and Co, respectively, with SCAN, while the corresponding \textit{U} values with r\textsuperscript{2}SCAN are 4.5, 0.9, 3.1, 5.4, and 4.9 eV (\textbf{Figure~{\ref{fig:enthalpyu}}b-f}). In the case of Mn, the optimal \textit{U} of 3.8~eV with SCAN and 3.1~eV with r\textsuperscript{2}SCAN are obtained by averaging \textit{U} for MnF\textsubscript{2} \(\rightarrow\) MnF\textsubscript{3} (1.9 eV with SCAN, 1.0~eV with r\textsuperscript{2}SCAN) and MnF\textsubscript{3} 
\(\rightarrow\) MnF\textsubscript{4} (5.6~eV, 5.1~eV) reactions (\textbf{Figure~S1}). Analogous averaging of \textit{U} values for the CrF\textsubscript{2} \(\rightarrow\) CrF\textsubscript{3} and CrF\textsubscript{3} \(\rightarrow\) CrF\textsubscript{4} reactions is used to determine the optimal \textit{U} for Cr with both SCAN and r\textsuperscript{2}SCAN. 

Notably, the optimal \textit{U} values obtained in the TMF chemical space are larger than the corresponding TMO chemical space \cite{swathilakshmi2023performance}, consistent with the higher degree of ionic bonding in fluorides than oxides, thereby requiring better electronic localization within the contracted 3\textit{d} orbitals. Also, the non-monotonic variation of optimal \textit{U}, across the 3$d$ series, in r\textsuperscript{2}SCAN is similar to SCAN, indicating that the differences between these two functionals are marginal with respect to their accuracy in calculating redox enthalpies. 

In the case of Ni and Cu fluorides, both metaGGA functionals underestimate and overestimate, respectively, the fluorination enthalpy compared to experimental values for the Ni\textsuperscript{+2}/Ni\textsuperscript{+3} and Cu\textsuperscript{+}/Cu\textsuperscript{+2} reactions (see \textbf{Figure S4}). The overestimation of fluorination enthalpy for the Cu\textsuperscript{+}/Cu\textsuperscript{+2} reaction by both SCAN and r\textsuperscript{2}SCAN is similar to our observation in the CuO/Cu\textsubscript{2}O system as well \cite{swathilakshmi2023performance}. Moreover, the addition of \textit{U} to both SCAN and r\textsuperscript{2}SCAN worsens the disparity between the calculated and experimental fluorination enthalpies for both Ni and Cu (\textbf{Figure S4}), unlike the behavior of other TMs in \textbf{Figure~{\ref{fig:enthalpyu}}}. As a result, we propose that the bare SCAN and r\textsuperscript{2}SCAN functionals provide better predictions of redox enthalpies in Ni and Cu-fluorides compared to the corresponding Hubbard \textit{U} corrected versions.

\subsection{Linear response theory calculations}

Note that the optimal \textit{U} calculated using experimental enthalpies at 0 K (see \textbf{Table~S2} of SI) and 298 K, are similar ($<$0.5~eV) or identical for Ti, Cr, Mn, and Co-fluorides with SCAN and r\textsuperscript{2}SCAN. Specifically, using 0~K experimental enthalpy data, the optimal \textit{U} with SCAN (r\textsuperscript{2}SCAN) are 4.5, 1.7, 3.9, and 3.7~eV (4.5, 1.1, 3.2, and 5.2~eV) for Ti, Cr, Mn and Co, respectively (\textbf{Table~S2}). These values are fairly similar to the corresponding optimal \textit{U} derived using 298~K enthalpy data with SCAN (r\textsuperscript{2}SCAN), namely, 4.5, 1.5, 3.8, and 4.0~eV (4.5, 0.9, 3.1, and 4.9~eV, \textbf{Figure~{\ref{fig:enthalpyu}}}) for Ti, Cr, Mn, and Co, respectively. 

However, the optimal \textit{U} for V- and Fe-fluorides can be significantly different ($>$1~eV), depending on whether 0~K or 298~K experimental enthalpies are considered. For instance, we obtain a \textit{U} value of 3.0~eV with experimental data at 0 K and 4.2~eV with data at 298 K for V-fluorides using SCAN, while the corresponding \textit{U} values with r\textsuperscript{2}SCAN are 3.3 (0 K) and 4.5~eV (298~K). Similarly, using 0~K and 298~K experimental data yield \textit{U} values of 3.6 (3.5) and 5.6 (5.4)~eV, respectively, with SCAN (r\textsuperscript{2}SCAN) for Fe-fluorides. 

\begin{figure*}
    \centering
    \includegraphics[width=1\textwidth]{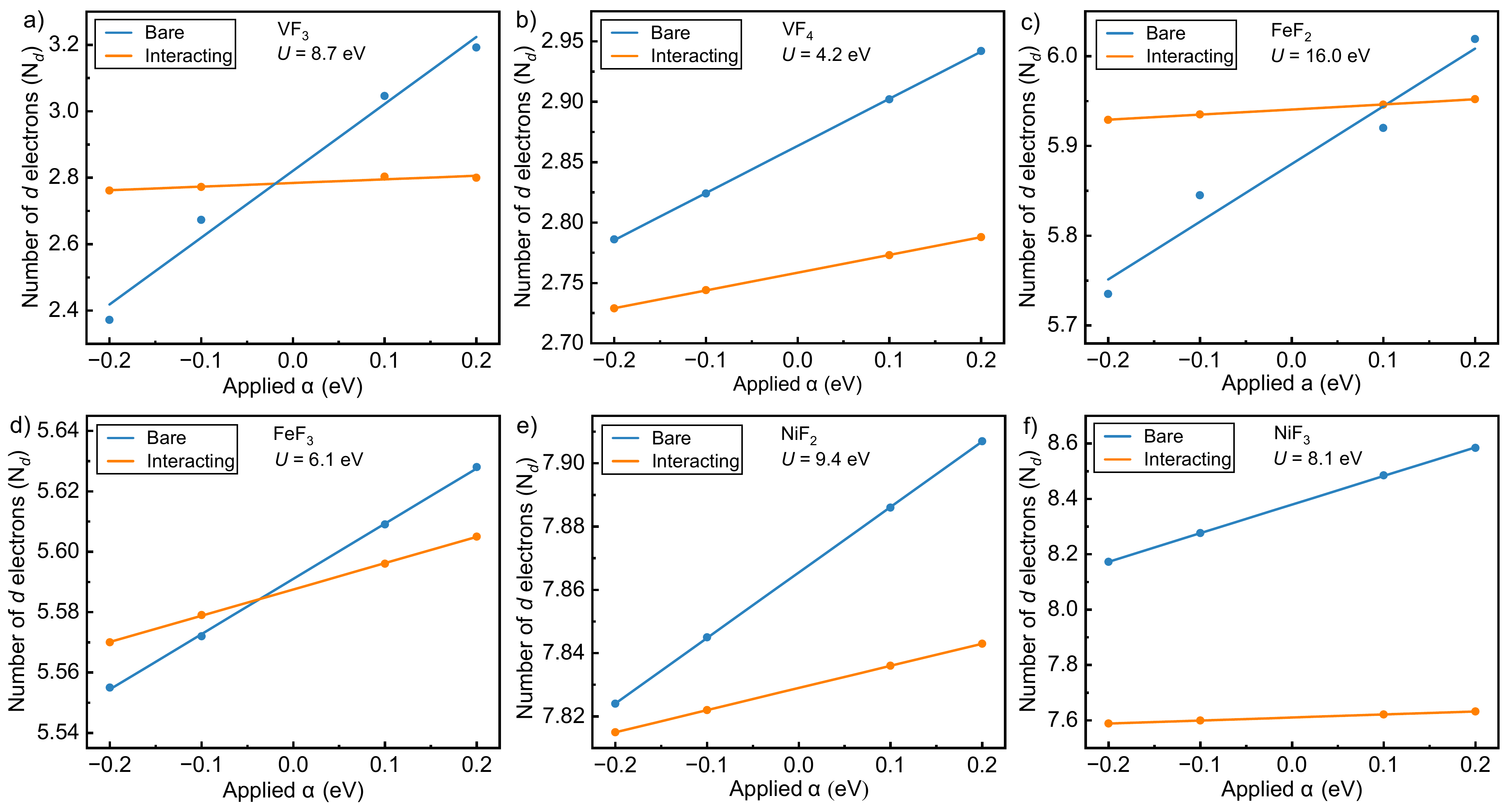}
    \caption{Determination of Hubbard \textit{U} using linear response theory and SCAN for (a) VF\textsubscript{3}, (b) VF\textsubscript{4}, (c) FeF\textsubscript{2}, (d) FeF\textsubscript{3}, (e) NiF\textsubscript{2}, and (f) NiF\textsubscript{3}. Each panel depicts the variation in the number of electrons (N\textit{\textsubscript{d}}) in \textit{d} orbitals of a single TM site as a function of the applied perturbation potential ($\alpha$ in eV). The blue line represents data obtained from non-self-consistent field (NSCF) or ``bare" calculations, while the orange line corresponds to SCF or ``interacting" calculations. The magnitude of \textit{U} is indicated as a text annotation in each panel.}
    \label{fig:lru}
\end{figure*}

Given this ambiguity over experimental enthalpies in V- and Fe-fluorides, we use the linear response theory to obtain a \textit{U} that is theory-derived and to provide a point-of-comparison to the experimental-derived \textit{U} values. Additionally, we used linear response theory in Ni-fluorides to verify the need for a \textit{U} correction, since experimental data indicates that a \textit{U} correction is not necessary for Ni, in contrast to observations in Ni-oxides \cite{swathilakshmi2023performance}. The NSCF or bare response and the SCF or interacting response with an applied potential (\(\alpha\)) and the SCAN functional, for VF\textsubscript{3}, VF\textsubscript{4}, FeF\textsubscript{2}, FeF\textsubscript{3}, NiF\textsubscript{2}, and NiF\textsubscript{3} are plotted in panels a-f of \textbf{Figure~{\ref{fig:lru}}}. The NSCF and SCF responses are quantified as the number of electrons in the \textit{d} orbitals (N\textit{\textsubscript{d}}). 

 While the linear-response-theory-derived \textit{U} for VF\textsubscript{3} (with SCAN) is 8.7 eV (\textbf{Figure~{\ref{fig:lru}}a}), which is higher than that either of the experimental-derived \textit{U} values (3.0-4.2 eV), the theory-derived \textit{U} for VF\textsubscript{4} is 4.2 eV (\textbf{Figure~{\ref{fig:lru}}b}), which matches with the experimental-derived \textit{U} using data at 298 K. Thus, the theory-derived \textit{U} value increases significantly (by $\sim$4.5 eV) with lowering the oxidation state of V from +4 to +3.  

Similarly, the theory-derived \textit{U} of Fe in FeF\textsubscript{2} and FeF\textsubscript{3} are 16 eV and 6.1 eV (\textbf{Figures~{\ref{fig:lru}}c} and \textbf{d}), respectively, where the latter is similar to the experimental-derived \textit{U} (5.6 eV), with the data at 298 K. As in V, the theory-derived \textit{U} in Fe also increases significantly (by $\sim$9.9 eV) with a reduction in Fe oxidation state. In the case of Ni, theory yields \textit{U} of 9.4 eV and 8.1 eV in NiF\textsubscript{2} and NiF\textsubscript{3} (\textbf{Figures~{\ref{fig:lru}}e} and \textbf{f}), respectively, significantly higher than the experimental-derived \textit{U} of 0~eV. The variation in theory-derived \textit{U} in Ni with reduction in Ni oxidation state is lower in Ni ($\sim$1.3 eV) compared to Fe and V, possibly because of the pairing of \textit{d} electrons in several orbitals resulting in lower differences in the SIEs incurred between Ni\textsuperscript{+2} and Ni\textsuperscript{+3}. Although we have performed linear response calculations with the SCAN functional, given the similarities in performance between SCAN and r\textsuperscript{2}SCAN, we believe that similar trends in theory-derived \textit{U} values will be encountered with r\textsuperscript{2}SCAN as well. 

Considering the substantial variation in theory-derived \textit{U} across different oxidation states in V, Fe, and Ni, and the large unphysical values of \textit{U} that we obtain for some oxidation states ($>$8 eV), we can conclude that linear response theory is not suitable for obtaining reliable \textit{U} values for TMFs. Note that linear response theory has been known to provide larger-than-necessary \textit{U} corrections in the TMO chemical space as well \cite{zhou2004first, jain2011formation, hsu2009first}. Given the unreliability of linear response theory, it does not provide a robust point-of-comparison for experimental-derived \textit{U} values, particularly for the cases of V- and Fe-fluorides. 

Nevertheless, we can use the following points to identify which experimental-derived \textit{U} value is better for V and Fe: \textit{i}) Both SCAN and r\textsuperscript{2}SCAN exhibit the highest optimal \textit{U} for Fe in TMOs given that Fe has the largest number of unpaired electrons in its +3 oxidation state, which is also quite stable compared to its +2 state \cite{swathilakshmi2023performance, gautam2018evaluating, long2020evaluating}. Also, both functionals display a lower optimal \textit{U} in V and Cr compared to Ti in TMOs \cite{swathilakshmi2023performance}. Given that fluorides are also ionically bonded compounds like oxides, we should expect similar trends to hold and the optimal \textit{U} to hit its maximum for Fe. \textit{ii}) We expect better localization of \textit{d} electrons in fluorides compared to oxides since F\textsuperscript{-} leads to more ionic bonds than O\textsuperscript{2-}. Hence, optimal \textit{U} values in fluorides should be larger than oxides. \textit{iii}) The calculated lattice parameters, on-site magnetic moments, and band gaps should have good agreement with experiments (see sub-sections below, \textbf{Table~S5} and \textbf{Figures~S13-14}). Thus, given the above considerations, we utilize the \textit{U} calculated using thermochemical data at 298~K for all further calculations and we believe that this \textit{U} value is more reliable than the one calculated using extrapolated thermochemical data at 0~K. 

\subsection{Lattice volumes}

\begin{figure*}
    \centering
    \includegraphics[width=1\textwidth]{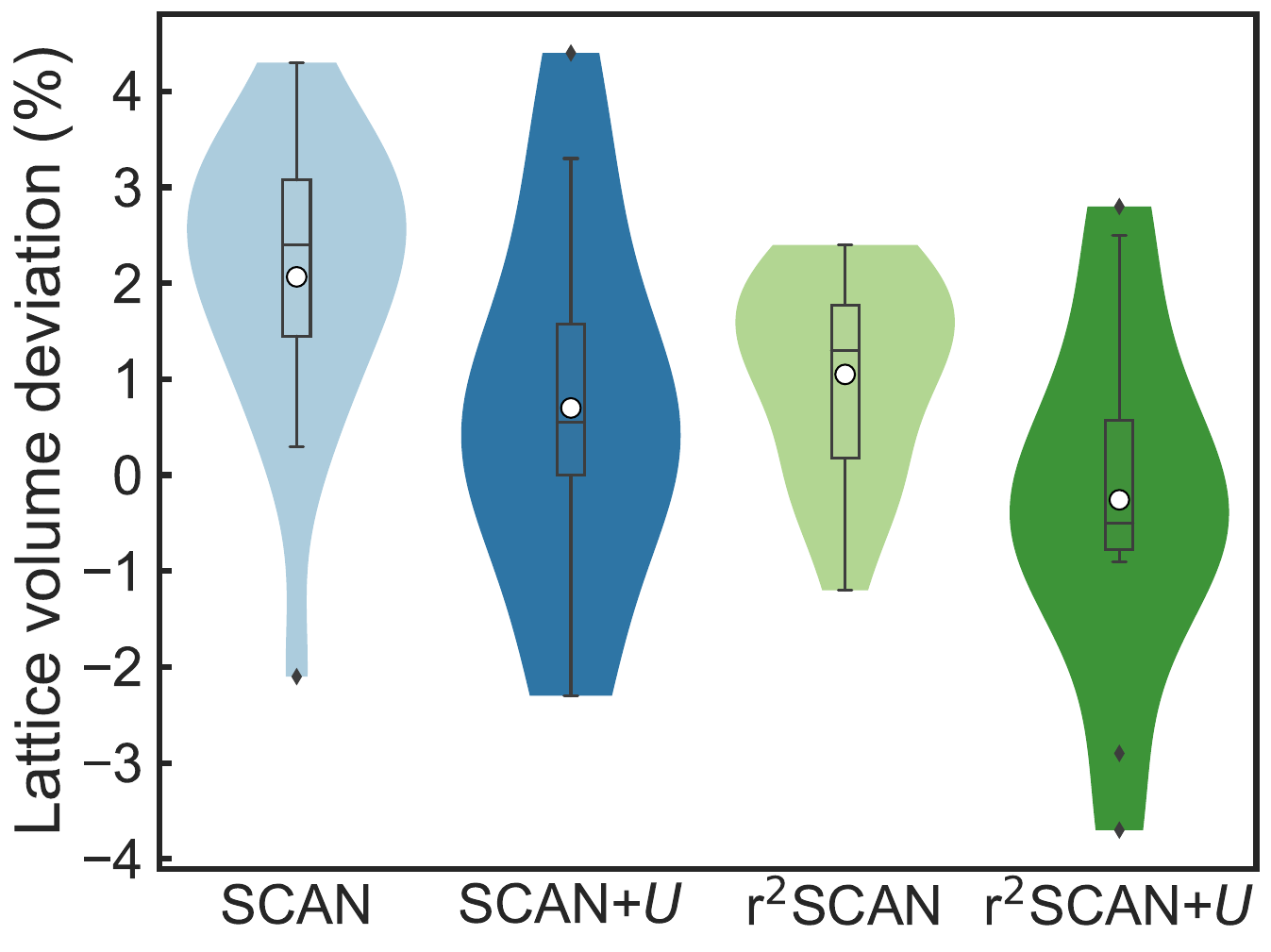}
    \caption{Violin plot presenting the percentage error in the deviation of SCAN (light blue), SCAN+\textit{U} (dark blue), r\textsuperscript{2}SCAN (light green), and r\textsuperscript{2}SCAN+\textit{U} (dark green) calculated lattice volumes with respect to the corresponding experimental volumes for different TMFs. Positive (negative) values on the y-axis indicate that the calculated volume is lower (higher) than the experimental value. The embedded inner boxes signify the lower to upper quartile range. Within each inner box, the mean and the median deviations are represented by the empty circle and the horizontal line, respectively. Black diamonds are outliers.}
    \label{fig:volume}
\end{figure*}

The SCAN(+\textit{U}) and r\textsuperscript{2}SCAN(+\textit{U}) calculated lattice parameters for all TMFs considered are compiled in \textbf{Table S3} of the SI. \textbf{Figure~{\ref{fig:volume}}} displays a violin plot of the percentage difference in calculated and experimental lattice volumes, for all four frameworks. Positive (negative) differences between the calculated and experimental lattice volume indicate that the calculation underestimates (overestimates) with respect to the experiment. The embedded inner boxes within each violin encompass values ranging from the lower to the upper quartiles, with the black diamonds indicating outliers. Mean and median of the percentage error is shown as the empty circle and the horizontal black line, respectively, within the embedded inner box for each functional.   

Overall, SCAN underestimates lattice volumes compared to experimental values, on average, with errors in the range of 0.3\% to 4.3\%, while the mean and median errors are 2.0\% and 2.4\%, respectively. The exception with SCAN is CuF, where SCAN overestimates the lattice volume by 2.1\%. Compared to SCAN, r\textsuperscript{2}SCAN provides lower lattice volume deviations, on average, with the calculated volumes underestimating experiments in the range of 0.9-2.4\%, with corresponding mean and median of 0.9\% and 1.3\%, respectively. Thus, r\textsuperscript{2}SCAN’s lattice volume predictions are on average better than SCAN in TMFs, consistent with trends observed by Kingsbury et al. \cite{kingsbury2022performance}. Note that there are several structures where r\textsuperscript{2}SCAN does overestimate experimental volumes, such as CrF\textsubscript{3} (0.1\%), FeF\textsubscript{3} (0.5\%), and CuF (1.2\%), compared to only CuF with SCAN.  

For several TMFs, adding the \textit{U} correction with both SCAN and r\textsuperscript{2}SCAN enhances the calculated lattice volumes, thereby reducing the extent to which volumes are underestimated by the non-corrected functionals, resulting in lower mean and median errors. For instance, SCAN+\textit{U} (r\textsuperscript{2}SCAN+\textit{U}) exhibits mean and median errors of 0.8\% (-0.2\%) and 1.1\% (-0.5\%), respectively, which are lower than the mean and median errors exhibited by SCAN (r\textsuperscript{2}SCAN). However, the range of errors observed in lattice volume calculations upon adding \textit{U} corrections also increases for both functionals, due to the presence of outliers. For example, the errors in calculated volumes range from -2.3\% to 3.3\% with SCAN+\textit{U}, and -0.9\% to 2.5\% for r\textsuperscript{2}SCAN+\textit{U}, which are higher than the range of errors observed with SCAN (0.3 to 4.3\%) and r\textsuperscript{2}SCAN (0.9 to 2.4\%), respectively. 

Notably, r\textsuperscript{2}SCAN+\textit{U} exhibits a smaller range of errors and lower (in magnitude) mean and median errors than SCAN+\textit{U}, similar to the comparison between r\textsuperscript{2}SCAN and SCAN. In terms of outliers, CoF\textsubscript{3} (error of 4.4\%) is a significant outlier with SCAN+\textit{U}, while CrF\textsubscript{4} (2.8\%), TiF\textsubscript{3}(-2.9\%), and VF\textsubscript{3} (-3.7\%) are the primary outliers with r\textsuperscript{2}SCAN+\textit{U}. Thus, the addition of Hubbard \textit{U} to both SCAN and r\textsuperscript{2}SCAN leads to inconsistent improvements in the lattice parameter estimations compared to the non-corrected functionals. Nevertheless, the absolute magnitude of errors made by all functionals against experimental values are quite reasonable ($<$5\% deviations), and the accuracy of one functional over another is marginal.

\subsection{On-site magnetic moments}

\begin{figure*}
    \centering
    \includegraphics[width=1\textwidth]{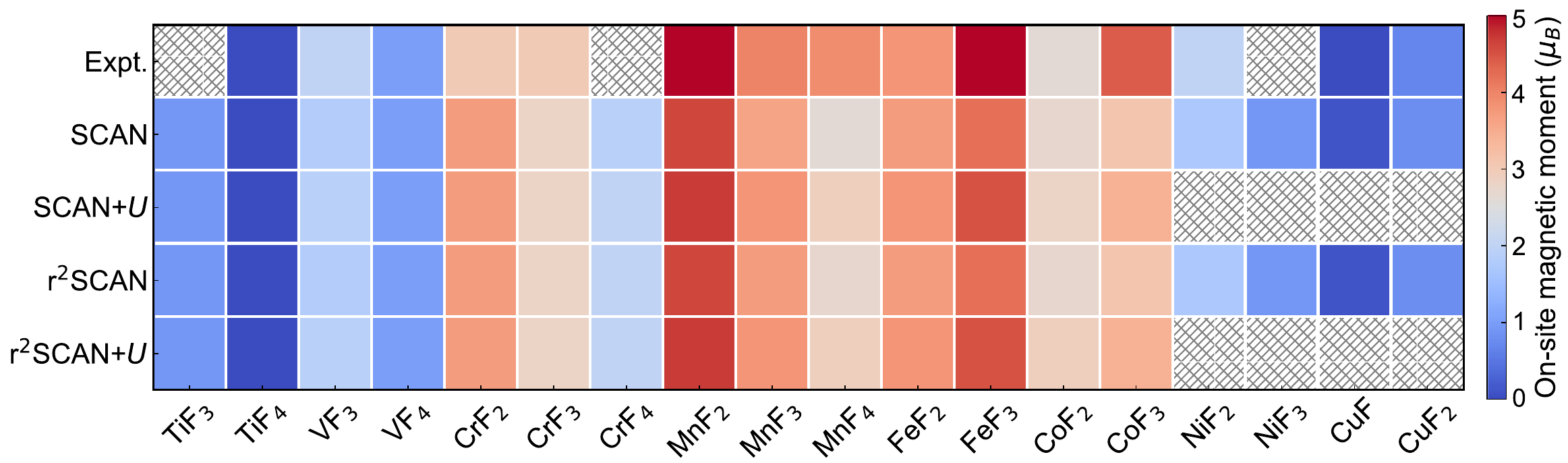}
    \caption{On-site magnetic moments (in units of  \(\mu_B\)) of 3\textit{d} TM in fluoride frameworks. Top (or first) row is experimental data, while second-fifth rows are calculated data, using SCAN (second), SCAN+\textit{U}(third), r\textsuperscript{2}SCAN (fourth), and r\textsuperscript{2}SCAN+\textit{U} (fifth). Hatched squares in the first row indicate absence of experimental data, while in third and fifth rows indicate absence of SCAN+\textit{U} and r\textsuperscript{2}SCAN+\textit{U} values, since a \textit{U} correction is not required for Ni and Cu fluorides.}
    \label{fig:magnetic}
\end{figure*}

\textbf{Figure~{\ref{fig:magnetic}}}depicts the experimental\cite{wollan1958antiferromagnetic, gossard1974magnetic, chatterji2011magnCrF2_CuF2, strempfer2004magnetic, lutar1988krf2, yang2012structural, chatterji2010CoF2, chatterji2010magnetoelastic, fischer1974magnetic} and calculated on-site magnetic moments of the 3\textit{d} TMs in the TMFs considered in this study, with the corresponding values also listed in \textbf{Table~S2}. The magnetic moments in \textbf{Figure~{\ref{fig:magnetic}}} are over a range of 0 \(\mu_B\) (dark blue) to 5 \(\mu_B\) (red). For fluorides without any experimental magnetic moment data, such as TiF\textsubscript{3}, CrF\textsubscript{4}, and NiF\textsubscript{3}, and for Ni- and Cu-fluorides that don’t require a \textit{U} correction, the lack of an experimental and/or calculated on-site magnetic moment is indicated by the hatched boxes in \textbf{Figure~{\ref{fig:magnetic}}}. Given the \textit{d}\textsuperscript{0} and \textit{d}\textsuperscript{10} electronic configurations on the Ti and Cu ions in TiF\textsubscript{4} and CuF, respectively, we have assigned the experimental on-site magnetic moments for these compounds to be 0. Note that the plotted on-site magnetic moments are averaged over all TMs present in a given structure. For AFM structures, we took the average of the absolute on-site magnetic moments on individual TMs.    

Interestingly, the calculated on-site magnetic moments vary quite marginally across all functionals used within each TMF. For instance, the calculated on-site magnetic moments of TiF\textsubscript{3} (0.9 \(\mu_B\)), TiF\textsubscript{4} (0.0 \(\mu_B\)), VF\textsubscript{4} (1.0 \(\mu_B\)), CrF\textsubscript{2} (3.7 \(\mu_B\)), CrF\textsubscript{3} (2.8 \(\mu_B\)), CrF\textsubscript{4} (2.0 \(\mu_B\)), NiF\textsubscript{2} (1.7 \(\mu_B\)), NiF\textsubscript{3} (1.6 \(\mu_B\)), CuF (0.1 \(\mu_B\)), and CuF\textsubscript{2} (0.8 \(\mu_B\)) are identical for all XC frameworks. There are also structures where the calculated moments are identical or similar for SCAN and r\textsuperscript{2}SCAN, and/or for SCAN+\textit{U} and r\textsuperscript{2}SCAN+\textit{U}. For example, SCAN and r\textsuperscript{2}SCAN calculated magnetic moments of VF\textsubscript{3} (1.8 \(\mu_B\)), MnF\textsubscript{2} (4.6 \(\mu_B\)), FeF\textsubscript{2} (3.7 \(\mu_B\)), FeF\textsubscript{3} (4.2 \(\mu_B\)), CoF\textsubscript{2} (2.7 \(\mu_B\)), and CoF\textsubscript{3} (3.1 \(\mu_B\)) are identical, while for MnF\textsubscript{3} and MnF\textsubscript{4}, the difference is marginal (by $\sim$0.1 \(\mu_B\)). Similarly, SCAN+\textit{U} and r\textsuperscript{2}SCAN+\textit{U} calculated on-site magnetic moments are identical for VF\textsubscript{3} (1.9 \(\mu_B\)), MnF\textsubscript{2} (4.7 \(\mu_B\)), MnF\textsubscript{3} (3.8 \(\mu_B\)), MnF\textsubscript{4} (2.9 \(\mu_B\)), FeF\textsubscript{2} (3.8 \(\mu_B\)), and FeF\textsubscript{3} (4.5 \(\mu_B\)), with a marginal difference of 0.1 \(\mu_B\) in CoF\textsubscript{3} ($\sim$3.4 \(\mu_B\)).  

In terms of comparison to experimental data, the calculated on-site magnetic moments of most TMFs is in good agreement with experimental values, across the four XC frameworks, with computed values marginally underestimating experiments. As an exception, the calculated moments of MnF\textsubscript{4} (2.6-2.9 \(\mu_B\)) and CoF\textsubscript{4} (3.1-3.4 \(\mu_B\)) are significantly less than the experimental value of 3.9 \(\mu_B\) and 4.4 \(\mu_B\), respectively, suggesting that all four XC frameworks favour a different spin configuration of Mn and Co as the electronic ground states compared to the experimental observations. Overall, the predictions of on-site magnetic moments do not change significantly ($<$0.3 \(\mu_B\)) across the four XC frameworks for all the TMFs considered, highlighting similar levels of accuracy in predicting magnetic moments by all four XC frameworks compared to experiments.

\subsection{Band gaps}

\begin{figure*}
    \centering
    \includegraphics[width=1\textwidth]{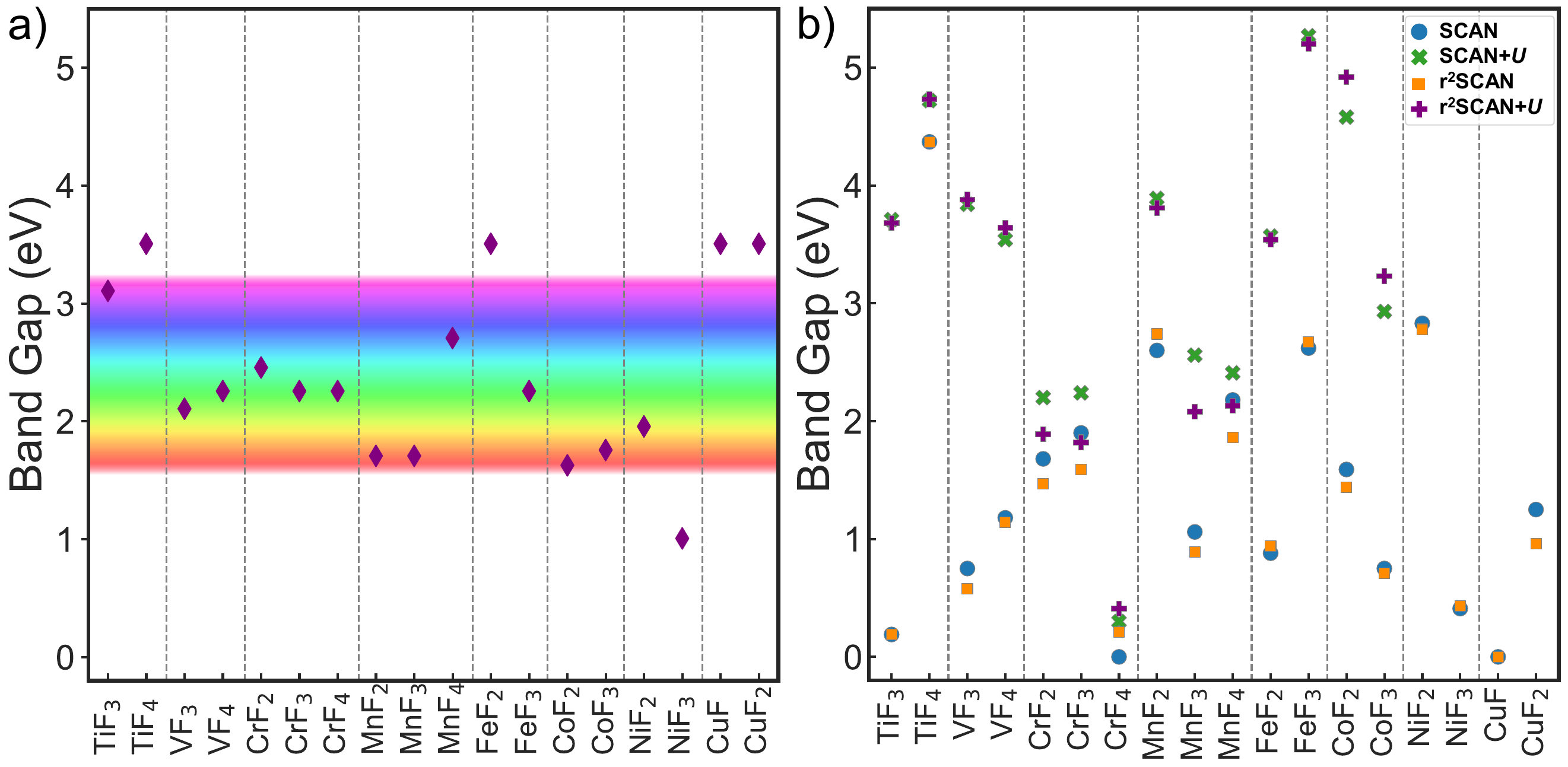}
    \caption{a) The estimated experimental band gap (purple diamond symbol) of various transition metal fluorides, based on their observed powder colors. The visible light spectrum (1.6-3.2 eV) is highlighted. A material with black (white) powder colour is categorized as having a low band gap or metallic (high band gap or insulating) character, with band gap of $<$ 1.6 eV ($>$ 3.2 eV). b) Schematic showing the calculated band gaps of TMFs using SCAN (blue symbols), SCAN+\textit{U} (green), r\textsuperscript{2}SCAN (orange), r\textsuperscript{2}SCAN+\textit{U} (purple) frameworks.}
    \label{fig:bandgap}
\end{figure*}

To check the performance of the XC frameworks in predicting electronic properties, we calculated the electronic DOS for all the TMFs considered. All calculated band gaps are compiled in \textbf{Figures S5-S12} in SI, with the corresponding band gap values listed in \textbf{Table~S2}. Note that experimental measurements of band gaps, either via optical or photoemission/inverse photoemission spectroscopy, in binary TMFs are quite scarce. However, the colors of the powdered samples of all considered TMFs have been documented in literature \cite{lide2004crc, kraus2020synthesis, nikitin2019thermochemistry}. Given the relationship between band gap and the color of a given compound \cite{hummel2011electronic}, we approximated the experimental band gaps of all TMFs, and used these for a qualitative comparison with the calculated band gaps. 

The approximate experimental band gaps of TMFs are indicated by purple diamonds in \textbf{Figure~{\ref{fig:bandgap}}a}, while the background indicates the visual spectrum (1.6-3.2 eV). We assumed all black (white) powder colored samples to exhibit low (high) band gap, i.e., $<$ 1.6 eV ($>$ 3.2 eV) eV, and specifically assigned a band gap value of 1 (3.5) eV. Note that all TMFs considered, except NiF\textsubscript{3}, TiF\textsubscript{4}, FeF\textsubscript{2}, CuF, and CuF\textsubscript{2}, should have a band gap in the 1.6-3.2 eV range, given their observed powder colors. Thus, except NiF\textsubscript{3} (which is a black powder), all TMFs considered are expected to be semiconductors or insulators. The literature data on the colors and our numerically approximated band gaps are tabulated in \textbf{Table~S4}.    

The calculated band gaps of the TMFs, using SCAN (blue circles), SCAN+\textit{U} (green crosses), r\textsuperscript{2}SCAN (orange squares), and r\textsuperscript{2}SCAN+\textit{U} (purple pluses) are displayed in \textbf{Figure~{\ref{fig:bandgap}}b}. The band gap of SCAN+\textit{U} and r\textsuperscript{2}SCAN+\textit{U} for Ni- and Cu-fluorides are not included, as the addition of \textit{U} is not required. Importantly, the SCAN and r\textsuperscript{2}SCAN calculated band gaps are quite similar for several TMFs ($<$ 1.1 eV deviation), and are in the range of $\sim$0-2.8 eV, except for TiF\textsubscript{4} with a significantly higher band gap of $\sim$4.4 eV. 

The band gaps of all TMFs, calculated using SCAN+\textit{U} (r\textsuperscript{2}SCAN+\textit{U}) lie within a range of $\sim$2.2-4.7 eV ($\sim$1.8-4.7 eV), except the low band gap of CrF\textsubscript{4} ($\sim$0.3 eV with SCAN+\textit{U} and $\sim$0.4 eV with r\textsuperscript{2}SCAN+\textit{U}). The trends in band gap estimates are quite similar among SCAN+\textit{U} and r\textsuperscript{2}SCAN+\textit{U}, with deviations $<$1.3 eV, similar to the observed trends among SCAN and r\textsuperscript{2}SCAN calculations. Notably, upon \textit{U} addition, the calculated band gaps of TMFs significantly increased compared to the corresponding non-\textit{U}-corrected versions, consistent with observations in other chemistries \cite{swathilakshmi2023performance, gautam2018evaluating, long2020evaluating, devey2009electronic, rodriguez2020first, seo2015calibrating}. 

Clearly, the estimated empirical band gap of most TMFs tend to fall within the range between the (underestimated) predictions of SCAN/r\textsuperscript{2}SCAN and the (overestimated) predictions of SCAN+\textit{U}/r\textsuperscript{2}SCAN+\textit{U}, as illustrated in \textbf{Figure~{\ref{fig:bandgap}}}. Specifically, we find that both SCAN and r\textsuperscript{2}SCAN underestimate the gap of several compounds, namely TiF\textsubscript{3}, VF\textsubscript{3}, VF\textsubscript{4}, CrF\textsubscript{4}, FeF\textsubscript{2}, CuF, and CuF\textsubscript{2}, with deviations from empirical estimates exceeding $\sim$1.0 eV. MnF\textsubscript{2} is an exception to this trend, as its band gap is overestimated (by $\sim$1.0 eV) by both SCAN and r\textsuperscript{2}SCAN. In some TMFs, both SCAN and r\textsuperscript{2}SCAN yield accurate gaps (with marginal underestimation), as in TiF\textsubscript{4}, CrF\textsubscript{3}, MnF\textsubscript{3}, MnF\textsubscript{4}, FeF\textsubscript{3}, CoF\textsubscript{2}, NiF\textsubscript{2}, and NiF\textsubscript{3}, with deviations $<$0.6 eV from empirical estimates. 

In several structures, the band gap calculated with the \textit{U}-corrected framework is significantly higher than the empirical estimates, which is unusual for a ground state DFT+\textit{U} theory. For example, both SCAN+\textit{U} and r\textsuperscript{2}SCAN+\textit{U} overestimate the gap (by $\sim$1.3~eV) in VF\textsubscript{3}, VF\textsubscript{4}, MnF\textsubscript{2}, FeF\textsubscript{3}, CoF\textsubscript{2}, and CoF\textsubscript{3}. CrF\textsubscript{4} is an exception in both \textit{U} corrected frameworks, given the significantly underestimated gap (by $\sim$1.9 eV). In some structures, both SCAN+\textit{U} and r\textsuperscript{2}SCAN+\textit{U} demonstrate reasonable accuracy, with marginal overestimation, in predicting gaps, including TiF\textsubscript{3}, TiF\textsubscript{4}, CrF\textsubscript{2}, CrF\textsubscript{3}, MnF\textsubscript{3}, MnF\textsubscript{4}, and FeF\textsubscript{2}, where deviations from empirical values are $< \sim$0.6 eV. 

Given the scarcity of experimentally measured band gaps in TMFs, it remains challenging to definitively declare one functional as more accurate than the other, emphasizing the need for additional experimental data for a comprehensive assessment. It is important to note that the \textit{U} parameter optimized with respect to experimental enthalpies may not result in accurate band gaps. Also, both DFT and DFT+\textit{U} frameworks are ground state theories and band gap predictions are expected to underestimate experimental values. Given that there are several structures where the \textit{U} corrected frameworks unphysically overestimate the empirical gaps, while the non-\textit{U}-corrected frameworks consistently underestimate the empirical values, we recommend using either SCAN or r\textsuperscript{2}SCAN for band gap calculations in TMFs, till better experimental data is available.

\subsection{Transferability check}

\begin{table}[b]
\caption{\label{tab:voltages}
Calculated average voltages using SCAN, SCAN+\textit{U}, r\textsuperscript{2}SCAN, r\textsuperscript{2}SCAN+\textit{U} in Mn, Fe, Co, and Ni fluorides. Expt. indicates the experimental voltage measured in the Na\textsubscript{x}FeF\textsubscript{3} system (0$\leq$x$\leq$1) \cite{dimov2013transition}.  V$^{\prime}$ denotes the voltage calculated at a composition between Na\textsubscript{0.75}MF\textsubscript{3} and NaMF\textsubscript{3} (M = Mn, Co, and Ni), or for partial Na extraction. V$^{\prime\prime}$ represents voltage calculated between compositions of MF\textsubscript{3} and NaMF\textsubscript{3} (M = Mn, Fe, Co, and Ni), or for full Na extraction.}

\begin{ruledtabular}
\begin{tabular}{cccc}
Compositions& Source & \multicolumn{2}{l}{Voltage (V)} \\
\cline{3-4}
(Space group) &   &  \\
 &  & V$^{\prime}$ & V$^{\prime\prime}$ \\
 \hline
  Na\textsubscript{x}MnF\textsubscript{3}& SCAN & 2.8 & 3.2 \\
  (\textit{Pnma})& SCAN+\textit{U} & 4.0 & 4.1 \\
  & r\textsuperscript{2}SCAN & 2.3 & 2.7 \\
  & r\textsuperscript{2}SCAN+\textit{U} & 3.3 & 3.5 \\
\hline
Na\textsubscript{x}FeF\textsubscript{3} & Expt. &  & 3.2 \\
  (\textit{Pnma}) & SCAN &  & 2.8 \\
  & SCAN+\textit{U} &  & 3.1 \\
  & r\textsuperscript{2}SCAN &  & 2.8 \\
  & r\textsuperscript{2}SCAN+\textit{U} &  & 3.1 \\
  \hline
   Na\textsubscript{x}CoF\textsubscript{3}& SCAN & 4.1 & 4.5 \\
  (\textit{Pnma})& SCAN+\textit{U} & 5.0 & 5.0 \\
  & r\textsuperscript{2}SCAN & 3.2 & 3.8 \\
  & r\textsuperscript{2}SCAN+\textit{U} & 4.3 & 4.3 \\
\hline
   Na\textsubscript{x}NiF\textsubscript{3}& SCAN & 4.9 & 5.1 \\
  (\textit{Pnma})& r\textsuperscript{2}SCAN & 4.2 & 4.5 \\
\end{tabular}
\end{ruledtabular}
\end{table}

It is important to verify the transferability of the optimal \textit{U} determined using experimental data in binary systems to other binary or higher-component fluorides that were not taken into account during the determination of the optimal \textit{U}. However, the lack of reliable experimental data in higher-component fluorides, except the case of Fe-fluorides, provides a significant challenge in our efforts to investigate the transferability of \textit{U} values. Consequently, we primarily focus on checking the transferability of the optimal \textit{U} for Fe in Fe-fluorides. We utilize the experimentally measured topotactic Na (de)intercalation voltage ($\sim$3.2~V vs. Na\cite{dimov2013transition}) from NaFeF\textsubscript{3}, i.e., for the full extraction of Na, or NaFeF\textsubscript{3} $\rightarrow$ Na + FeF\textsubscript{3}, as a point-of-comparison with calculated values. Structure of NaFeF\textsubscript{3} is displayed in \textbf{Figure S15}. A description of calculating the average voltage from experimental enthalpies is given in the SI.


Importantly, the calculated enthalpy for the Na + FeF\textsubscript{3} \(\rightarrow\) NaFeF\textsubscript{3} with both SCAN and r\textsuperscript{2}SCAN is -2.8 eV. This reaction enthalpy is equivalent to an average Na (de)intercalation voltage of 2.8~V vs. Na, which is $\sim$13\% lower than the experimental value. Remarkably, the average Na (de)intercalation voltage calculated with SCAN+\textit{U}/r\textsuperscript{2}SCAN+\textit{U}, with our optimal \textit{U} values, is 3.1~V, in excellent agreement with the experimental voltage. Thus, we believe that our optimal \textit{U} value, especially in the case of Fe, should be suitable in modelling redox reactions across other Fe-fluorides. 

Demov et al.\cite{dimov2013transition} measured voltages for partial extraction of Na (from x~=~1) in Na\textsubscript{x}MnF\textsubscript{3}, Na\textsubscript{x}CoF\textsubscript{3}, and Na\textsubscript{x}NiF\textsubscript{3}, all of which adopt structures similar to NaFeF\textsubscript{3}. Partial desodiation in the Mn, Co, and Ni fluorides also caused the formation of some impurity phases. Since modelling partial Na extraction will require considering various Na-vacancy configurations, and modelling impurities is not a trivial task computationally, we are unable to rigorously validate the measured partial desodiation voltages from experiment. However, we have calculated the average (de)intercalation voltage of Na, over the entire Na composition range (i.e., 0$\leq$x$\leq$1), and over a partial composition range (0.75$\leq$x$\leq$1) in the Mn, Co, and Ni fluorides with all four XC frameworks. The calculated voltages are compiled in \textbf{Table~{\ref{tab:voltages}}}. 

From the calculated trends in voltages, we suspect that r\textsuperscript{2}SCAN+\textit{U} can yield better accuracy than the other XC frameworks for the following reasons. First, SCAN+\textit{U}, and also SCAN to an extent, has been shown to overestimate intercalation voltages in oxides,\cite{long2021assessing} and we can expect similar behavior to hold for fluorides. Second, while intercalation voltages, across the entire Na content, are expected to increase from Fe to Co, a $\sim$2~V increase predicted by SCAN(+\textit{U}) is likely not physical.

\section{Discussion}
In this work, we have evaluated the accuracy of metaGGA (i.e., SCAN and r\textsuperscript{2}SCAN) functionals in describing the redox thermodynamics, lattice parameters, on-site magnetic moments, and band gaps of TMFs. While both SCAN and r\textsuperscript{2}SCAN don't exhibit a significant error in the binding of F\textsubscript{2} molecule, we found both functionals to exhibit significant errors in fluorination reaction enthalpies compared to experiments (\textbf{Figure~\ref{fig:enthalpyu}}). Such errors in fluorination (i.e., oxidation) enthalpies can be attributed to SIEs of the 3$d$ electrons, necessitating the addition of \textit{U} corrections to both functionals. Subsequently, we determined optimal \textit{U} values based on experimental fluorination enthalpies for both SCAN and r\textsuperscript{2}SCAN. Importantly, we found marginal changes in lattice parameter and on-site magnetic moment predictions, while band gaps increased significantly upon adding the \textit{U} correction to both SCAN and r\textsuperscript{2}SCAN (\textbf{Figures~{\ref{fig:volume}}}, \textbf{\ref{fig:magnetic}}, and \textbf{\ref{fig:bandgap}}, and \textbf{Table~S3}). Finally, we looked at predictions of average Na intercalation voltages in ternary fluorides as transferability checks of our optimal \textit{U} values in Mn, Fe, Co, and Ni systems (\textbf{Table~{\ref{tab:voltages}}}).  

Based on experimental enthalpies, we obtained an optimal \textit{U} of 4.5 (Ti), 4.2 (V), 1.5 (Cr), 3.8 (Mn), 5.6 (Fe), 4.0 eV (Co) with SCAN+\textit{U}, and 4.5 (Ti), 4.5 (V), 0.9 (Cr), 3.1 (Mn), 5.4 (Fe), 4.9 eV (Co) with r\textsuperscript{2}SCAN+\textit{U}. Also, we determined that the \textit{U} correction is not required for Ni and Cu-fluorides, based on the (dis)agreement with experimental enthalpies. Notably, we did not obtain physically reasonable \textit{U} values, particularly for the lower oxidation states, in V, Fe, and Ni-fluorides using linear response theory (\textbf{Figure~{\ref{fig:lru}}}). Such unphysical \textit{U} values indicate potential limitations in the applicability of the linear response theory with metaGGA functionals. Note that there are other ways of reducing SIEs in semi-local functionals, such as the DFT+\textit{U}+\textit{V} framework \cite{mahajan2021importance, timrov2022accurate,ricca2020phyrevl, tancogne2020parameter}, which require more tunable parameters that are not known \textit{a priori} compared to the DFT+\textit{U} framework.  

Optimal \textit{U} values in this work are nominally higher than the reported values derived from oxidation enthalpies for oxides \cite{gautam2018evaluating,long2020evaluating,swathilakshmi2023performance}. This is expected given that fluorides are more ionic than oxides, resulting in more localized \textit{d} electrons, thus making fluorides more susceptible to SIEs. However, the exception to this trend is Ni, where we do not observe the need for a \textit{U} correction using redox enthalpies (\textbf{Figure~S4}). One reason could be the low band gap of NiF\textsubscript{3} (black powder color experimentally), giving rise to an electronic ground state that has delocalized \textit{d} electrons, which is better described by SCAN and r\textsuperscript{2}SCAN. Another possibility is a one-off cancellation of SIEs for both SCAN and r\textsuperscript{2}SCAN while taking energy differences between NiF\textsubscript{2} and NiF\textsubscript{3}. Nevertheless, more experimental data, both thermochemical and electronic, is required to fully understand the source of this discrepancy.

For lattice volume and on-site magnetic moment predictions, there is practically no difference between Hubbard \textit{U} corrected and non-corrected functionals. On the surface, this may give an indication that the residual SIEs don’t really impact property predictions in the non-corrected functionals. However, the significant variation in the redox enthalpies indicate that even marginal changes in the electronic structure or geometry can have a significant impact in the redox behavior, in the TMF chemical space. 

The lack of reliable experimental data, specifically thermochemical data, makes  comparisons with theoretical predictions difficult. For example, a cursory usage of thermochemical data given by Aykol and Wolverton \cite{aykol2014local} and tabulated in JANAF \cite{allison1998nist} or Barin \cite{barin1995thermochemical} tables can give rise to dramatically different \textit{U} values. This is at least the case for V and Fe fluorides, where we obtain significantly different \textit{U} based on thermochemical data at 0 K and 298 K (\textbf{Table~S2}). The discrepancies observed in the experimental data between 0~K and 298~K may be attributed to erroneous measurement at 298~K, spurious extrapolation of data to 0~K, and/or the occurrence of an unreported phase transitions in the bulk fluorides. 

Similarly, it is hard to benchmark the band gaps in TMFs since robust measurements are not available. Note that colors of powdered samples can also be influenced by defects present in the sample – hence any band gap estimate can represent a significant approximation. This may be a source of error for cases where experimental band gaps (based on colors) do not even align qualitatively with computed band gaps (e.g., CuF, CuF\textsubscript{2}, and CrF\textsubscript{4}). Thus, more experimental data is required to increase confidence in theoretical predictions. An alternative is to employ computationally expensive but accurate techniques, such as GW calculations \cite{tiago2004effect, oshikiri1999band, aryasetiawan1995electronic}, to get an estimate of band gaps and examine their comparison with DFT+\textit{U} values.

\section{Conclusion}

TMFs, which have a wide range of applications including energy storage, catalysis, and magnetic devices, are highly-correlated electronic systems, which are susceptible to SIEs when described using semi-local metaGGA functionals, such as SCAN and r\textsuperscript{2}SCAN. Hence, we have systematically examined the accuracy of SCAN and r\textsuperscript{2}SCAN functionals in estimating the redox enthalpies, lattice parameters, on-site magnetic moments, and band gaps of binary TMFs. Importantly, we revealed that metaGGA functionals do overestimate (i.e., more negative) fluorination enthalpies among binary TMFs, which can be primarily attributed to the SIEs among $d$ electrons since both SCAN and r\textsuperscript{2}SCAN bind F\textsubscript{2} precisely. Given that SIEs can be mitigated by the addition of a Hubbard \textit{U} correction, we subsequently derived optimal \textit{U} values for different TMs based on experimental fluorination enthalpies. While the \textit{U}-corrected frameworks increased the calculated band gaps significantly compared to the non-\textit{U}-corrected functionals, the lattice parameters and on-site magnetic moments were only marginally different. Also, we examined the transferability of the optimal \textit{U} values determined in this work via comparison with available Na-intercalation voltages in ternary TMFs, in Mn, Fe, Co, and Ni systems. Additionally, we observed that the linear response theory, when used in conjunction with metaGGA functionals, can give rise to unphysical \textit{U} corrections, particularly for TMs with low oxidation states. Overall, our study signifies the importance of adding an optimal \textit{U} correction to both SCAN and r\textsuperscript{2}SCAN functionals to enhance the accuracy of predicting redox behavior in TMFs, while the non-\textit{U}-corrected frameworks may still provide reasonable estimates for calculated band gaps. We hope that our study will spur screening studies with higher accuracy in the chemical space of TMFs, which can result in the identification of novel materials for various applications.

\begin{acknowledgments}
G.S.G. acknowledges financial support from the Indian Institute of Science (IISc) Seed Grant, SG/MHRD/20/0020 and SR/MHRD/20/0013, and support from the Science and Engineering Research Board (SERB) of Government of India, under Sanction Numbers SRG/2021/000201 and IPA/2021/000007. D.B.T. thanks IISc for financial assistance. The authors acknowledge the computational resources provided by the Supercomputer Education and Research Centre (SERC), IISc. A portion of the calculations in this work used computational resources of the supercomputer Fugaku provided by RIKEN through the HPCI
System Research Project (Project ID hp220393). We acknowledge National Supercomputing Mission (NSM) for providing computing resources of ‘PARAM Siddhi-AI’, under National PARAM Supercomputing Facility (NPSF), C-DAC, Pune and supported by the Ministry of Electronics and Information Technology (MeitY) and Department of Science and Technology (DST), Government of India. All computed data that has been presented in this work are available to the public in our \href{https://github.com/sai-mat-group/fluorides-benchmarking}{GitHub repository}. 
\end{acknowledgments}

\bibliography{bibfile}

\begin{thebibliography}{135}%
\makeatletter
\providecommand \@ifxundefined [1]{%
 \@ifx{#1\undefined}
}%
\providecommand \@ifnum [1]{%
 \ifnum #1\expandafter \@firstoftwo
 \else \expandafter \@secondoftwo
 \fi
}%
\providecommand \@ifx [1]{%
 \ifx #1\expandafter \@firstoftwo
 \else \expandafter \@secondoftwo
 \fi
}%
\providecommand \natexlab [1]{#1}%
\providecommand \enquote  [1]{``#1''}%
\providecommand \bibnamefont  [1]{#1}%
\providecommand \bibfnamefont [1]{#1}%
\providecommand \citenamefont [1]{#1}%
\providecommand \href@noop [0]{\@secondoftwo}%
\providecommand \href [0]{\begingroup \@sanitize@url \@href}%
\providecommand \@href[1]{\@@startlink{#1}\@@href}%
\providecommand \@@href[1]{\endgroup#1\@@endlink}%
\providecommand \@sanitize@url [0]{\catcode `\\12\catcode `\$12\catcode
  `\&12\catcode `\#12\catcode `\^12\catcode `\_12\catcode `\%12\relax}%
\providecommand \@@startlink[1]{}%
\providecommand \@@endlink[0]{}%
\providecommand \url  [0]{\begingroup\@sanitize@url \@url }%
\providecommand \@url [1]{\endgroup\@href {#1}{\urlprefix }}%
\providecommand \urlprefix  [0]{URL }%
\providecommand \Eprint [0]{\href }%
\providecommand \doibase [0]{https://doi.org/}%
\providecommand \selectlanguage [0]{\@gobble}%
\providecommand \bibinfo  [0]{\@secondoftwo}%
\providecommand \bibfield  [0]{\@secondoftwo}%
\providecommand \translation [1]{[#1]}%
\providecommand \BibitemOpen [0]{}%
\providecommand \bibitemStop [0]{}%
\providecommand \bibitemNoStop [0]{.\EOS\space}%
\providecommand \EOS [0]{\spacefactor3000\relax}%
\providecommand \BibitemShut  [1]{\csname bibitem#1\endcsname}%
\let\auto@bib@innerbib\@empty
\bibitem [{\citenamefont {Dubrovin}\ \emph {et~al.}(2020)\citenamefont
  {Dubrovin}, \citenamefont {Alyabyeva}, \citenamefont {Siverin}, \citenamefont
  {Gorshunov}, \citenamefont {Novikova}, \citenamefont {Boldyrev},\ and\
  \citenamefont {Pisarev}}]{dubrovin2020incipient}%
  \BibitemOpen
  \bibfield  {author} {\bibinfo {author} {\bibfnamefont {R.}~\bibnamefont
  {Dubrovin}}, \bibinfo {author} {\bibfnamefont {L.}~\bibnamefont {Alyabyeva}},
  \bibinfo {author} {\bibfnamefont {N.}~\bibnamefont {Siverin}}, \bibinfo
  {author} {\bibfnamefont {B.}~\bibnamefont {Gorshunov}}, \bibinfo {author}
  {\bibfnamefont {N.}~\bibnamefont {Novikova}}, \bibinfo {author}
  {\bibfnamefont {K.}~\bibnamefont {Boldyrev}},\ and\ \bibinfo {author}
  {\bibfnamefont {R.}~\bibnamefont {Pisarev}},\ }\bibfield  {title} {\bibinfo
  {title} {Incipient multiferroicity in p n m a fluoroperovskite namnf 3},\
  }\href@noop {} {\bibfield  {journal} {\bibinfo  {journal} {Physical Review
  B}\ }\textbf {\bibinfo {volume} {101}},\ \bibinfo {pages} {180403} (\bibinfo
  {year} {2020})}\BibitemShut {NoStop}%
\bibitem [{\citenamefont {Dubrovin}\ \emph {et~al.}(2018)\citenamefont
  {Dubrovin}, \citenamefont {Kizhaev}, \citenamefont {Syrnikov}, \citenamefont
  {Gesland},\ and\ \citenamefont {Pisarev}}]{dubrovin2018unveiling}%
  \BibitemOpen
  \bibfield  {author} {\bibinfo {author} {\bibfnamefont {R.}~\bibnamefont
  {Dubrovin}}, \bibinfo {author} {\bibfnamefont {S.}~\bibnamefont {Kizhaev}},
  \bibinfo {author} {\bibfnamefont {P.}~\bibnamefont {Syrnikov}}, \bibinfo
  {author} {\bibfnamefont {J.-Y.}\ \bibnamefont {Gesland}},\ and\ \bibinfo
  {author} {\bibfnamefont {R.}~\bibnamefont {Pisarev}},\ }\bibfield  {title}
  {\bibinfo {title} {Unveiling hidden structural instabilities and
  magnetodielectric effect in manganese fluoroperovskites a mnf 3},\
  }\href@noop {} {\bibfield  {journal} {\bibinfo  {journal} {Physical Review
  B}\ }\textbf {\bibinfo {volume} {98}},\ \bibinfo {pages} {060403} (\bibinfo
  {year} {2018})}\BibitemShut {NoStop}%
\bibitem [{\citenamefont {Zheng}\ \emph {et~al.}(2021)\citenamefont {Zheng},
  \citenamefont {Wang}, \citenamefont {Xiao}, \citenamefont {Cheng},
  \citenamefont {Zhang},\ and\ \citenamefont {Chen}}]{zheng2021improved}%
  \BibitemOpen
  \bibfield  {author} {\bibinfo {author} {\bibfnamefont {J.}~\bibnamefont
  {Zheng}}, \bibinfo {author} {\bibfnamefont {X.}~\bibnamefont {Wang}},
  \bibinfo {author} {\bibfnamefont {X.}~\bibnamefont {Xiao}}, \bibinfo {author}
  {\bibfnamefont {H.}~\bibnamefont {Cheng}}, \bibinfo {author} {\bibfnamefont
  {L.}~\bibnamefont {Zhang}},\ and\ \bibinfo {author} {\bibfnamefont
  {L.}~\bibnamefont {Chen}},\ }\bibfield  {title} {\bibinfo {title} {Improved
  reversible dehydrogenation properties of mg (bh4) 2 catalyzed by dual-cation
  transition metal fluorides k2tif6 and k2nbf7},\ }\href@noop {} {\bibfield
  {journal} {\bibinfo  {journal} {Chemical Engineering Journal}\ }\textbf
  {\bibinfo {volume} {412}},\ \bibinfo {pages} {128738} (\bibinfo {year}
  {2021})}\BibitemShut {NoStop}%
\bibitem [{\citenamefont {Wang}\ \emph {et~al.}(2023)\citenamefont {Wang},
  \citenamefont {Wang}, \citenamefont {Shan}, \citenamefont {Wang},
  \citenamefont {Qiu}, \citenamefont {Song},\ and\ \citenamefont
  {Li}}]{wang2023fluorine}%
  \BibitemOpen
  \bibfield  {author} {\bibinfo {author} {\bibfnamefont {M.}~\bibnamefont
  {Wang}}, \bibinfo {author} {\bibfnamefont {Z.}~\bibnamefont {Wang}}, \bibinfo
  {author} {\bibfnamefont {M.}~\bibnamefont {Shan}}, \bibinfo {author}
  {\bibfnamefont {J.}~\bibnamefont {Wang}}, \bibinfo {author} {\bibfnamefont
  {Z.}~\bibnamefont {Qiu}}, \bibinfo {author} {\bibfnamefont {J.}~\bibnamefont
  {Song}},\ and\ \bibinfo {author} {\bibfnamefont {Z.}~\bibnamefont {Li}},\
  }\bibfield  {title} {\bibinfo {title} {Fluorine-substituted donor--acceptor
  covalent organic frameworks for efficient photocatalyst hydrogen evolution},\
  }\href@noop {} {\bibfield  {journal} {\bibinfo  {journal} {Chemistry of
  Materials}\ }\textbf {\bibinfo {volume} {35}},\ \bibinfo {pages} {5368}
  (\bibinfo {year} {2023})}\BibitemShut {NoStop}%
\bibitem [{\citenamefont {Ali}\ \emph {et~al.}(2022)\citenamefont {Ali},
  \citenamefont {Amjad}, \citenamefont {Shafique}, \citenamefont {Naseer},
  \citenamefont {Al-Rashida}, \citenamefont {Sindhu}, \citenamefont {Iftikhar},
  \citenamefont {Shah}, \citenamefont {Hameed},\ and\ \citenamefont
  {Iqbal}}]{ali2022utilization}%
  \BibitemOpen
  \bibfield  {author} {\bibinfo {author} {\bibfnamefont {D.}~\bibnamefont
  {Ali}}, \bibinfo {author} {\bibfnamefont {S.~T.}\ \bibnamefont {Amjad}},
  \bibinfo {author} {\bibfnamefont {Z.}~\bibnamefont {Shafique}}, \bibinfo
  {author} {\bibfnamefont {M.~M.}\ \bibnamefont {Naseer}}, \bibinfo {author}
  {\bibfnamefont {M.}~\bibnamefont {Al-Rashida}}, \bibinfo {author}
  {\bibfnamefont {T.~A.}\ \bibnamefont {Sindhu}}, \bibinfo {author}
  {\bibfnamefont {S.}~\bibnamefont {Iftikhar}}, \bibinfo {author}
  {\bibfnamefont {M.~R.}\ \bibnamefont {Shah}}, \bibinfo {author}
  {\bibfnamefont {A.}~\bibnamefont {Hameed}},\ and\ \bibinfo {author}
  {\bibfnamefont {J.}~\bibnamefont {Iqbal}},\ }\bibfield  {title} {\bibinfo
  {title} {Utilization of transition metal fluoride-based solid support
  catalysts for the synthesis of sulfonamides: carbonic anhydrase inhibitory
  activity and in silico study},\ }\href@noop {} {\bibfield  {journal}
  {\bibinfo  {journal} {RSC advances}\ }\textbf {\bibinfo {volume} {12}},\
  \bibinfo {pages} {3165} (\bibinfo {year} {2022})}\BibitemShut {NoStop}%
\bibitem [{\citenamefont {Han}\ \emph {et~al.}(2019)\citenamefont {Han},
  \citenamefont {Woo}, \citenamefont {Hong}, \citenamefont {Chung},\ and\
  \citenamefont {Mhin}}]{han2019polarized}%
  \BibitemOpen
  \bibfield  {author} {\bibinfo {author} {\bibfnamefont {H.}~\bibnamefont
  {Han}}, \bibinfo {author} {\bibfnamefont {J.}~\bibnamefont {Woo}}, \bibinfo
  {author} {\bibfnamefont {Y.-R.}\ \bibnamefont {Hong}}, \bibinfo {author}
  {\bibfnamefont {Y.-C.}\ \bibnamefont {Chung}},\ and\ \bibinfo {author}
  {\bibfnamefont {S.}~\bibnamefont {Mhin}},\ }\bibfield  {title} {\bibinfo
  {title} {Polarized electronic configuration in transition metal--fluoride
  oxide hollow nanoprism for highly efficient and robust water splitting},\
  }\href@noop {} {\bibfield  {journal} {\bibinfo  {journal} {ACS Applied Energy
  Materials}\ }\textbf {\bibinfo {volume} {2}},\ \bibinfo {pages} {3999}
  (\bibinfo {year} {2019})}\BibitemShut {NoStop}%
\bibitem [{\citenamefont {Garcia-Castro}\ \emph {et~al.}(2016)\citenamefont
  {Garcia-Castro}, \citenamefont {Romero},\ and\ \citenamefont
  {Bousquet}}]{garcia2016strain}%
  \BibitemOpen
  \bibfield  {author} {\bibinfo {author} {\bibfnamefont {A.~C.}\ \bibnamefont
  {Garcia-Castro}}, \bibinfo {author} {\bibfnamefont {A.~H.}\ \bibnamefont
  {Romero}},\ and\ \bibinfo {author} {\bibfnamefont {E.}~\bibnamefont
  {Bousquet}},\ }\bibfield  {title} {\bibinfo {title} {Strain-engineered
  multiferroicity in p n m a namnf 3 fluoroperovskite},\ }\href@noop {}
  {\bibfield  {journal} {\bibinfo  {journal} {Physical Review Letters}\
  }\textbf {\bibinfo {volume} {116}},\ \bibinfo {pages} {117202} (\bibinfo
  {year} {2016})}\BibitemShut {NoStop}%
\bibitem [{\citenamefont {Borisov}\ \emph {et~al.}(2016)\citenamefont
  {Borisov}, \citenamefont {Johnson}, \citenamefont {Garc{\'\i}a-Castro},
  \citenamefont {KC}, \citenamefont {Schrecongost}, \citenamefont {Cen},
  \citenamefont {Romero},\ and\ \citenamefont
  {Lederman}}]{borisov2016multiferroic}%
  \BibitemOpen
  \bibfield  {author} {\bibinfo {author} {\bibfnamefont {P.}~\bibnamefont
  {Borisov}}, \bibinfo {author} {\bibfnamefont {T.~A.}\ \bibnamefont
  {Johnson}}, \bibinfo {author} {\bibfnamefont {A.~C.}\ \bibnamefont
  {Garc{\'\i}a-Castro}}, \bibinfo {author} {\bibfnamefont {A.}~\bibnamefont
  {KC}}, \bibinfo {author} {\bibfnamefont {D.}~\bibnamefont {Schrecongost}},
  \bibinfo {author} {\bibfnamefont {C.}~\bibnamefont {Cen}}, \bibinfo {author}
  {\bibfnamefont {A.~H.}\ \bibnamefont {Romero}},\ and\ \bibinfo {author}
  {\bibfnamefont {D.}~\bibnamefont {Lederman}},\ }\bibfield  {title} {\bibinfo
  {title} {Multiferroic bacof4 in thin film form: ferroelectricity, magnetic
  ordering, and strain},\ }\href@noop {} {\bibfield  {journal} {\bibinfo
  {journal} {ACS applied materials \& interfaces}\ }\textbf {\bibinfo {volume}
  {8}},\ \bibinfo {pages} {2694} (\bibinfo {year} {2016})}\BibitemShut
  {NoStop}%
\bibitem [{\citenamefont {Reisinger}\ \emph {et~al.}(2011)\citenamefont
  {Reisinger}, \citenamefont {Leblanc}, \citenamefont {Mercier}, \citenamefont
  {Tang}, \citenamefont {Parker}, \citenamefont {Morrison},\ and\ \citenamefont
  {Lightfoot}}]{reisinger2011phase}%
  \BibitemOpen
  \bibfield  {author} {\bibinfo {author} {\bibfnamefont {S.~A.}\ \bibnamefont
  {Reisinger}}, \bibinfo {author} {\bibfnamefont {M.}~\bibnamefont {Leblanc}},
  \bibinfo {author} {\bibfnamefont {A.-M.}\ \bibnamefont {Mercier}}, \bibinfo
  {author} {\bibfnamefont {C.~C.}\ \bibnamefont {Tang}}, \bibinfo {author}
  {\bibfnamefont {J.~E.}\ \bibnamefont {Parker}}, \bibinfo {author}
  {\bibfnamefont {F.~D.}\ \bibnamefont {Morrison}},\ and\ \bibinfo {author}
  {\bibfnamefont {P.}~\bibnamefont {Lightfoot}},\ }\bibfield  {title} {\bibinfo
  {title} {Phase separation and phase transitions in multiferroic k0. 58fef3
  with the tetragonal tungsten bronze structure},\ }\href@noop {} {\bibfield
  {journal} {\bibinfo  {journal} {Chemistry of Materials}\ }\textbf {\bibinfo
  {volume} {23}},\ \bibinfo {pages} {5440} (\bibinfo {year}
  {2011})}\BibitemShut {NoStop}%
\bibitem [{\citenamefont {Wang}\ \emph {et~al.}(2015)\citenamefont {Wang},
  \citenamefont {Kim}, \citenamefont {Seo}, \citenamefont {Kang}, \citenamefont
  {Wang}, \citenamefont {Su}, \citenamefont {Vajo}, \citenamefont {Wang},\ and\
  \citenamefont {Graetz}}]{wang2015ternary}%
  \BibitemOpen
  \bibfield  {author} {\bibinfo {author} {\bibfnamefont {F.}~\bibnamefont
  {Wang}}, \bibinfo {author} {\bibfnamefont {S.-W.}\ \bibnamefont {Kim}},
  \bibinfo {author} {\bibfnamefont {D.-H.}\ \bibnamefont {Seo}}, \bibinfo
  {author} {\bibfnamefont {K.}~\bibnamefont {Kang}}, \bibinfo {author}
  {\bibfnamefont {L.}~\bibnamefont {Wang}}, \bibinfo {author} {\bibfnamefont
  {D.}~\bibnamefont {Su}}, \bibinfo {author} {\bibfnamefont {J.~J.}\
  \bibnamefont {Vajo}}, \bibinfo {author} {\bibfnamefont {J.}~\bibnamefont
  {Wang}},\ and\ \bibinfo {author} {\bibfnamefont {J.}~\bibnamefont {Graetz}},\
  }\bibfield  {title} {\bibinfo {title} {Ternary metal fluorides as high-energy
  cathodes with low cycling hysteresis},\ }\href@noop {} {\bibfield  {journal}
  {\bibinfo  {journal} {Nature communications}\ }\textbf {\bibinfo {volume}
  {6}},\ \bibinfo {pages} {6668} (\bibinfo {year} {2015})}\BibitemShut
  {NoStop}%
\bibitem [{\citenamefont {Zhang}\ \emph {et~al.}(2019)\citenamefont {Zhang},
  \citenamefont {Xiao},\ and\ \citenamefont {Pang}}]{zhang2019transition}%
  \BibitemOpen
  \bibfield  {author} {\bibinfo {author} {\bibfnamefont {N.}~\bibnamefont
  {Zhang}}, \bibinfo {author} {\bibfnamefont {X.}~\bibnamefont {Xiao}},\ and\
  \bibinfo {author} {\bibfnamefont {H.}~\bibnamefont {Pang}},\ }\bibfield
  {title} {\bibinfo {title} {Transition metal (fe, co, ni) fluoride-based
  materials for electrochemical energy storage},\ }\href@noop {} {\bibfield
  {journal} {\bibinfo  {journal} {Nanoscale Horizons}\ }\textbf {\bibinfo
  {volume} {4}},\ \bibinfo {pages} {99} (\bibinfo {year} {2019})}\BibitemShut
  {NoStop}%
\bibitem [{\citenamefont {Park}\ \emph {et~al.}(2021)\citenamefont {Park},
  \citenamefont {Lee}, \citenamefont {Cho}, \citenamefont {Kang}, \citenamefont
  {Ko}, \citenamefont {Jung}, \citenamefont {Jeon}, \citenamefont {Hong},
  \citenamefont {Kim}, \citenamefont {Myung} \emph {et~al.}}]{park20212}%
  \BibitemOpen
  \bibfield  {author} {\bibinfo {author} {\bibfnamefont {H.}~\bibnamefont
  {Park}}, \bibinfo {author} {\bibfnamefont {Y.}~\bibnamefont {Lee}}, \bibinfo
  {author} {\bibfnamefont {M.-k.}\ \bibnamefont {Cho}}, \bibinfo {author}
  {\bibfnamefont {J.}~\bibnamefont {Kang}}, \bibinfo {author} {\bibfnamefont
  {W.}~\bibnamefont {Ko}}, \bibinfo {author} {\bibfnamefont {Y.~H.}\
  \bibnamefont {Jung}}, \bibinfo {author} {\bibfnamefont {T.-Y.}\ \bibnamefont
  {Jeon}}, \bibinfo {author} {\bibfnamefont {J.}~\bibnamefont {Hong}}, \bibinfo
  {author} {\bibfnamefont {H.}~\bibnamefont {Kim}}, \bibinfo {author}
  {\bibfnamefont {S.-T.}\ \bibnamefont {Myung}}, \emph {et~al.},\ }\bibfield
  {title} {\bibinfo {title} {Na 2 fe 2 f 7: a fluoride-based cathode for high
  power and long life na-ion batteries},\ }\href@noop {} {\bibfield  {journal}
  {\bibinfo  {journal} {Energy \& Environmental Science}\ }\textbf {\bibinfo
  {volume} {14}},\ \bibinfo {pages} {1469} (\bibinfo {year}
  {2021})}\BibitemShut {NoStop}%
\bibitem [{\citenamefont {Foley}\ \emph {et~al.}(2023)\citenamefont {Foley},
  \citenamefont {Wu}, \citenamefont {Jin}, \citenamefont {Cui}, \citenamefont
  {Yoshida}, \citenamefont {Manche},\ and\ \citenamefont
  {Cl{\'e}ment}}]{foley2023polymorphism}%
  \BibitemOpen
  \bibfield  {author} {\bibinfo {author} {\bibfnamefont {E.~E.}\ \bibnamefont
  {Foley}}, \bibinfo {author} {\bibfnamefont {V.~C.}\ \bibnamefont {Wu}},
  \bibinfo {author} {\bibfnamefont {W.}~\bibnamefont {Jin}}, \bibinfo {author}
  {\bibfnamefont {W.}~\bibnamefont {Cui}}, \bibinfo {author} {\bibfnamefont
  {E.}~\bibnamefont {Yoshida}}, \bibinfo {author} {\bibfnamefont
  {A.}~\bibnamefont {Manche}},\ and\ \bibinfo {author} {\bibfnamefont {R.~J.}\
  \bibnamefont {Cl{\'e}ment}},\ }\bibfield  {title} {\bibinfo {title}
  {Polymorphism in weberite na2fe2f7 and its effects on electrochemical
  properties as a na-ion cathode},\ }\href@noop {} {\bibfield  {journal}
  {\bibinfo  {journal} {Chemistry of Materials}\ }\textbf {\bibinfo {volume}
  {35}},\ \bibinfo {pages} {3614} (\bibinfo {year} {2023})}\BibitemShut
  {NoStop}%
\bibitem [{\citenamefont {Lu}\ \emph {et~al.}(2023)\citenamefont {Lu},
  \citenamefont {Meng},\ and\ \citenamefont {Liu}}]{lu2023weberite}%
  \BibitemOpen
  \bibfield  {author} {\bibinfo {author} {\bibfnamefont {T.}~\bibnamefont
  {Lu}}, \bibinfo {author} {\bibfnamefont {S.}~\bibnamefont {Meng}},\ and\
  \bibinfo {author} {\bibfnamefont {M.}~\bibnamefont {Liu}},\ }\bibfield
  {title} {\bibinfo {title} {Weberite na $ \_2 $ mm'f $ \_7 $(m, m'=
  redox-active metal) as promising fluoride-based sodium-ion battery
  cathodes},\ }\href@noop {} {\bibfield  {journal} {\bibinfo  {journal} {arXiv
  preprint arXiv:2310.04222}\ } (\bibinfo {year} {2023})}\BibitemShut {NoStop}%
\bibitem [{\citenamefont {Dey}\ \emph {et~al.}(2019)\citenamefont {Dey},
  \citenamefont {Barman}, \citenamefont {Ghosh}, \citenamefont {Sarkar},
  \citenamefont {Peter},\ and\ \citenamefont
  {Senguttuvan}}]{dey2019topochemical}%
  \BibitemOpen
  \bibfield  {author} {\bibinfo {author} {\bibfnamefont {U.~K.}\ \bibnamefont
  {Dey}}, \bibinfo {author} {\bibfnamefont {N.}~\bibnamefont {Barman}},
  \bibinfo {author} {\bibfnamefont {S.}~\bibnamefont {Ghosh}}, \bibinfo
  {author} {\bibfnamefont {S.}~\bibnamefont {Sarkar}}, \bibinfo {author}
  {\bibfnamefont {S.~C.}\ \bibnamefont {Peter}},\ and\ \bibinfo {author}
  {\bibfnamefont {P.}~\bibnamefont {Senguttuvan}},\ }\bibfield  {title}
  {\bibinfo {title} {Topochemical bottom-up synthesis of 2d-and 3d-sodium iron
  fluoride frameworks},\ }\href@noop {} {\bibfield  {journal} {\bibinfo
  {journal} {Chemistry of Materials}\ }\textbf {\bibinfo {volume} {31}},\
  \bibinfo {pages} {295} (\bibinfo {year} {2019})}\BibitemShut {NoStop}%
\bibitem [{\citenamefont {Hwang}\ \emph {et~al.}(2017)\citenamefont {Hwang},
  \citenamefont {Jung}, \citenamefont {Jeong}, \citenamefont {Kim},
  \citenamefont {Cho}, \citenamefont {Ku}, \citenamefont {Kim}, \citenamefont
  {Yoon},\ and\ \citenamefont {Kang}}]{hwang2017naf}%
  \BibitemOpen
  \bibfield  {author} {\bibinfo {author} {\bibfnamefont {I.}~\bibnamefont
  {Hwang}}, \bibinfo {author} {\bibfnamefont {S.-K.}\ \bibnamefont {Jung}},
  \bibinfo {author} {\bibfnamefont {E.-S.}\ \bibnamefont {Jeong}}, \bibinfo
  {author} {\bibfnamefont {H.}~\bibnamefont {Kim}}, \bibinfo {author}
  {\bibfnamefont {S.-P.}\ \bibnamefont {Cho}}, \bibinfo {author} {\bibfnamefont
  {K.}~\bibnamefont {Ku}}, \bibinfo {author} {\bibfnamefont {H.}~\bibnamefont
  {Kim}}, \bibinfo {author} {\bibfnamefont {W.-S.}\ \bibnamefont {Yoon}},\ and\
  \bibinfo {author} {\bibfnamefont {K.}~\bibnamefont {Kang}},\ }\bibfield
  {title} {\bibinfo {title} {Naf--fef 2 nanocomposite: New type of na-ion
  battery cathode material},\ }\href@noop {} {\bibfield  {journal} {\bibinfo
  {journal} {Nano Research}\ }\textbf {\bibinfo {volume} {10}},\ \bibinfo
  {pages} {4388} (\bibinfo {year} {2017})}\BibitemShut {NoStop}%
\bibitem [{\citenamefont {Kim}\ \emph {et~al.}(2010)\citenamefont {Kim},
  \citenamefont {Seo}, \citenamefont {Gwon}, \citenamefont {Kim},\ and\
  \citenamefont {Kang}}]{kim2010fabrication}%
  \BibitemOpen
  \bibfield  {author} {\bibinfo {author} {\bibfnamefont {S.-W.}\ \bibnamefont
  {Kim}}, \bibinfo {author} {\bibfnamefont {D.-H.}\ \bibnamefont {Seo}},
  \bibinfo {author} {\bibfnamefont {H.}~\bibnamefont {Gwon}}, \bibinfo {author}
  {\bibfnamefont {J.}~\bibnamefont {Kim}},\ and\ \bibinfo {author}
  {\bibfnamefont {K.}~\bibnamefont {Kang}},\ }\bibfield  {title} {\bibinfo
  {title} {Fabrication of fef3 nanoflowers on cnt branches and their
  application to high power lithium rechargeable batteries},\ }\href@noop {}
  {\bibfield  {journal} {\bibinfo  {journal} {Advanced Materials}\ }\textbf
  {\bibinfo {volume} {22}},\ \bibinfo {pages} {5260} (\bibinfo {year}
  {2010})}\BibitemShut {NoStop}%
\bibitem [{\citenamefont {Hua}\ \emph {et~al.}(2021)\citenamefont {Hua},
  \citenamefont {Eggeman}, \citenamefont {Castillo-Mart{\'\i}nez},
  \citenamefont {Robert}, \citenamefont {Geddes}, \citenamefont {Lu},
  \citenamefont {Pickard}, \citenamefont {Meng}, \citenamefont {Wiaderek},
  \citenamefont {Pereira} \emph {et~al.}}]{hua2021revisiting}%
  \BibitemOpen
  \bibfield  {author} {\bibinfo {author} {\bibfnamefont {X.}~\bibnamefont
  {Hua}}, \bibinfo {author} {\bibfnamefont {A.~S.}\ \bibnamefont {Eggeman}},
  \bibinfo {author} {\bibfnamefont {E.}~\bibnamefont {Castillo-Mart{\'\i}nez}},
  \bibinfo {author} {\bibfnamefont {R.}~\bibnamefont {Robert}}, \bibinfo
  {author} {\bibfnamefont {H.~S.}\ \bibnamefont {Geddes}}, \bibinfo {author}
  {\bibfnamefont {Z.}~\bibnamefont {Lu}}, \bibinfo {author} {\bibfnamefont
  {C.~J.}\ \bibnamefont {Pickard}}, \bibinfo {author} {\bibfnamefont
  {W.}~\bibnamefont {Meng}}, \bibinfo {author} {\bibfnamefont {K.~M.}\
  \bibnamefont {Wiaderek}}, \bibinfo {author} {\bibfnamefont {N.}~\bibnamefont
  {Pereira}}, \emph {et~al.},\ }\bibfield  {title} {\bibinfo {title}
  {Revisiting metal fluorides as lithium-ion battery cathodes},\ }\href@noop {}
  {\bibfield  {journal} {\bibinfo  {journal} {Nature materials}\ }\textbf
  {\bibinfo {volume} {20}},\ \bibinfo {pages} {841} (\bibinfo {year}
  {2021})}\BibitemShut {NoStop}%
\bibitem [{\citenamefont {Manthiram}\ and\ \citenamefont
  {Goodenough}(1989)}]{manthiram1989lithium}%
  \BibitemOpen
  \bibfield  {author} {\bibinfo {author} {\bibfnamefont {A.}~\bibnamefont
  {Manthiram}}\ and\ \bibinfo {author} {\bibfnamefont {J.}~\bibnamefont
  {Goodenough}},\ }\bibfield  {title} {\bibinfo {title} {Lithium insertion into
  fe2 (so4) 3 frameworks},\ }\href@noop {} {\bibfield  {journal} {\bibinfo
  {journal} {Journal of Power Sources}\ }\textbf {\bibinfo {volume} {26}},\
  \bibinfo {pages} {403} (\bibinfo {year} {1989})}\BibitemShut {NoStop}%
\bibitem [{\citenamefont {Fan}\ \emph {et~al.}(2018)\citenamefont {Fan},
  \citenamefont {Hu}, \citenamefont {Ji}, \citenamefont {Zhu}, \citenamefont
  {Han}, \citenamefont {Hwang}, \citenamefont {Liu}, \citenamefont {Bak},
  \citenamefont {Ma}, \citenamefont {Gao} \emph {et~al.}}]{fan2018high}%
  \BibitemOpen
  \bibfield  {author} {\bibinfo {author} {\bibfnamefont {X.}~\bibnamefont
  {Fan}}, \bibinfo {author} {\bibfnamefont {E.}~\bibnamefont {Hu}}, \bibinfo
  {author} {\bibfnamefont {X.}~\bibnamefont {Ji}}, \bibinfo {author}
  {\bibfnamefont {Y.}~\bibnamefont {Zhu}}, \bibinfo {author} {\bibfnamefont
  {F.}~\bibnamefont {Han}}, \bibinfo {author} {\bibfnamefont {S.}~\bibnamefont
  {Hwang}}, \bibinfo {author} {\bibfnamefont {J.}~\bibnamefont {Liu}}, \bibinfo
  {author} {\bibfnamefont {S.}~\bibnamefont {Bak}}, \bibinfo {author}
  {\bibfnamefont {Z.}~\bibnamefont {Ma}}, \bibinfo {author} {\bibfnamefont
  {T.}~\bibnamefont {Gao}}, \emph {et~al.},\ }\bibfield  {title} {\bibinfo
  {title} {High energy-density and reversibility of iron fluoride cathode
  enabled via an intercalation-extrusion reaction},\ }\href@noop {} {\bibfield
  {journal} {\bibinfo  {journal} {Nature communications}\ }\textbf {\bibinfo
  {volume} {9}},\ \bibinfo {pages} {2324} (\bibinfo {year} {2018})}\BibitemShut
  {NoStop}%
\bibitem [{\citenamefont {Martin}\ \emph {et~al.}(2018)\citenamefont {Martin},
  \citenamefont {Doublet}, \citenamefont {Kemnitz},\ and\ \citenamefont
  {Pinna}}]{martin2018reversible}%
  \BibitemOpen
  \bibfield  {author} {\bibinfo {author} {\bibfnamefont {A.}~\bibnamefont
  {Martin}}, \bibinfo {author} {\bibfnamefont {M.-L.}\ \bibnamefont {Doublet}},
  \bibinfo {author} {\bibfnamefont {E.}~\bibnamefont {Kemnitz}},\ and\ \bibinfo
  {author} {\bibfnamefont {N.}~\bibnamefont {Pinna}},\ }\bibfield  {title}
  {\bibinfo {title} {Reversible sodium and lithium insertion in iron fluoride
  perovskites},\ }\href@noop {} {\bibfield  {journal} {\bibinfo  {journal}
  {Advanced Functional Materials}\ }\textbf {\bibinfo {volume} {28}},\ \bibinfo
  {pages} {1802057} (\bibinfo {year} {2018})}\BibitemShut {NoStop}%
\bibitem [{\citenamefont {Cao}\ \emph {et~al.}(2017)\citenamefont {Cao},
  \citenamefont {Yin}, \citenamefont {Shi}, \citenamefont {Fu}, \citenamefont
  {Zhang},\ and\ \citenamefont {Li}}]{cao2017cubic}%
  \BibitemOpen
  \bibfield  {author} {\bibinfo {author} {\bibfnamefont {D.}~\bibnamefont
  {Cao}}, \bibinfo {author} {\bibfnamefont {C.}~\bibnamefont {Yin}}, \bibinfo
  {author} {\bibfnamefont {D.}~\bibnamefont {Shi}}, \bibinfo {author}
  {\bibfnamefont {Z.}~\bibnamefont {Fu}}, \bibinfo {author} {\bibfnamefont
  {J.}~\bibnamefont {Zhang}},\ and\ \bibinfo {author} {\bibfnamefont
  {C.}~\bibnamefont {Li}},\ }\bibfield  {title} {\bibinfo {title} {Cubic
  perovskite fluoride as open framework cathode for na-ion batteries},\
  }\href@noop {} {\bibfield  {journal} {\bibinfo  {journal} {Advanced
  Functional Materials}\ }\textbf {\bibinfo {volume} {27}},\ \bibinfo {pages}
  {1701130} (\bibinfo {year} {2017})}\BibitemShut {NoStop}%
\bibitem [{\citenamefont {Kitajou}\ \emph {et~al.}(2017)\citenamefont
  {Kitajou}, \citenamefont {Ishado}, \citenamefont {Yamashita}, \citenamefont
  {Momida}, \citenamefont {Oguchi},\ and\ \citenamefont
  {Okada}}]{kitajou2017cathode}%
  \BibitemOpen
  \bibfield  {author} {\bibinfo {author} {\bibfnamefont {A.}~\bibnamefont
  {Kitajou}}, \bibinfo {author} {\bibfnamefont {Y.}~\bibnamefont {Ishado}},
  \bibinfo {author} {\bibfnamefont {T.}~\bibnamefont {Yamashita}}, \bibinfo
  {author} {\bibfnamefont {H.}~\bibnamefont {Momida}}, \bibinfo {author}
  {\bibfnamefont {T.}~\bibnamefont {Oguchi}},\ and\ \bibinfo {author}
  {\bibfnamefont {S.}~\bibnamefont {Okada}},\ }\bibfield  {title} {\bibinfo
  {title} {Cathode properties of perovskite-type namf3 (m= fe, mn, and co)
  prepared by mechanical ball milling for sodium-ion battery},\ }\href@noop {}
  {\bibfield  {journal} {\bibinfo  {journal} {Electrochimica Acta}\ }\textbf
  {\bibinfo {volume} {245}},\ \bibinfo {pages} {424} (\bibinfo {year}
  {2017})}\BibitemShut {NoStop}%
\bibitem [{\citenamefont {Byeon}\ \emph {et~al.}(2023)\citenamefont {Byeon},
  \citenamefont {Gong}, \citenamefont {Cai}, \citenamefont {Sun}, \citenamefont
  {Szymanski}, \citenamefont {Bai}, \citenamefont {Seo},\ and\ \citenamefont
  {Kim}}]{byeon2023effects}%
  \BibitemOpen
  \bibfield  {author} {\bibinfo {author} {\bibfnamefont {Y.-W.}\ \bibnamefont
  {Byeon}}, \bibinfo {author} {\bibfnamefont {M.-J.}\ \bibnamefont {Gong}},
  \bibinfo {author} {\bibfnamefont {Z.}~\bibnamefont {Cai}}, \bibinfo {author}
  {\bibfnamefont {Y.}~\bibnamefont {Sun}}, \bibinfo {author} {\bibfnamefont
  {N.~J.}\ \bibnamefont {Szymanski}}, \bibinfo {author} {\bibfnamefont
  {J.}~\bibnamefont {Bai}}, \bibinfo {author} {\bibfnamefont {D.-H.}\
  \bibnamefont {Seo}},\ and\ \bibinfo {author} {\bibfnamefont {H.}~\bibnamefont
  {Kim}},\ }\bibfield  {title} {\bibinfo {title} {Effects of cation and anion
  substitution in kvpo4f for k-ion batteries},\ }\href@noop {} {\bibfield
  {journal} {\bibinfo  {journal} {Energy Storage Materials}\ }\textbf {\bibinfo
  {volume} {57}},\ \bibinfo {pages} {81} (\bibinfo {year} {2023})}\BibitemShut
  {NoStop}%
\bibitem [{\citenamefont {Li}\ \emph {et~al.}(2010)\citenamefont {Li},
  \citenamefont {Li}, \citenamefont {Cao}, \citenamefont {Ai},\ and\
  \citenamefont {Yang}}]{li2010reversible}%
  \BibitemOpen
  \bibfield  {author} {\bibinfo {author} {\bibfnamefont {T.}~\bibnamefont
  {Li}}, \bibinfo {author} {\bibfnamefont {L.}~\bibnamefont {Li}}, \bibinfo
  {author} {\bibfnamefont {Y.~L.}\ \bibnamefont {Cao}}, \bibinfo {author}
  {\bibfnamefont {X.~P.}\ \bibnamefont {Ai}},\ and\ \bibinfo {author}
  {\bibfnamefont {H.~X.}\ \bibnamefont {Yang}},\ }\bibfield  {title} {\bibinfo
  {title} {Reversible three-electron redox behaviors of fef3 nanocrystals as
  high-capacity cathode-active materials for li-ion batteries},\ }\href@noop {}
  {\bibfield  {journal} {\bibinfo  {journal} {The Journal of Physical Chemistry
  C}\ }\textbf {\bibinfo {volume} {114}},\ \bibinfo {pages} {3190} (\bibinfo
  {year} {2010})}\BibitemShut {NoStop}%
\bibitem [{\citenamefont {Kim}\ \emph {et~al.}(2018)\citenamefont {Kim},
  \citenamefont {Kim},\ and\ \citenamefont {Kang}}]{kim2018conversion}%
  \BibitemOpen
  \bibfield  {author} {\bibinfo {author} {\bibfnamefont {J.}~\bibnamefont
  {Kim}}, \bibinfo {author} {\bibfnamefont {H.}~\bibnamefont {Kim}},\ and\
  \bibinfo {author} {\bibfnamefont {K.}~\bibnamefont {Kang}},\ }\bibfield
  {title} {\bibinfo {title} {Conversion-based cathode materials for
  rechargeable sodium batteries},\ }\href@noop {} {\bibfield  {journal}
  {\bibinfo  {journal} {Advanced Energy Materials}\ }\textbf {\bibinfo {volume}
  {8}},\ \bibinfo {pages} {1702646} (\bibinfo {year} {2018})}\BibitemShut
  {NoStop}%
\bibitem [{\citenamefont {Fan}\ \emph {et~al.}(2016)\citenamefont {Fan},
  \citenamefont {Zhu}, \citenamefont {Luo}, \citenamefont {Gao}, \citenamefont
  {Suo}, \citenamefont {Liou}, \citenamefont {Xu},\ and\ \citenamefont
  {Wang}}]{fan2016situ}%
  \BibitemOpen
  \bibfield  {author} {\bibinfo {author} {\bibfnamefont {X.}~\bibnamefont
  {Fan}}, \bibinfo {author} {\bibfnamefont {Y.}~\bibnamefont {Zhu}}, \bibinfo
  {author} {\bibfnamefont {C.}~\bibnamefont {Luo}}, \bibinfo {author}
  {\bibfnamefont {T.}~\bibnamefont {Gao}}, \bibinfo {author} {\bibfnamefont
  {L.}~\bibnamefont {Suo}}, \bibinfo {author} {\bibfnamefont {S.-C.}\
  \bibnamefont {Liou}}, \bibinfo {author} {\bibfnamefont {K.}~\bibnamefont
  {Xu}},\ and\ \bibinfo {author} {\bibfnamefont {C.}~\bibnamefont {Wang}},\
  }\bibfield  {title} {\bibinfo {title} {In situ lithiated fef3/c nanocomposite
  as high energy conversion-reaction cathode for lithium-ion batteries},\
  }\href@noop {} {\bibfield  {journal} {\bibinfo  {journal} {Journal of Power
  Sources}\ }\textbf {\bibinfo {volume} {307}},\ \bibinfo {pages} {435}
  (\bibinfo {year} {2016})}\BibitemShut {NoStop}%
\bibitem [{\citenamefont {Jung}\ \emph {et~al.}(2018)\citenamefont {Jung},
  \citenamefont {Hwang}, \citenamefont {Cho}, \citenamefont {Oh}, \citenamefont
  {Ku}, \citenamefont {Choi},\ and\ \citenamefont {Kang}}]{jung2018new}%
  \BibitemOpen
  \bibfield  {author} {\bibinfo {author} {\bibfnamefont {S.-K.}\ \bibnamefont
  {Jung}}, \bibinfo {author} {\bibfnamefont {I.}~\bibnamefont {Hwang}},
  \bibinfo {author} {\bibfnamefont {S.-P.}\ \bibnamefont {Cho}}, \bibinfo
  {author} {\bibfnamefont {K.}~\bibnamefont {Oh}}, \bibinfo {author}
  {\bibfnamefont {K.}~\bibnamefont {Ku}}, \bibinfo {author} {\bibfnamefont
  {I.~R.}\ \bibnamefont {Choi}},\ and\ \bibinfo {author} {\bibfnamefont
  {K.}~\bibnamefont {Kang}},\ }\bibfield  {title} {\bibinfo {title} {New
  iron-based intercalation host for lithium-ion batteries},\ }\href@noop {}
  {\bibfield  {journal} {\bibinfo  {journal} {Chemistry of Materials}\ }\textbf
  {\bibinfo {volume} {30}},\ \bibinfo {pages} {1956} (\bibinfo {year}
  {2018})}\BibitemShut {NoStop}%
\bibitem [{\citenamefont {Tekliye}\ \emph {et~al.}(2022)\citenamefont
  {Tekliye}, \citenamefont {Kumar}, \citenamefont {Weihang}, \citenamefont
  {Mercy}, \citenamefont {Canepa},\ and\ \citenamefont
  {Sai~Gautam}}]{tekliye2022exploration}%
  \BibitemOpen
  \bibfield  {author} {\bibinfo {author} {\bibfnamefont {D.~B.}\ \bibnamefont
  {Tekliye}}, \bibinfo {author} {\bibfnamefont {A.}~\bibnamefont {Kumar}},
  \bibinfo {author} {\bibfnamefont {X.}~\bibnamefont {Weihang}}, \bibinfo
  {author} {\bibfnamefont {T.~D.}\ \bibnamefont {Mercy}}, \bibinfo {author}
  {\bibfnamefont {P.}~\bibnamefont {Canepa}},\ and\ \bibinfo {author}
  {\bibfnamefont {G.}~\bibnamefont {Sai~Gautam}},\ }\bibfield  {title}
  {\bibinfo {title} {Exploration of nasicon frameworks as calcium-ion battery
  electrodes},\ }\href@noop {} {\bibfield  {journal} {\bibinfo  {journal}
  {Chemistry of Materials}\ }\textbf {\bibinfo {volume} {34}},\ \bibinfo
  {pages} {10133} (\bibinfo {year} {2022})}\BibitemShut {NoStop}%
\bibitem [{\citenamefont {Lu}\ \emph {et~al.}(2021)\citenamefont {Lu},
  \citenamefont {Wang}, \citenamefont {Sai~Gautam},\ and\ \citenamefont
  {Canepa}}]{lu2021searching}%
  \BibitemOpen
  \bibfield  {author} {\bibinfo {author} {\bibfnamefont {W.}~\bibnamefont
  {Lu}}, \bibinfo {author} {\bibfnamefont {J.}~\bibnamefont {Wang}}, \bibinfo
  {author} {\bibfnamefont {G.}~\bibnamefont {Sai~Gautam}},\ and\ \bibinfo
  {author} {\bibfnamefont {P.}~\bibnamefont {Canepa}},\ }\bibfield  {title}
  {\bibinfo {title} {Searching ternary oxides and chalcogenides as positive
  electrodes for calcium batteries},\ }\href@noop {} {\bibfield  {journal}
  {\bibinfo  {journal} {Chemistry of Materials}\ }\textbf {\bibinfo {volume}
  {33}},\ \bibinfo {pages} {5809} (\bibinfo {year} {2021})}\BibitemShut
  {NoStop}%
\bibitem [{\citenamefont {Cheng}\ \emph {et~al.}(2015)\citenamefont {Cheng},
  \citenamefont {Assary}, \citenamefont {Qu}, \citenamefont {Jain},
  \citenamefont {Ong}, \citenamefont {Rajput}, \citenamefont {Persson},\ and\
  \citenamefont {Curtiss}}]{cheng2015accelerating}%
  \BibitemOpen
  \bibfield  {author} {\bibinfo {author} {\bibfnamefont {L.}~\bibnamefont
  {Cheng}}, \bibinfo {author} {\bibfnamefont {R.~S.}\ \bibnamefont {Assary}},
  \bibinfo {author} {\bibfnamefont {X.}~\bibnamefont {Qu}}, \bibinfo {author}
  {\bibfnamefont {A.}~\bibnamefont {Jain}}, \bibinfo {author} {\bibfnamefont
  {S.~P.}\ \bibnamefont {Ong}}, \bibinfo {author} {\bibfnamefont {N.~N.}\
  \bibnamefont {Rajput}}, \bibinfo {author} {\bibfnamefont {K.}~\bibnamefont
  {Persson}},\ and\ \bibinfo {author} {\bibfnamefont {L.~A.}\ \bibnamefont
  {Curtiss}},\ }\bibfield  {title} {\bibinfo {title} {Accelerating electrolyte
  discovery for energy storage with high-throughput screening},\ }\href@noop {}
  {\bibfield  {journal} {\bibinfo  {journal} {The journal of physical chemistry
  letters}\ }\textbf {\bibinfo {volume} {6}},\ \bibinfo {pages} {283} (\bibinfo
  {year} {2015})}\BibitemShut {NoStop}%
\bibitem [{\citenamefont {Sai~Gautam}\ \emph {et~al.}(2020)\citenamefont
  {Sai~Gautam}, \citenamefont {Stechel},\ and\ \citenamefont
  {Carter}}]{sai2020exploring}%
  \BibitemOpen
  \bibfield  {author} {\bibinfo {author} {\bibfnamefont {G.}~\bibnamefont
  {Sai~Gautam}}, \bibinfo {author} {\bibfnamefont {E.~B.}\ \bibnamefont
  {Stechel}},\ and\ \bibinfo {author} {\bibfnamefont {E.~A.}\ \bibnamefont
  {Carter}},\ }\bibfield  {title} {\bibinfo {title} {Exploring ca--ce--m--o (m=
  3d transition metal) oxide perovskites for solar thermochemical
  applications},\ }\href@noop {} {\bibfield  {journal} {\bibinfo  {journal}
  {Chemistry of Materials}\ }\textbf {\bibinfo {volume} {32}},\ \bibinfo
  {pages} {9964} (\bibinfo {year} {2020})}\BibitemShut {NoStop}%
\bibitem [{\citenamefont {Sun}\ \emph {et~al.}(2019)\citenamefont {Sun},
  \citenamefont {Bartel}, \citenamefont {Arca}, \citenamefont {Bauers},
  \citenamefont {Matthews}, \citenamefont {Orva{\~n}anos}, \citenamefont
  {Chen}, \citenamefont {Toney}, \citenamefont {Schelhas}, \citenamefont
  {Tumas} \emph {et~al.}}]{sun2019map}%
  \BibitemOpen
  \bibfield  {author} {\bibinfo {author} {\bibfnamefont {W.}~\bibnamefont
  {Sun}}, \bibinfo {author} {\bibfnamefont {C.~J.}\ \bibnamefont {Bartel}},
  \bibinfo {author} {\bibfnamefont {E.}~\bibnamefont {Arca}}, \bibinfo {author}
  {\bibfnamefont {S.~R.}\ \bibnamefont {Bauers}}, \bibinfo {author}
  {\bibfnamefont {B.}~\bibnamefont {Matthews}}, \bibinfo {author}
  {\bibfnamefont {B.}~\bibnamefont {Orva{\~n}anos}}, \bibinfo {author}
  {\bibfnamefont {B.-R.}\ \bibnamefont {Chen}}, \bibinfo {author}
  {\bibfnamefont {M.~F.}\ \bibnamefont {Toney}}, \bibinfo {author}
  {\bibfnamefont {L.~T.}\ \bibnamefont {Schelhas}}, \bibinfo {author}
  {\bibfnamefont {W.}~\bibnamefont {Tumas}}, \emph {et~al.},\ }\bibfield
  {title} {\bibinfo {title} {A map of the inorganic ternary metal nitrides},\
  }\href@noop {} {\bibfield  {journal} {\bibinfo  {journal} {Nature materials}\
  }\textbf {\bibinfo {volume} {18}},\ \bibinfo {pages} {732} (\bibinfo {year}
  {2019})}\BibitemShut {NoStop}%
\bibitem [{\citenamefont {Chakraborty}\ \emph {et~al.}(2017)\citenamefont
  {Chakraborty}, \citenamefont {Xie}, \citenamefont {Mathews}, \citenamefont
  {Sherburne}, \citenamefont {Ahuja}, \citenamefont {Asta},\ and\ \citenamefont
  {Mhaisalkar}}]{chakraborty2017rational}%
  \BibitemOpen
  \bibfield  {author} {\bibinfo {author} {\bibfnamefont {S.}~\bibnamefont
  {Chakraborty}}, \bibinfo {author} {\bibfnamefont {W.}~\bibnamefont {Xie}},
  \bibinfo {author} {\bibfnamefont {N.}~\bibnamefont {Mathews}}, \bibinfo
  {author} {\bibfnamefont {M.}~\bibnamefont {Sherburne}}, \bibinfo {author}
  {\bibfnamefont {R.}~\bibnamefont {Ahuja}}, \bibinfo {author} {\bibfnamefont
  {M.}~\bibnamefont {Asta}},\ and\ \bibinfo {author} {\bibfnamefont {S.~G.}\
  \bibnamefont {Mhaisalkar}},\ }\bibfield  {title} {\bibinfo {title} {Rational
  design: A high-throughput computational screening and experimental validation
  methodology for lead-free and emergent hybrid perovskites},\ }\href@noop {}
  {\bibfield  {journal} {\bibinfo  {journal} {ACS Energy Letters}\ }\textbf
  {\bibinfo {volume} {2}},\ \bibinfo {pages} {837} (\bibinfo {year}
  {2017})}\BibitemShut {NoStop}%
\bibitem [{\citenamefont {Hohenberg}\ and\ \citenamefont
  {Kohn}(1964)}]{hohenberg1964inhomogeneous}%
  \BibitemOpen
  \bibfield  {author} {\bibinfo {author} {\bibfnamefont {P.}~\bibnamefont
  {Hohenberg}}\ and\ \bibinfo {author} {\bibfnamefont {W.}~\bibnamefont
  {Kohn}},\ }\bibfield  {title} {\bibinfo {title} {Inhomogeneous electron
  gas},\ }\href@noop {} {\bibfield  {journal} {\bibinfo  {journal} {Physical
  review}\ }\textbf {\bibinfo {volume} {136}},\ \bibinfo {pages} {B864}
  (\bibinfo {year} {1964})}\BibitemShut {NoStop}%
\bibitem [{\citenamefont {Kohn}\ and\ \citenamefont
  {Sham}(1965)}]{kohn1965self}%
  \BibitemOpen
  \bibfield  {author} {\bibinfo {author} {\bibfnamefont {W.}~\bibnamefont
  {Kohn}}\ and\ \bibinfo {author} {\bibfnamefont {L.~J.}\ \bibnamefont
  {Sham}},\ }\bibfield  {title} {\bibinfo {title} {Self-consistent equations
  including exchange and correlation effects},\ }\href@noop {} {\bibfield
  {journal} {\bibinfo  {journal} {Physical review}\ }\textbf {\bibinfo {volume}
  {140}},\ \bibinfo {pages} {A1133} (\bibinfo {year} {1965})}\BibitemShut
  {NoStop}%
\bibitem [{\citenamefont {Perdew}\ \emph {et~al.}(1996)\citenamefont {Perdew},
  \citenamefont {Burke},\ and\ \citenamefont
  {Ernzerhof}}]{perdew1996generalized}%
  \BibitemOpen
  \bibfield  {author} {\bibinfo {author} {\bibfnamefont {J.~P.}\ \bibnamefont
  {Perdew}}, \bibinfo {author} {\bibfnamefont {K.}~\bibnamefont {Burke}},\ and\
  \bibinfo {author} {\bibfnamefont {M.}~\bibnamefont {Ernzerhof}},\ }\bibfield
  {title} {\bibinfo {title} {Generalized gradient approximation made simple},\
  }\href@noop {} {\bibfield  {journal} {\bibinfo  {journal} {Physical review
  letters}\ }\textbf {\bibinfo {volume} {77}},\ \bibinfo {pages} {3865}
  (\bibinfo {year} {1996})}\BibitemShut {NoStop}%
\bibitem [{\citenamefont {Sun}\ \emph {et~al.}(2015)\citenamefont {Sun},
  \citenamefont {Ruzsinszky},\ and\ \citenamefont {Perdew}}]{sun2015strongly}%
  \BibitemOpen
  \bibfield  {author} {\bibinfo {author} {\bibfnamefont {J.}~\bibnamefont
  {Sun}}, \bibinfo {author} {\bibfnamefont {A.}~\bibnamefont {Ruzsinszky}},\
  and\ \bibinfo {author} {\bibfnamefont {J.~P.}\ \bibnamefont {Perdew}},\
  }\bibfield  {title} {\bibinfo {title} {Strongly constrained and appropriately
  normed semilocal density functional},\ }\href@noop {} {\bibfield  {journal}
  {\bibinfo  {journal} {Physical review letters}\ }\textbf {\bibinfo {volume}
  {115}},\ \bibinfo {pages} {036402} (\bibinfo {year} {2015})}\BibitemShut
  {NoStop}%
\bibitem [{\citenamefont {Bart{\'o}k}\ and\ \citenamefont
  {Yates}(2019)}]{bartok2019regularized}%
  \BibitemOpen
  \bibfield  {author} {\bibinfo {author} {\bibfnamefont {A.~P.}\ \bibnamefont
  {Bart{\'o}k}}\ and\ \bibinfo {author} {\bibfnamefont {J.~R.}\ \bibnamefont
  {Yates}},\ }\bibfield  {title} {\bibinfo {title} {Regularized scan
  functional},\ }\href@noop {} {\bibfield  {journal} {\bibinfo  {journal} {The
  Journal of chemical physics}\ }\textbf {\bibinfo {volume} {150}},\ \bibinfo
  {pages} {161101} (\bibinfo {year} {2019})}\BibitemShut {NoStop}%
\bibitem [{\citenamefont {Furness}\ \emph {et~al.}(2020)\citenamefont
  {Furness}, \citenamefont {Kaplan}, \citenamefont {Ning}, \citenamefont
  {Perdew},\ and\ \citenamefont {Sun}}]{furness2020accurate}%
  \BibitemOpen
  \bibfield  {author} {\bibinfo {author} {\bibfnamefont {J.~W.}\ \bibnamefont
  {Furness}}, \bibinfo {author} {\bibfnamefont {A.~D.}\ \bibnamefont {Kaplan}},
  \bibinfo {author} {\bibfnamefont {J.}~\bibnamefont {Ning}}, \bibinfo {author}
  {\bibfnamefont {J.~P.}\ \bibnamefont {Perdew}},\ and\ \bibinfo {author}
  {\bibfnamefont {J.}~\bibnamefont {Sun}},\ }\bibfield  {title} {\bibinfo
  {title} {Accurate and numerically efficient r2scan meta-generalized gradient
  approximation},\ }\href@noop {} {\bibfield  {journal} {\bibinfo  {journal}
  {The journal of physical chemistry letters}\ }\textbf {\bibinfo {volume}
  {11}},\ \bibinfo {pages} {8208} (\bibinfo {year} {2020})}\BibitemShut
  {NoStop}%
\bibitem [{\citenamefont {Car}(2016)}]{car2016fixing}%
  \BibitemOpen
  \bibfield  {author} {\bibinfo {author} {\bibfnamefont {R.}~\bibnamefont
  {Car}},\ }\bibfield  {title} {\bibinfo {title} {Fixing jacob's ladder},\
  }\href@noop {} {\bibfield  {journal} {\bibinfo  {journal} {Nature chemistry}\
  }\textbf {\bibinfo {volume} {8}},\ \bibinfo {pages} {820} (\bibinfo {year}
  {2016})}\BibitemShut {NoStop}%
\bibitem [{\citenamefont {Yang}\ \emph {et~al.}(2019)\citenamefont {Yang},
  \citenamefont {Kitchaev},\ and\ \citenamefont
  {Ceder}}]{yang2019rationalizing}%
  \BibitemOpen
  \bibfield  {author} {\bibinfo {author} {\bibfnamefont {J.~H.}\ \bibnamefont
  {Yang}}, \bibinfo {author} {\bibfnamefont {D.~A.}\ \bibnamefont {Kitchaev}},\
  and\ \bibinfo {author} {\bibfnamefont {G.}~\bibnamefont {Ceder}},\ }\bibfield
   {title} {\bibinfo {title} {Rationalizing accurate structure prediction in
  the meta-gga scan functional},\ }\href@noop {} {\bibfield  {journal}
  {\bibinfo  {journal} {Physical Review B}\ }\textbf {\bibinfo {volume}
  {100}},\ \bibinfo {pages} {035132} (\bibinfo {year} {2019})}\BibitemShut
  {NoStop}%
\bibitem [{\citenamefont {Yao}\ and\ \citenamefont
  {Kanai}(2017)}]{yao2017plane}%
  \BibitemOpen
  \bibfield  {author} {\bibinfo {author} {\bibfnamefont {Y.}~\bibnamefont
  {Yao}}\ and\ \bibinfo {author} {\bibfnamefont {Y.}~\bibnamefont {Kanai}},\
  }\bibfield  {title} {\bibinfo {title} {Plane-wave pseudopotential
  implementation and performance of scan meta-gga exchange-correlation
  functional for extended systems},\ }\href@noop {} {\bibfield  {journal}
  {\bibinfo  {journal} {The Journal of chemical physics}\ }\textbf {\bibinfo
  {volume} {146}} (\bibinfo {year} {2017})}\BibitemShut {NoStop}%
\bibitem [{\citenamefont {Sa{\ss}nick}\ and\ \citenamefont
  {Cocchi}(2021)}]{sassnick2021electronic}%
  \BibitemOpen
  \bibfield  {author} {\bibinfo {author} {\bibfnamefont {H.-D.}\ \bibnamefont
  {Sa{\ss}nick}}\ and\ \bibinfo {author} {\bibfnamefont {C.}~\bibnamefont
  {Cocchi}},\ }\bibfield  {title} {\bibinfo {title} {Electronic structure of
  cesium-based photocathode materials from density functional theory:
  Performance of pbe, scan, and hse06 functionals},\ }\href@noop {} {\bibfield
  {journal} {\bibinfo  {journal} {Electronic Structure}\ }\textbf {\bibinfo
  {volume} {3}},\ \bibinfo {pages} {027001} (\bibinfo {year}
  {2021})}\BibitemShut {NoStop}%
\bibitem [{\citenamefont {Fu}\ and\ \citenamefont
  {Singh}(2019)}]{fu2019density}%
  \BibitemOpen
  \bibfield  {author} {\bibinfo {author} {\bibfnamefont {Y.}~\bibnamefont
  {Fu}}\ and\ \bibinfo {author} {\bibfnamefont {D.~J.}\ \bibnamefont {Singh}},\
  }\bibfield  {title} {\bibinfo {title} {Density functional methods for the
  magnetism of transition metals: Scan in relation to other functionals},\
  }\href@noop {} {\bibfield  {journal} {\bibinfo  {journal} {Physical Review
  B}\ }\textbf {\bibinfo {volume} {100}},\ \bibinfo {pages} {045126} (\bibinfo
  {year} {2019})}\BibitemShut {NoStop}%
\bibitem [{\citenamefont {Kingsbury}\ \emph {et~al.}(2022)\citenamefont
  {Kingsbury}, \citenamefont {Gupta}, \citenamefont {Bartel}, \citenamefont
  {Munro}, \citenamefont {Dwaraknath}, \citenamefont {Horton},\ and\
  \citenamefont {Persson}}]{kingsbury2022performance}%
  \BibitemOpen
  \bibfield  {author} {\bibinfo {author} {\bibfnamefont {R.}~\bibnamefont
  {Kingsbury}}, \bibinfo {author} {\bibfnamefont {A.~S.}\ \bibnamefont
  {Gupta}}, \bibinfo {author} {\bibfnamefont {C.~J.}\ \bibnamefont {Bartel}},
  \bibinfo {author} {\bibfnamefont {J.~M.}\ \bibnamefont {Munro}}, \bibinfo
  {author} {\bibfnamefont {S.}~\bibnamefont {Dwaraknath}}, \bibinfo {author}
  {\bibfnamefont {M.}~\bibnamefont {Horton}},\ and\ \bibinfo {author}
  {\bibfnamefont {K.~A.}\ \bibnamefont {Persson}},\ }\bibfield  {title}
  {\bibinfo {title} {Performance comparison of r 2 scan and scan metagga
  density functionals for solid materials via an automated, high-throughput
  computational workflow},\ }\href@noop {} {\bibfield  {journal} {\bibinfo
  {journal} {Physical Review Materials}\ }\textbf {\bibinfo {volume} {6}},\
  \bibinfo {pages} {013801} (\bibinfo {year} {2022})}\BibitemShut {NoStop}%
\bibitem [{\citenamefont {Swathilakshmi}\ \emph {et~al.}(2023)\citenamefont
  {Swathilakshmi}, \citenamefont {Devi},\ and\ \citenamefont
  {Sai~Gautam}}]{swathilakshmi2023performance}%
  \BibitemOpen
  \bibfield  {author} {\bibinfo {author} {\bibfnamefont {S.}~\bibnamefont
  {Swathilakshmi}}, \bibinfo {author} {\bibfnamefont {R.}~\bibnamefont
  {Devi}},\ and\ \bibinfo {author} {\bibfnamefont {G.}~\bibnamefont
  {Sai~Gautam}},\ }\bibfield  {title} {\bibinfo {title} {Performance of the
  r2scan functional in transition metal oxides},\ }\href@noop {} {\bibfield
  {journal} {\bibinfo  {journal} {Journal of Chemical Theory and Computation}\
  } (\bibinfo {year} {2023})}\BibitemShut {NoStop}%
\bibitem [{\citenamefont {DelloStritto}\ \emph {et~al.}(2023)\citenamefont
  {DelloStritto}, \citenamefont {Kaplan}, \citenamefont {Perdew},\ and\
  \citenamefont {Klein}}]{dellostritto2023predicting}%
  \BibitemOpen
  \bibfield  {author} {\bibinfo {author} {\bibfnamefont {M.~J.}\ \bibnamefont
  {DelloStritto}}, \bibinfo {author} {\bibfnamefont {A.~D.}\ \bibnamefont
  {Kaplan}}, \bibinfo {author} {\bibfnamefont {J.~P.}\ \bibnamefont {Perdew}},\
  and\ \bibinfo {author} {\bibfnamefont {M.~L.}\ \bibnamefont {Klein}},\
  }\bibfield  {title} {\bibinfo {title} {Predicting the properties of nio with
  density functional theory: Impact of exchange and correlation approximations
  and validation of the r2scan functional},\ }\href@noop {} {\bibfield
  {journal} {\bibinfo  {journal} {APL Materials}\ }\textbf {\bibinfo {volume}
  {11}} (\bibinfo {year} {2023})}\BibitemShut {NoStop}%
\bibitem [{\citenamefont {Sun}\ \emph {et~al.}(2017)\citenamefont {Sun},
  \citenamefont {Holder}, \citenamefont {Orva{\~n}anos}, \citenamefont {Arca},
  \citenamefont {Zakutayev}, \citenamefont {Lany},\ and\ \citenamefont
  {Ceder}}]{sun2017thermodynamic}%
  \BibitemOpen
  \bibfield  {author} {\bibinfo {author} {\bibfnamefont {W.}~\bibnamefont
  {Sun}}, \bibinfo {author} {\bibfnamefont {A.}~\bibnamefont {Holder}},
  \bibinfo {author} {\bibfnamefont {B.}~\bibnamefont {Orva{\~n}anos}}, \bibinfo
  {author} {\bibfnamefont {E.}~\bibnamefont {Arca}}, \bibinfo {author}
  {\bibfnamefont {A.}~\bibnamefont {Zakutayev}}, \bibinfo {author}
  {\bibfnamefont {S.}~\bibnamefont {Lany}},\ and\ \bibinfo {author}
  {\bibfnamefont {G.}~\bibnamefont {Ceder}},\ }\bibfield  {title} {\bibinfo
  {title} {Thermodynamic routes to novel metastable nitrogen-rich nitrides},\
  }\href@noop {} {\bibfield  {journal} {\bibinfo  {journal} {Chemistry of
  Materials}\ }\textbf {\bibinfo {volume} {29}},\ \bibinfo {pages} {6936}
  (\bibinfo {year} {2017})}\BibitemShut {NoStop}%
\bibitem [{\citenamefont {Deng}\ \emph {et~al.}(2022)\citenamefont {Deng},
  \citenamefont {Mishra}, \citenamefont {Mahayoni}, \citenamefont {Ma},
  \citenamefont {Tieu}, \citenamefont {Guillon}, \citenamefont {Chotard},
  \citenamefont {Seznec}, \citenamefont {Cheetham}, \citenamefont {Masquelier}
  \emph {et~al.}}]{deng2022fundamental}%
  \BibitemOpen
  \bibfield  {author} {\bibinfo {author} {\bibfnamefont {Z.}~\bibnamefont
  {Deng}}, \bibinfo {author} {\bibfnamefont {T.~P.}\ \bibnamefont {Mishra}},
  \bibinfo {author} {\bibfnamefont {E.}~\bibnamefont {Mahayoni}}, \bibinfo
  {author} {\bibfnamefont {Q.}~\bibnamefont {Ma}}, \bibinfo {author}
  {\bibfnamefont {A.~J.~K.}\ \bibnamefont {Tieu}}, \bibinfo {author}
  {\bibfnamefont {O.}~\bibnamefont {Guillon}}, \bibinfo {author} {\bibfnamefont
  {J.-N.}\ \bibnamefont {Chotard}}, \bibinfo {author} {\bibfnamefont
  {V.}~\bibnamefont {Seznec}}, \bibinfo {author} {\bibfnamefont {A.~K.}\
  \bibnamefont {Cheetham}}, \bibinfo {author} {\bibfnamefont {C.}~\bibnamefont
  {Masquelier}}, \emph {et~al.},\ }\bibfield  {title} {\bibinfo {title}
  {Fundamental investigations on the sodium-ion transport properties of mixed
  polyanion solid-state battery electrolytes},\ }\href@noop {} {\bibfield
  {journal} {\bibinfo  {journal} {Nature communications}\ }\textbf {\bibinfo
  {volume} {13}},\ \bibinfo {pages} {4470} (\bibinfo {year}
  {2022})}\BibitemShut {NoStop}%
\bibitem [{\citenamefont {Ning}\ \emph {et~al.}(2022)\citenamefont {Ning},
  \citenamefont {Furness},\ and\ \citenamefont {Sun}}]{ning2022reliable}%
  \BibitemOpen
  \bibfield  {author} {\bibinfo {author} {\bibfnamefont {J.}~\bibnamefont
  {Ning}}, \bibinfo {author} {\bibfnamefont {J.~W.}\ \bibnamefont {Furness}},\
  and\ \bibinfo {author} {\bibfnamefont {J.}~\bibnamefont {Sun}},\ }\bibfield
  {title} {\bibinfo {title} {Reliable lattice dynamics from an efficient
  density functional approximation},\ }\href@noop {} {\bibfield  {journal}
  {\bibinfo  {journal} {Chemistry of Materials}\ }\textbf {\bibinfo {volume}
  {34}},\ \bibinfo {pages} {2562} (\bibinfo {year} {2022})}\BibitemShut
  {NoStop}%
\bibitem [{\citenamefont {Kothakonda}\ \emph {et~al.}(2023)\citenamefont
  {Kothakonda}, \citenamefont {Zhu}, \citenamefont {Guan}, \citenamefont {He},
  \citenamefont {Kidd}, \citenamefont {Zhang}, \citenamefont {Ning},
  \citenamefont {Gopalan}, \citenamefont {Xie}, \citenamefont {Mao} \emph
  {et~al.}}]{kothakonda2023high}%
  \BibitemOpen
  \bibfield  {author} {\bibinfo {author} {\bibfnamefont {M.}~\bibnamefont
  {Kothakonda}}, \bibinfo {author} {\bibfnamefont {Y.}~\bibnamefont {Zhu}},
  \bibinfo {author} {\bibfnamefont {Y.}~\bibnamefont {Guan}}, \bibinfo {author}
  {\bibfnamefont {J.}~\bibnamefont {He}}, \bibinfo {author} {\bibfnamefont
  {J.}~\bibnamefont {Kidd}}, \bibinfo {author} {\bibfnamefont {R.}~\bibnamefont
  {Zhang}}, \bibinfo {author} {\bibfnamefont {J.}~\bibnamefont {Ning}},
  \bibinfo {author} {\bibfnamefont {V.}~\bibnamefont {Gopalan}}, \bibinfo
  {author} {\bibfnamefont {W.}~\bibnamefont {Xie}}, \bibinfo {author}
  {\bibfnamefont {Z.}~\bibnamefont {Mao}}, \emph {et~al.},\ }\bibfield  {title}
  {\bibinfo {title} {High-throughput screening assisted discovery of a stable
  layered anti-ferromagnetic semiconductor: Cdfep2se6},\ }\href@noop {}
  {\bibfield  {journal} {\bibinfo  {journal} {Advanced Functional Materials}\
  }\textbf {\bibinfo {volume} {33}},\ \bibinfo {pages} {2210965} (\bibinfo
  {year} {2023})}\BibitemShut {NoStop}%
\bibitem [{\citenamefont {Devi}\ \emph {et~al.}(2022)\citenamefont {Devi},
  \citenamefont {Singh}, \citenamefont {Canepa},\ and\ \citenamefont
  {Sai~Gautam}}]{devi2022effect}%
  \BibitemOpen
  \bibfield  {author} {\bibinfo {author} {\bibfnamefont {R.}~\bibnamefont
  {Devi}}, \bibinfo {author} {\bibfnamefont {B.}~\bibnamefont {Singh}},
  \bibinfo {author} {\bibfnamefont {P.}~\bibnamefont {Canepa}},\ and\ \bibinfo
  {author} {\bibfnamefont {G.}~\bibnamefont {Sai~Gautam}},\ }\bibfield  {title}
  {\bibinfo {title} {Effect of exchange-correlation functionals on the
  estimation of migration barriers in battery materials},\ }\href@noop {}
  {\bibfield  {journal} {\bibinfo  {journal} {npj Computational Materials}\
  }\textbf {\bibinfo {volume} {8}},\ \bibinfo {pages} {160} (\bibinfo {year}
  {2022})}\BibitemShut {NoStop}%
\bibitem [{\citenamefont {Jha}\ \emph {et~al.}(2023)\citenamefont {Jha},
  \citenamefont {Totade}, \citenamefont {Barpanda},\ and\ \citenamefont
  {Sai~Gautam}}]{jha2023evaluation}%
  \BibitemOpen
  \bibfield  {author} {\bibinfo {author} {\bibfnamefont {P.~K.}\ \bibnamefont
  {Jha}}, \bibinfo {author} {\bibfnamefont {S.~N.}\ \bibnamefont {Totade}},
  \bibinfo {author} {\bibfnamefont {P.}~\bibnamefont {Barpanda}},\ and\
  \bibinfo {author} {\bibfnamefont {G.}~\bibnamefont {Sai~Gautam}},\ }\bibfield
   {title} {\bibinfo {title} {Evaluation of p3-type layered oxides as k-ion
  battery cathodes},\ }\href@noop {} {\bibfield  {journal} {\bibinfo  {journal}
  {Inorganic Chemistry}\ }\textbf {\bibinfo {volume} {62}},\ \bibinfo {pages}
  {14971} (\bibinfo {year} {2023})}\BibitemShut {NoStop}%
\bibitem [{\citenamefont {Kumar}\ and\ \citenamefont
  {Gautam}(2023)}]{kumar2023study}%
  \BibitemOpen
  \bibfield  {author} {\bibinfo {author} {\bibfnamefont {J.}~\bibnamefont
  {Kumar}}\ and\ \bibinfo {author} {\bibfnamefont {G.~S.}\ \bibnamefont
  {Gautam}},\ }\bibfield  {title} {\bibinfo {title} {Study of pnictides for
  photovoltaic applications},\ }\href@noop {} {\bibfield  {journal} {\bibinfo
  {journal} {Physical Chemistry Chemical Physics}\ }\textbf {\bibinfo {volume}
  {25}},\ \bibinfo {pages} {9626} (\bibinfo {year} {2023})}\BibitemShut
  {NoStop}%
\bibitem [{\citenamefont {Perdew}\ and\ \citenamefont
  {Zunger}(1981)}]{perdew1981self}%
  \BibitemOpen
  \bibfield  {author} {\bibinfo {author} {\bibfnamefont {J.~P.}\ \bibnamefont
  {Perdew}}\ and\ \bibinfo {author} {\bibfnamefont {A.}~\bibnamefont
  {Zunger}},\ }\bibfield  {title} {\bibinfo {title} {Self-interaction
  correction to density-functional approximations for many-electron systems},\
  }\href@noop {} {\bibfield  {journal} {\bibinfo  {journal} {Physical Review
  B}\ }\textbf {\bibinfo {volume} {23}},\ \bibinfo {pages} {5048} (\bibinfo
  {year} {1981})}\BibitemShut {NoStop}%
\bibitem [{\citenamefont {Gautam}\ and\ \citenamefont
  {Carter}(2018)}]{gautam2018evaluating}%
  \BibitemOpen
  \bibfield  {author} {\bibinfo {author} {\bibfnamefont {G.~S.}\ \bibnamefont
  {Gautam}}\ and\ \bibinfo {author} {\bibfnamefont {E.~A.}\ \bibnamefont
  {Carter}},\ }\bibfield  {title} {\bibinfo {title} {Evaluating transition
  metal oxides within dft-scan and scan+ u frameworks for solar thermochemical
  applications},\ }\href@noop {} {\bibfield  {journal} {\bibinfo  {journal}
  {Physical Review Materials}\ }\textbf {\bibinfo {volume} {2}},\ \bibinfo
  {pages} {095401} (\bibinfo {year} {2018})}\BibitemShut {NoStop}%
\bibitem [{\citenamefont {Long}\ \emph {et~al.}(2020)\citenamefont {Long},
  \citenamefont {Gautam},\ and\ \citenamefont {Carter}}]{long2020evaluating}%
  \BibitemOpen
  \bibfield  {author} {\bibinfo {author} {\bibfnamefont {O.~Y.}\ \bibnamefont
  {Long}}, \bibinfo {author} {\bibfnamefont {G.~S.}\ \bibnamefont {Gautam}},\
  and\ \bibinfo {author} {\bibfnamefont {E.~A.}\ \bibnamefont {Carter}},\
  }\bibfield  {title} {\bibinfo {title} {Evaluating optimal u for 3 d
  transition-metal oxides within the scan+ u framework},\ }\href@noop {}
  {\bibfield  {journal} {\bibinfo  {journal} {Physical Review Materials}\
  }\textbf {\bibinfo {volume} {4}},\ \bibinfo {pages} {045401} (\bibinfo {year}
  {2020})}\BibitemShut {NoStop}%
\bibitem [{\citenamefont {Long}\ \emph {et~al.}(2021)\citenamefont {Long},
  \citenamefont {Gautam},\ and\ \citenamefont {Carter}}]{long2021assessing}%
  \BibitemOpen
  \bibfield  {author} {\bibinfo {author} {\bibfnamefont {O.~Y.}\ \bibnamefont
  {Long}}, \bibinfo {author} {\bibfnamefont {G.~S.}\ \bibnamefont {Gautam}},\
  and\ \bibinfo {author} {\bibfnamefont {E.~A.}\ \bibnamefont {Carter}},\
  }\bibfield  {title} {\bibinfo {title} {Assessing cathode property prediction
  via exchange-correlation functionals with and without long-range dispersion
  corrections},\ }\href@noop {} {\bibfield  {journal} {\bibinfo  {journal}
  {Physical Chemistry Chemical Physics}\ }\textbf {\bibinfo {volume} {23}},\
  \bibinfo {pages} {24726} (\bibinfo {year} {2021})}\BibitemShut {NoStop}%
\bibitem [{\citenamefont {Anisimov}\ \emph {et~al.}(1991)\citenamefont
  {Anisimov}, \citenamefont {Zaanen},\ and\ \citenamefont
  {Andersen}}]{anisimov1991band}%
  \BibitemOpen
  \bibfield  {author} {\bibinfo {author} {\bibfnamefont {V.~I.}\ \bibnamefont
  {Anisimov}}, \bibinfo {author} {\bibfnamefont {J.}~\bibnamefont {Zaanen}},\
  and\ \bibinfo {author} {\bibfnamefont {O.~K.}\ \bibnamefont {Andersen}},\
  }\bibfield  {title} {\bibinfo {title} {Band theory and mott insulators:
  Hubbard u instead of stoner i},\ }\href@noop {} {\bibfield  {journal}
  {\bibinfo  {journal} {Physical Review B}\ }\textbf {\bibinfo {volume} {44}},\
  \bibinfo {pages} {943} (\bibinfo {year} {1991})}\BibitemShut {NoStop}%
\bibitem [{\citenamefont {Timrov}\ \emph {et~al.}(2018)\citenamefont {Timrov},
  \citenamefont {Marzari},\ and\ \citenamefont
  {Cococcioni}}]{timrov2018hubbard}%
  \BibitemOpen
  \bibfield  {author} {\bibinfo {author} {\bibfnamefont {I.}~\bibnamefont
  {Timrov}}, \bibinfo {author} {\bibfnamefont {N.}~\bibnamefont {Marzari}},\
  and\ \bibinfo {author} {\bibfnamefont {M.}~\bibnamefont {Cococcioni}},\
  }\bibfield  {title} {\bibinfo {title} {Hubbard parameters from
  density-functional perturbation theory},\ }\href@noop {} {\bibfield
  {journal} {\bibinfo  {journal} {Physical Review B}\ }\textbf {\bibinfo
  {volume} {98}},\ \bibinfo {pages} {085127} (\bibinfo {year}
  {2018})}\BibitemShut {NoStop}%
\bibitem [{\citenamefont {Zhou}\ \emph {et~al.}(2004)\citenamefont {Zhou},
  \citenamefont {Cococcioni}, \citenamefont {Marianetti}, \citenamefont
  {Morgan},\ and\ \citenamefont {Ceder}}]{zhou2004first}%
  \BibitemOpen
  \bibfield  {author} {\bibinfo {author} {\bibfnamefont {F.}~\bibnamefont
  {Zhou}}, \bibinfo {author} {\bibfnamefont {M.}~\bibnamefont {Cococcioni}},
  \bibinfo {author} {\bibfnamefont {C.~A.}\ \bibnamefont {Marianetti}},
  \bibinfo {author} {\bibfnamefont {D.}~\bibnamefont {Morgan}},\ and\ \bibinfo
  {author} {\bibfnamefont {G.}~\bibnamefont {Ceder}},\ }\bibfield  {title}
  {\bibinfo {title} {First-principles prediction of redox potentials in
  transition-metal compounds with lda+ u},\ }\href@noop {} {\bibfield
  {journal} {\bibinfo  {journal} {Physical Review B}\ }\textbf {\bibinfo
  {volume} {70}},\ \bibinfo {pages} {235121} (\bibinfo {year}
  {2004})}\BibitemShut {NoStop}%
\bibitem [{\citenamefont {Moore}\ \emph {et~al.}(2022)\citenamefont {Moore},
  \citenamefont {Horton}, \citenamefont {Ganose}, \citenamefont {Siron},
  \citenamefont {Linscott}, \citenamefont {O'Regan},\ and\ \citenamefont
  {Persson}}]{moore2022high}%
  \BibitemOpen
  \bibfield  {author} {\bibinfo {author} {\bibfnamefont {G.~C.}\ \bibnamefont
  {Moore}}, \bibinfo {author} {\bibfnamefont {M.~K.}\ \bibnamefont {Horton}},
  \bibinfo {author} {\bibfnamefont {A.~M.}\ \bibnamefont {Ganose}}, \bibinfo
  {author} {\bibfnamefont {M.}~\bibnamefont {Siron}}, \bibinfo {author}
  {\bibfnamefont {E.}~\bibnamefont {Linscott}}, \bibinfo {author}
  {\bibfnamefont {D.~D.}\ \bibnamefont {O'Regan}},\ and\ \bibinfo {author}
  {\bibfnamefont {K.~A.}\ \bibnamefont {Persson}},\ }\bibfield  {title}
  {\bibinfo {title} {High-throughput determination of hubbard u and hund j
  values for transition metal oxides via linear response formalism},\
  }\href@noop {} {\bibfield  {journal} {\bibinfo  {journal} {arXiv preprint
  arXiv:2201.04213}\ } (\bibinfo {year} {2022})}\BibitemShut {NoStop}%
\bibitem [{\citenamefont {Lan}\ \emph {et~al.}(2018)\citenamefont {Lan},
  \citenamefont {Song},\ and\ \citenamefont {Yang}}]{lan2018linear}%
  \BibitemOpen
  \bibfield  {author} {\bibinfo {author} {\bibfnamefont {G.}~\bibnamefont
  {Lan}}, \bibinfo {author} {\bibfnamefont {J.}~\bibnamefont {Song}},\ and\
  \bibinfo {author} {\bibfnamefont {Z.}~\bibnamefont {Yang}},\ }\bibfield
  {title} {\bibinfo {title} {A linear response approach to determine hubbard u
  and its application to evaluate properties of y2b2o7, b= transition metals
  3d, 4d and 5d},\ }\href@noop {} {\bibfield  {journal} {\bibinfo  {journal}
  {Journal of Alloys and Compounds}\ }\textbf {\bibinfo {volume} {749}},\
  \bibinfo {pages} {909} (\bibinfo {year} {2018})}\BibitemShut {NoStop}%
\bibitem [{\citenamefont {Shishkin}\ and\ \citenamefont
  {Sato}(2016)}]{shishkin2016self}%
  \BibitemOpen
  \bibfield  {author} {\bibinfo {author} {\bibfnamefont {M.}~\bibnamefont
  {Shishkin}}\ and\ \bibinfo {author} {\bibfnamefont {H.}~\bibnamefont
  {Sato}},\ }\bibfield  {title} {\bibinfo {title} {Self-consistent
  parametrization of dft+ u framework using linear response approach:
  Application to evaluation of redox potentials of battery cathodes},\
  }\href@noop {} {\bibfield  {journal} {\bibinfo  {journal} {Physical Review
  B}\ }\textbf {\bibinfo {volume} {93}},\ \bibinfo {pages} {085135} (\bibinfo
  {year} {2016})}\BibitemShut {NoStop}%
\bibitem [{\citenamefont {Mosey}\ and\ \citenamefont
  {Carter}(2007)}]{mosey2007ab}%
  \BibitemOpen
  \bibfield  {author} {\bibinfo {author} {\bibfnamefont {N.~J.}\ \bibnamefont
  {Mosey}}\ and\ \bibinfo {author} {\bibfnamefont {E.~A.}\ \bibnamefont
  {Carter}},\ }\bibfield  {title} {\bibinfo {title} {Ab initio evaluation of
  coulomb and exchange parameters for dft+ u calculations},\ }\href@noop {}
  {\bibfield  {journal} {\bibinfo  {journal} {Physical Review B}\ }\textbf
  {\bibinfo {volume} {76}},\ \bibinfo {pages} {155123} (\bibinfo {year}
  {2007})}\BibitemShut {NoStop}%
\bibitem [{\citenamefont {Mosey}\ \emph {et~al.}(2008)\citenamefont {Mosey},
  \citenamefont {Liao},\ and\ \citenamefont {Carter}}]{mosey2008rotationally}%
  \BibitemOpen
  \bibfield  {author} {\bibinfo {author} {\bibfnamefont {N.~J.}\ \bibnamefont
  {Mosey}}, \bibinfo {author} {\bibfnamefont {P.}~\bibnamefont {Liao}},\ and\
  \bibinfo {author} {\bibfnamefont {E.~A.}\ \bibnamefont {Carter}},\ }\bibfield
   {title} {\bibinfo {title} {Rotationally invariant ab initio evaluation of
  coulomb and exchange parameters for dft+ u calculations},\ }\href@noop {}
  {\bibfield  {journal} {\bibinfo  {journal} {The Journal of chemical physics}\
  }\textbf {\bibinfo {volume} {129}} (\bibinfo {year} {2008})}\BibitemShut
  {NoStop}%
\bibitem [{\citenamefont {Yu}\ \emph {et~al.}(2020)\citenamefont {Yu},
  \citenamefont {Yang}, \citenamefont {Wu},\ and\ \citenamefont
  {Marom}}]{yu2020machine}%
  \BibitemOpen
  \bibfield  {author} {\bibinfo {author} {\bibfnamefont {M.}~\bibnamefont
  {Yu}}, \bibinfo {author} {\bibfnamefont {S.}~\bibnamefont {Yang}}, \bibinfo
  {author} {\bibfnamefont {C.}~\bibnamefont {Wu}},\ and\ \bibinfo {author}
  {\bibfnamefont {N.}~\bibnamefont {Marom}},\ }\bibfield  {title} {\bibinfo
  {title} {Machine learning the hubbard u parameter in dft+ u using bayesian
  optimization},\ }\href@noop {} {\bibfield  {journal} {\bibinfo  {journal}
  {npj computational materials}\ }\textbf {\bibinfo {volume} {6}},\ \bibinfo
  {pages} {180} (\bibinfo {year} {2020})}\BibitemShut {NoStop}%
\bibitem [{\citenamefont {Artrith}\ \emph {et~al.}(2022)\citenamefont
  {Artrith}, \citenamefont {Garrido~Torres}, \citenamefont {Urban},\ and\
  \citenamefont {Hybertsen}}]{Artrith@2022data}%
  \BibitemOpen
  \bibfield  {author} {\bibinfo {author} {\bibfnamefont {N.}~\bibnamefont
  {Artrith}}, \bibinfo {author} {\bibfnamefont {J.~A.}\ \bibnamefont
  {Garrido~Torres}}, \bibinfo {author} {\bibfnamefont {A.}~\bibnamefont
  {Urban}},\ and\ \bibinfo {author} {\bibfnamefont {M.~S.}\ \bibnamefont
  {Hybertsen}},\ }\bibfield  {title} {\bibinfo {title} {Data-driven approach to
  parameterize $\mathrm{SCAN}+u$ for an accurate description of $3d$ transition
  metal oxide thermochemistry},\ }\href@noop {} {\bibfield  {journal} {\bibinfo
   {journal} {Phys. Rev. Mater.}\ }\textbf {\bibinfo {volume} {6}},\ \bibinfo
  {pages} {035003} (\bibinfo {year} {2022})}\BibitemShut {NoStop}%
\bibitem [{\citenamefont {Wang}\ \emph {et~al.}(2006)\citenamefont {Wang},
  \citenamefont {Maxisch},\ and\ \citenamefont {Ceder}}]{wang2006oxidation}%
  \BibitemOpen
  \bibfield  {author} {\bibinfo {author} {\bibfnamefont {L.}~\bibnamefont
  {Wang}}, \bibinfo {author} {\bibfnamefont {T.}~\bibnamefont {Maxisch}},\ and\
  \bibinfo {author} {\bibfnamefont {G.}~\bibnamefont {Ceder}},\ }\bibfield
  {title} {\bibinfo {title} {Oxidation energies of transition metal oxides
  within the gga+ u framework},\ }\href@noop {} {\bibfield  {journal} {\bibinfo
   {journal} {Physical Review B}\ }\textbf {\bibinfo {volume} {73}},\ \bibinfo
  {pages} {195107} (\bibinfo {year} {2006})}\BibitemShut {NoStop}%
\bibitem [{\citenamefont {Jain}\ \emph {et~al.}(2011)\citenamefont {Jain},
  \citenamefont {Hautier}, \citenamefont {Ong}, \citenamefont {Moore},
  \citenamefont {Fischer}, \citenamefont {Persson},\ and\ \citenamefont
  {Ceder}}]{jain2011formation}%
  \BibitemOpen
  \bibfield  {author} {\bibinfo {author} {\bibfnamefont {A.}~\bibnamefont
  {Jain}}, \bibinfo {author} {\bibfnamefont {G.}~\bibnamefont {Hautier}},
  \bibinfo {author} {\bibfnamefont {S.~P.}\ \bibnamefont {Ong}}, \bibinfo
  {author} {\bibfnamefont {C.~J.}\ \bibnamefont {Moore}}, \bibinfo {author}
  {\bibfnamefont {C.~C.}\ \bibnamefont {Fischer}}, \bibinfo {author}
  {\bibfnamefont {K.~A.}\ \bibnamefont {Persson}},\ and\ \bibinfo {author}
  {\bibfnamefont {G.}~\bibnamefont {Ceder}},\ }\bibfield  {title} {\bibinfo
  {title} {Formation enthalpies by mixing gga and gga+ u calculations},\
  }\href@noop {} {\bibfield  {journal} {\bibinfo  {journal} {Physical Review
  B}\ }\textbf {\bibinfo {volume} {84}},\ \bibinfo {pages} {045115} (\bibinfo
  {year} {2011})}\BibitemShut {NoStop}%
\bibitem [{\citenamefont {Lutfalla}\ \emph {et~al.}(2011)\citenamefont
  {Lutfalla}, \citenamefont {Shapovalov},\ and\ \citenamefont
  {Bell}}]{lutfalla2011calibration}%
  \BibitemOpen
  \bibfield  {author} {\bibinfo {author} {\bibfnamefont {S.}~\bibnamefont
  {Lutfalla}}, \bibinfo {author} {\bibfnamefont {V.}~\bibnamefont
  {Shapovalov}},\ and\ \bibinfo {author} {\bibfnamefont {A.~T.}\ \bibnamefont
  {Bell}},\ }\bibfield  {title} {\bibinfo {title} {Calibration of the dft/gga+
  u method for determination of reduction energies for transition and rare
  earth metal oxides of ti, v, mo, and ce},\ }\href@noop {} {\bibfield
  {journal} {\bibinfo  {journal} {Journal of chemical theory and computation}\
  }\textbf {\bibinfo {volume} {7}},\ \bibinfo {pages} {2218} (\bibinfo {year}
  {2011})}\BibitemShut {NoStop}%
\bibitem [{\citenamefont {Loschen}\ \emph {et~al.}(2007)\citenamefont
  {Loschen}, \citenamefont {Carrasco}, \citenamefont {Neyman},\ and\
  \citenamefont {Illas}}]{loschen2007first}%
  \BibitemOpen
  \bibfield  {author} {\bibinfo {author} {\bibfnamefont {C.}~\bibnamefont
  {Loschen}}, \bibinfo {author} {\bibfnamefont {J.}~\bibnamefont {Carrasco}},
  \bibinfo {author} {\bibfnamefont {K.~M.}\ \bibnamefont {Neyman}},\ and\
  \bibinfo {author} {\bibfnamefont {F.}~\bibnamefont {Illas}},\ }\bibfield
  {title} {\bibinfo {title} {First-principles lda+ u and gga+ u study of cerium
  oxides: Dependence on the effective u parameter},\ }\href@noop {} {\bibfield
  {journal} {\bibinfo  {journal} {Physical Review B}\ }\textbf {\bibinfo
  {volume} {75}},\ \bibinfo {pages} {035115} (\bibinfo {year}
  {2007})}\BibitemShut {NoStop}%
\bibitem [{\citenamefont {Curtarolo}\ \emph {et~al.}(2013)\citenamefont
  {Curtarolo}, \citenamefont {Hart}, \citenamefont {Nardelli}, \citenamefont
  {Mingo}, \citenamefont {Sanvito},\ and\ \citenamefont
  {Levy}}]{curtarolo2013high}%
  \BibitemOpen
  \bibfield  {author} {\bibinfo {author} {\bibfnamefont {S.}~\bibnamefont
  {Curtarolo}}, \bibinfo {author} {\bibfnamefont {G.~L.}\ \bibnamefont {Hart}},
  \bibinfo {author} {\bibfnamefont {M.~B.}\ \bibnamefont {Nardelli}}, \bibinfo
  {author} {\bibfnamefont {N.}~\bibnamefont {Mingo}}, \bibinfo {author}
  {\bibfnamefont {S.}~\bibnamefont {Sanvito}},\ and\ \bibinfo {author}
  {\bibfnamefont {O.}~\bibnamefont {Levy}},\ }\bibfield  {title} {\bibinfo
  {title} {The high-throughput highway to computational materials design},\
  }\href@noop {} {\bibfield  {journal} {\bibinfo  {journal} {Nature materials}\
  }\textbf {\bibinfo {volume} {12}},\ \bibinfo {pages} {191} (\bibinfo {year}
  {2013})}\BibitemShut {NoStop}%
\bibitem [{\citenamefont {Lun}\ \emph {et~al.}(2021)\citenamefont {Lun},
  \citenamefont {Ouyang}, \citenamefont {Kwon}, \citenamefont {Ha},
  \citenamefont {Foley}, \citenamefont {Huang}, \citenamefont {Cai},
  \citenamefont {Kim}, \citenamefont {Balasubramanian}, \citenamefont {Sun}
  \emph {et~al.}}]{lun2021cation}%
  \BibitemOpen
  \bibfield  {author} {\bibinfo {author} {\bibfnamefont {Z.}~\bibnamefont
  {Lun}}, \bibinfo {author} {\bibfnamefont {B.}~\bibnamefont {Ouyang}},
  \bibinfo {author} {\bibfnamefont {D.-H.}\ \bibnamefont {Kwon}}, \bibinfo
  {author} {\bibfnamefont {Y.}~\bibnamefont {Ha}}, \bibinfo {author}
  {\bibfnamefont {E.~E.}\ \bibnamefont {Foley}}, \bibinfo {author}
  {\bibfnamefont {T.-Y.}\ \bibnamefont {Huang}}, \bibinfo {author}
  {\bibfnamefont {Z.}~\bibnamefont {Cai}}, \bibinfo {author} {\bibfnamefont
  {H.}~\bibnamefont {Kim}}, \bibinfo {author} {\bibfnamefont {M.}~\bibnamefont
  {Balasubramanian}}, \bibinfo {author} {\bibfnamefont {Y.}~\bibnamefont
  {Sun}}, \emph {et~al.},\ }\bibfield  {title} {\bibinfo {title}
  {Cation-disordered rocksalt-type high-entropy cathodes for li-ion
  batteries},\ }\href@noop {} {\bibfield  {journal} {\bibinfo  {journal}
  {Nature materials}\ }\textbf {\bibinfo {volume} {20}},\ \bibinfo {pages}
  {214} (\bibinfo {year} {2021})}\BibitemShut {NoStop}%
\bibitem [{\citenamefont {Bartel}\ \emph {et~al.}(2019)\citenamefont {Bartel},
  \citenamefont {Weimer}, \citenamefont {Lany}, \citenamefont {Musgrave},\ and\
  \citenamefont {Holder}}]{bartel2019role}%
  \BibitemOpen
  \bibfield  {author} {\bibinfo {author} {\bibfnamefont {C.~J.}\ \bibnamefont
  {Bartel}}, \bibinfo {author} {\bibfnamefont {A.~W.}\ \bibnamefont {Weimer}},
  \bibinfo {author} {\bibfnamefont {S.}~\bibnamefont {Lany}}, \bibinfo {author}
  {\bibfnamefont {C.~B.}\ \bibnamefont {Musgrave}},\ and\ \bibinfo {author}
  {\bibfnamefont {A.~M.}\ \bibnamefont {Holder}},\ }\bibfield  {title}
  {\bibinfo {title} {The role of decomposition reactions in assessing
  first-principles predictions of solid stability},\ }\href@noop {} {\bibfield
  {journal} {\bibinfo  {journal} {npj Computational Materials}\ }\textbf
  {\bibinfo {volume} {5}},\ \bibinfo {pages} {4} (\bibinfo {year}
  {2019})}\BibitemShut {NoStop}%
\bibitem [{\citenamefont {Zhong}\ \emph {et~al.}(2023)\citenamefont {Zhong},
  \citenamefont {Xie}, \citenamefont {Barroso-Luque}, \citenamefont {Huang},\
  and\ \citenamefont {Ceder}}]{zhong2023modeling}%
  \BibitemOpen
  \bibfield  {author} {\bibinfo {author} {\bibfnamefont {P.}~\bibnamefont
  {Zhong}}, \bibinfo {author} {\bibfnamefont {F.}~\bibnamefont {Xie}}, \bibinfo
  {author} {\bibfnamefont {L.}~\bibnamefont {Barroso-Luque}}, \bibinfo {author}
  {\bibfnamefont {L.}~\bibnamefont {Huang}},\ and\ \bibinfo {author}
  {\bibfnamefont {G.}~\bibnamefont {Ceder}},\ }\bibfield  {title} {\bibinfo
  {title} {Modeling intercalation chemistry with multiredox reactions by sparse
  lattice models in disordered rocksalt cathodes},\ }\href@noop {} {\bibfield
  {journal} {\bibinfo  {journal} {PRX Energy}\ }\textbf {\bibinfo {volume}
  {2}},\ \bibinfo {pages} {043005} (\bibinfo {year} {2023})}\BibitemShut
  {NoStop}%
\bibitem [{\citenamefont {Kresse}\ and\ \citenamefont
  {Hafner}(1993)}]{kresse1993abinitio}%
  \BibitemOpen
  \bibfield  {author} {\bibinfo {author} {\bibfnamefont {G.}~\bibnamefont
  {Kresse}}\ and\ \bibinfo {author} {\bibfnamefont {J.}~\bibnamefont
  {Hafner}},\ }\bibfield  {title} {\bibinfo {title} {Ab initio molecular
  dynamics for liquid metals},\ }\href@noop {} {\bibfield  {journal} {\bibinfo
  {journal} {Phys. Rev. B}\ }\textbf {\bibinfo {volume} {47}},\ \bibinfo
  {pages} {558} (\bibinfo {year} {1993})}\BibitemShut {NoStop}%
\bibitem [{\citenamefont {Kresse}\ and\ \citenamefont
  {Furthm{\"u}ller}(1996)}]{kresse1996efficient}%
  \BibitemOpen
  \bibfield  {author} {\bibinfo {author} {\bibfnamefont {G.}~\bibnamefont
  {Kresse}}\ and\ \bibinfo {author} {\bibfnamefont {J.}~\bibnamefont
  {Furthm{\"u}ller}},\ }\bibfield  {title} {\bibinfo {title} {Efficient
  iterative schemes for ab initio total-energy calculations using a plane-wave
  basis set},\ }\href@noop {} {\bibfield  {journal} {\bibinfo  {journal}
  {Physical review B}\ }\textbf {\bibinfo {volume} {54}},\ \bibinfo {pages}
  {11169} (\bibinfo {year} {1996})}\BibitemShut {NoStop}%
\bibitem [{\citenamefont {Kresse}\ and\ \citenamefont
  {Joubert}(1999)}]{kresse1999ultrasoft}%
  \BibitemOpen
  \bibfield  {author} {\bibinfo {author} {\bibfnamefont {G.}~\bibnamefont
  {Kresse}}\ and\ \bibinfo {author} {\bibfnamefont {D.}~\bibnamefont
  {Joubert}},\ }\bibfield  {title} {\bibinfo {title} {From ultrasoft
  pseudopotentials to the projector augmented-wave method},\ }\href@noop {}
  {\bibfield  {journal} {\bibinfo  {journal} {Physical review b}\ }\textbf
  {\bibinfo {volume} {59}},\ \bibinfo {pages} {1758} (\bibinfo {year}
  {1999})}\BibitemShut {NoStop}%
\bibitem [{\citenamefont {Bl{\"o}chl}(1994)}]{blochl1994projector}%
  \BibitemOpen
  \bibfield  {author} {\bibinfo {author} {\bibfnamefont {P.~E.}\ \bibnamefont
  {Bl{\"o}chl}},\ }\bibfield  {title} {\bibinfo {title} {Projector
  augmented-wave method},\ }\href@noop {} {\bibfield  {journal} {\bibinfo
  {journal} {Physical review B}\ }\textbf {\bibinfo {volume} {50}},\ \bibinfo
  {pages} {17953} (\bibinfo {year} {1994})}\BibitemShut {NoStop}%
\bibitem [{\citenamefont {Dudarev}\ \emph {et~al.}(1998)\citenamefont
  {Dudarev}, \citenamefont {Botton}, \citenamefont {Savrasov}, \citenamefont
  {Humphreys},\ and\ \citenamefont {Sutton}}]{dudarev1998electron}%
  \BibitemOpen
  \bibfield  {author} {\bibinfo {author} {\bibfnamefont {S.~L.}\ \bibnamefont
  {Dudarev}}, \bibinfo {author} {\bibfnamefont {G.~A.}\ \bibnamefont {Botton}},
  \bibinfo {author} {\bibfnamefont {S.~Y.}\ \bibnamefont {Savrasov}}, \bibinfo
  {author} {\bibfnamefont {C.}~\bibnamefont {Humphreys}},\ and\ \bibinfo
  {author} {\bibfnamefont {A.~P.}\ \bibnamefont {Sutton}},\ }\bibfield  {title}
  {\bibinfo {title} {Electron-energy-loss spectra and the structural stability
  of nickel oxide: An lsda+ u study},\ }\href@noop {} {\bibfield  {journal}
  {\bibinfo  {journal} {Physical Review B}\ }\textbf {\bibinfo {volume} {57}},\
  \bibinfo {pages} {1505} (\bibinfo {year} {1998})}\BibitemShut {NoStop}%
\bibitem [{\citenamefont {Monkhorst}\ and\ \citenamefont
  {Pack}(1976)}]{monkhorst1976special}%
  \BibitemOpen
  \bibfield  {author} {\bibinfo {author} {\bibfnamefont {H.~J.}\ \bibnamefont
  {Monkhorst}}\ and\ \bibinfo {author} {\bibfnamefont {J.~D.}\ \bibnamefont
  {Pack}},\ }\bibfield  {title} {\bibinfo {title} {Special points for
  brillouin-zone integrations},\ }\href@noop {} {\bibfield  {journal} {\bibinfo
   {journal} {Physical review B}\ }\textbf {\bibinfo {volume} {13}},\ \bibinfo
  {pages} {5188} (\bibinfo {year} {1976})}\BibitemShut {NoStop}%
\bibitem [{\citenamefont {Hellenbrandt}(2004)}]{hellenbrandt2004inorganic}%
  \BibitemOpen
  \bibfield  {author} {\bibinfo {author} {\bibfnamefont {M.}~\bibnamefont
  {Hellenbrandt}},\ }\bibfield  {title} {\bibinfo {title} {The inorganic
  crystal structure database (icsd)—present and future},\ }\href@noop {}
  {\bibfield  {journal} {\bibinfo  {journal} {Crystallography Reviews}\
  }\textbf {\bibinfo {volume} {10}},\ \bibinfo {pages} {17} (\bibinfo {year}
  {2004})}\BibitemShut {NoStop}%
\bibitem [{\citenamefont {Wollan}\ \emph {et~al.}(1958)\citenamefont {Wollan},
  \citenamefont {Child}, \citenamefont {Koehler},\ and\ \citenamefont
  {Wilkinson}}]{wollan1958antiferromagnetic}%
  \BibitemOpen
  \bibfield  {author} {\bibinfo {author} {\bibfnamefont {E.}~\bibnamefont
  {Wollan}}, \bibinfo {author} {\bibfnamefont {H.}~\bibnamefont {Child}},
  \bibinfo {author} {\bibfnamefont {W.}~\bibnamefont {Koehler}},\ and\ \bibinfo
  {author} {\bibfnamefont {M.}~\bibnamefont {Wilkinson}},\ }\bibfield  {title}
  {\bibinfo {title} {Antiferromagnetic properties of the iron group
  trifluorides},\ }\href@noop {} {\bibfield  {journal} {\bibinfo  {journal}
  {Physical Review}\ }\textbf {\bibinfo {volume} {112}},\ \bibinfo {pages}
  {1132} (\bibinfo {year} {1958})}\BibitemShut {NoStop}%
\bibitem [{\citenamefont {Perdew}\ \emph {et~al.}(2017)\citenamefont {Perdew},
  \citenamefont {Yang}, \citenamefont {Burke}, \citenamefont {Yang},
  \citenamefont {Gross}, \citenamefont {Scheffler}, \citenamefont {Scuseria},
  \citenamefont {Henderson}, \citenamefont {Zhang}, \citenamefont {Ruzsinszky}
  \emph {et~al.}}]{perdew2017understanding}%
  \BibitemOpen
  \bibfield  {author} {\bibinfo {author} {\bibfnamefont {J.~P.}\ \bibnamefont
  {Perdew}}, \bibinfo {author} {\bibfnamefont {W.}~\bibnamefont {Yang}},
  \bibinfo {author} {\bibfnamefont {K.}~\bibnamefont {Burke}}, \bibinfo
  {author} {\bibfnamefont {Z.}~\bibnamefont {Yang}}, \bibinfo {author}
  {\bibfnamefont {E.~K.}\ \bibnamefont {Gross}}, \bibinfo {author}
  {\bibfnamefont {M.}~\bibnamefont {Scheffler}}, \bibinfo {author}
  {\bibfnamefont {G.~E.}\ \bibnamefont {Scuseria}}, \bibinfo {author}
  {\bibfnamefont {T.~M.}\ \bibnamefont {Henderson}}, \bibinfo {author}
  {\bibfnamefont {I.~Y.}\ \bibnamefont {Zhang}}, \bibinfo {author}
  {\bibfnamefont {A.}~\bibnamefont {Ruzsinszky}}, \emph {et~al.},\ }\bibfield
  {title} {\bibinfo {title} {Understanding band gaps of solids in generalized
  kohn--sham theory},\ }\href@noop {} {\bibfield  {journal} {\bibinfo
  {journal} {Proceedings of the national academy of sciences}\ }\textbf
  {\bibinfo {volume} {114}},\ \bibinfo {pages} {2801} (\bibinfo {year}
  {2017})}\BibitemShut {NoStop}%
\bibitem [{\citenamefont {Allison}(1998)}]{allison1998nist}%
  \BibitemOpen
  \bibfield  {author} {\bibinfo {author} {\bibfnamefont {T.~C.}\ \bibnamefont
  {Allison}},\ }\bibfield  {title} {\bibinfo {title} {Nist-janaf thermochemical
  tables - srd 13}\ }(\bibinfo {year} {1998})\BibitemShut {NoStop}%
\bibitem [{\citenamefont {O.~Kubaschewski}(1980)}]{kubaschewski}%
  \BibitemOpen
  \bibfield  {author} {\bibinfo {author} {\bibfnamefont {C.~B.~A.}\
  \bibnamefont {O.~Kubaschewski}},\ }\bibfield  {title} {\bibinfo {title}
  {Metallurgical thermochemistry. 5th edition revised and enlarged.
  international series on materials science and technology, vol. 24, g. v.
  raynor (ed.). pergamon press oxford, new york, toronto, sidney, paris,
  frankfurt 1979 500 seiten mit 180 tabellen, 118 abbildungen und 984
  literaturzitate. preis: Broschiert us \$ 20,00},\ }\href@noop {} {\bibfield
  {journal} {\bibinfo  {journal} {Kristall und Technik}\ }\textbf {\bibinfo
  {volume} {15}},\ \bibinfo {pages} {176} (\bibinfo {year} {1980})}\BibitemShut
  {NoStop}%
\bibitem [{\citenamefont {Barin}(1995)}]{barin1995thermochemical}%
  \BibitemOpen
  \bibfield  {author} {\bibinfo {author} {\bibfnamefont {I.}~\bibnamefont
  {Barin}},\ }\href@noop {} {\emph {\bibinfo {title} {Thermochemical Data of
  Pure Substances}}},\ \bibinfo {edition} {3rd}\ ed.\ (\bibinfo  {publisher}
  {John Wiley \& Sons, Ltd},\ \bibinfo {year} {1995})\BibitemShut {NoStop}%
\bibitem [{\citenamefont {Wagman}(1982)}]{wagman1982nbs}%
  \BibitemOpen
  \bibfield  {author} {\bibinfo {author} {\bibfnamefont {D.~D.}\ \bibnamefont
  {Wagman}},\ }\bibfield  {title} {\bibinfo {title} {Nbs tables of chemical
  thermodynamic properties},\ }\href@noop {} {\bibfield  {journal} {\bibinfo
  {journal} {J. Phys. Chem. Ref. Data}\ }\textbf {\bibinfo {volume} {11}}
  (\bibinfo {year} {1982})}\BibitemShut {NoStop}%
\bibitem [{\citenamefont {Johnson}(1981)}]{johnson1981enthalpy}%
  \BibitemOpen
  \bibfield  {author} {\bibinfo {author} {\bibfnamefont {G.~K.}\ \bibnamefont
  {Johnson}},\ }\bibfield  {title} {\bibinfo {title} {The enthalpy of formation
  of fef3 by fluorine bomb calorimetry},\ }\href@noop {} {\bibfield  {journal}
  {\bibinfo  {journal} {The Journal of Chemical Thermodynamics}\ }\textbf
  {\bibinfo {volume} {13}},\ \bibinfo {pages} {465} (\bibinfo {year}
  {1981})}\BibitemShut {NoStop}%
\bibitem [{\citenamefont {Solov’ev}\ \emph {et~al.}(2012)\citenamefont
  {Solov’ev}, \citenamefont {Korunov}, \citenamefont {Zubkov},\ and\
  \citenamefont {Firer}}]{solov2012standard}%
  \BibitemOpen
  \bibfield  {author} {\bibinfo {author} {\bibfnamefont {S.}~\bibnamefont
  {Solov’ev}}, \bibinfo {author} {\bibfnamefont {A.}~\bibnamefont {Korunov}},
  \bibinfo {author} {\bibfnamefont {K.}~\bibnamefont {Zubkov}},\ and\ \bibinfo
  {author} {\bibfnamefont {A.}~\bibnamefont {Firer}},\ }\bibfield  {title}
  {\bibinfo {title} {Standard enthalpy of formation of nickel trifluoride by
  isothermal calorimetry},\ }\href@noop {} {\bibfield  {journal} {\bibinfo
  {journal} {Russian Journal of Physical Chemistry A}\ }\textbf {\bibinfo
  {volume} {86}},\ \bibinfo {pages} {516} (\bibinfo {year} {2012})}\BibitemShut
  {NoStop}%
\bibitem [{\citenamefont {Aykol}\ and\ \citenamefont
  {Wolverton}(2014)}]{aykol2014local}%
  \BibitemOpen
  \bibfield  {author} {\bibinfo {author} {\bibfnamefont {M.}~\bibnamefont
  {Aykol}}\ and\ \bibinfo {author} {\bibfnamefont {C.}~\bibnamefont
  {Wolverton}},\ }\bibfield  {title} {\bibinfo {title} {Local environment
  dependent gga+ u method for accurate thermochemistry of transition metal
  compounds},\ }\href@noop {} {\bibfield  {journal} {\bibinfo  {journal}
  {Physical Review B}\ }\textbf {\bibinfo {volume} {90}},\ \bibinfo {pages}
  {115105} (\bibinfo {year} {2014})}\BibitemShut {NoStop}%
\bibitem [{\citenamefont {Cococcioni}\ and\ \citenamefont
  {De~Gironcoli}(2005)}]{cococcioni2005linear}%
  \BibitemOpen
  \bibfield  {author} {\bibinfo {author} {\bibfnamefont {M.}~\bibnamefont
  {Cococcioni}}\ and\ \bibinfo {author} {\bibfnamefont {S.}~\bibnamefont
  {De~Gironcoli}},\ }\bibfield  {title} {\bibinfo {title} {Linear response
  approach to the calculation of the effective interaction parameters in the
  lda+ u method},\ }\href@noop {} {\bibfield  {journal} {\bibinfo  {journal}
  {Physical Review B}\ }\textbf {\bibinfo {volume} {71}},\ \bibinfo {pages}
  {035105} (\bibinfo {year} {2005})}\BibitemShut {NoStop}%
\bibitem [{\citenamefont {Sheets}(2023)}]{sheets2023frustrated}%
  \BibitemOpen
  \bibfield  {author} {\bibinfo {author} {\bibfnamefont {D.}~\bibnamefont
  {Sheets}},\ }\bibfield  {title} {\bibinfo {title} {Frustrated magnetic state
  of mott insulating tif3},\ }\href@noop {} {\bibfield  {journal} {\bibinfo
  {journal} {Bulletin of the American Physical Society}\ } (\bibinfo {year}
  {2023})}\BibitemShut {NoStop}%
\bibitem [{\citenamefont {Gossard}\ \emph {et~al.}(1974)\citenamefont
  {Gossard}, \citenamefont {Di~Salvo}, \citenamefont {Falconer}, \citenamefont
  {Rice}, \citenamefont {Voorhoeve},\ and\ \citenamefont
  {Yasuoka}}]{gossard1974magnetic}%
  \BibitemOpen
  \bibfield  {author} {\bibinfo {author} {\bibfnamefont {A.}~\bibnamefont
  {Gossard}}, \bibinfo {author} {\bibfnamefont {F.}~\bibnamefont {Di~Salvo}},
  \bibinfo {author} {\bibfnamefont {W.}~\bibnamefont {Falconer}}, \bibinfo
  {author} {\bibfnamefont {T.}~\bibnamefont {Rice}}, \bibinfo {author}
  {\bibfnamefont {J.}~\bibnamefont {Voorhoeve}},\ and\ \bibinfo {author}
  {\bibfnamefont {H.}~\bibnamefont {Yasuoka}},\ }\bibfield  {title} {\bibinfo
  {title} {Magnetic ordering of a d1 compound: Vf4},\ }\href@noop {} {\bibfield
   {journal} {\bibinfo  {journal} {Solid State Communications}\ }\textbf
  {\bibinfo {volume} {14}},\ \bibinfo {pages} {1207} (\bibinfo {year}
  {1974})}\BibitemShut {NoStop}%
\bibitem [{\citenamefont {Chatterji}\ and\ \citenamefont
  {Hansen}(2011)}]{chatterji2011magnCrF2_CuF2}%
  \BibitemOpen
  \bibfield  {author} {\bibinfo {author} {\bibfnamefont {T.}~\bibnamefont
  {Chatterji}}\ and\ \bibinfo {author} {\bibfnamefont {T.~C.}\ \bibnamefont
  {Hansen}},\ }\bibfield  {title} {\bibinfo {title} {Magnetoelastic effects in
  jahn--teller distorted crf2 and cuf2 studied by neutron powder diffraction},\
  }\href@noop {} {\bibfield  {journal} {\bibinfo  {journal} {Journal of
  Physics: Condensed Matter}\ }\textbf {\bibinfo {volume} {23}},\ \bibinfo
  {pages} {276007} (\bibinfo {year} {2011})}\BibitemShut {NoStop}%
\bibitem [{\citenamefont {Strempfer}\ \emph {et~al.}(2004)\citenamefont
  {Strempfer}, \citenamefont {R{\"u}tt}, \citenamefont {Bayrakci},
  \citenamefont {Br{\"u}ckel},\ and\ \citenamefont
  {Jauch}}]{strempfer2004magnetic}%
  \BibitemOpen
  \bibfield  {author} {\bibinfo {author} {\bibfnamefont {J.}~\bibnamefont
  {Strempfer}}, \bibinfo {author} {\bibfnamefont {U.}~\bibnamefont {R{\"u}tt}},
  \bibinfo {author} {\bibfnamefont {S.}~\bibnamefont {Bayrakci}}, \bibinfo
  {author} {\bibfnamefont {T.}~\bibnamefont {Br{\"u}ckel}},\ and\ \bibinfo
  {author} {\bibfnamefont {W.}~\bibnamefont {Jauch}},\ }\bibfield  {title}
  {\bibinfo {title} {Magnetic properties of transition metal fluorides m f 2
  (m= mn, fe, co, ni) via high-energy photon diffraction},\ }\href@noop {}
  {\bibfield  {journal} {\bibinfo  {journal} {Physical Review B}\ }\textbf
  {\bibinfo {volume} {69}},\ \bibinfo {pages} {014417} (\bibinfo {year}
  {2004})}\BibitemShut {NoStop}%
\bibitem [{\citenamefont {Lutar}\ \emph {et~al.}(1988)\citenamefont {Lutar},
  \citenamefont {Jesih},\ and\ \citenamefont {{\v{Z}}emva}}]{lutar1988krf2}%
  \BibitemOpen
  \bibfield  {author} {\bibinfo {author} {\bibfnamefont {K.}~\bibnamefont
  {Lutar}}, \bibinfo {author} {\bibfnamefont {A.}~\bibnamefont {Jesih}},\ and\
  \bibinfo {author} {\bibfnamefont {B.}~\bibnamefont {{\v{Z}}emva}},\
  }\bibfield  {title} {\bibinfo {title} {Krf2/mnf4 adducts from krf2/mnf2
  interaction in hf as a route to high purity mnf4},\ }\href@noop {} {\bibfield
   {journal} {\bibinfo  {journal} {Polyhedron}\ }\textbf {\bibinfo {volume}
  {7}},\ \bibinfo {pages} {1217} (\bibinfo {year} {1988})}\BibitemShut
  {NoStop}%
\bibitem [{\citenamefont {YANG}\ \emph {et~al.}(2012)\citenamefont {YANG},
  \citenamefont {WANG}, \citenamefont {Li},\ and\ \citenamefont
  {SU}}]{yang2012structural}%
  \BibitemOpen
  \bibfield  {author} {\bibinfo {author} {\bibfnamefont {Z.-h.}\ \bibnamefont
  {YANG}}, \bibinfo {author} {\bibfnamefont {X.-y.}\ \bibnamefont {WANG}},
  \bibinfo {author} {\bibfnamefont {L.}~\bibnamefont {Li}},\ and\ \bibinfo
  {author} {\bibfnamefont {X.-p.}\ \bibnamefont {SU}},\ }\bibfield  {title}
  {\bibinfo {title} {Structural, magnetic and electronic properties of fef2 by
  first-principle calculation},\ }\href@noop {} {\bibfield  {journal} {\bibinfo
   {journal} {Transactions of Nonferrous Metals Society of China}\ }\textbf
  {\bibinfo {volume} {22}},\ \bibinfo {pages} {386} (\bibinfo {year}
  {2012})}\BibitemShut {NoStop}%
\bibitem [{\citenamefont {Chatterji}\ \emph
  {et~al.}(2010{\natexlab{a}})\citenamefont {Chatterji}, \citenamefont
  {Ouladdiaf},\ and\ \citenamefont {Hansen}}]{chatterji2010CoF2}%
  \BibitemOpen
  \bibfield  {author} {\bibinfo {author} {\bibfnamefont {T.}~\bibnamefont
  {Chatterji}}, \bibinfo {author} {\bibfnamefont {B.}~\bibnamefont
  {Ouladdiaf}},\ and\ \bibinfo {author} {\bibfnamefont {T.~C.}\ \bibnamefont
  {Hansen}},\ }\bibfield  {title} {\bibinfo {title} {The magnetoelastic effect
  in cof2 investigated by means of neutron powder diffraction},\ }\href@noop {}
  {\bibfield  {journal} {\bibinfo  {journal} {Journal of Physics: Condensed
  Matter}\ }\textbf {\bibinfo {volume} {22}},\ \bibinfo {pages} {096001}
  (\bibinfo {year} {2010}{\natexlab{a}})}\BibitemShut {NoStop}%
\bibitem [{\citenamefont {Chatterji}\ \emph
  {et~al.}(2010{\natexlab{b}})\citenamefont {Chatterji}, \citenamefont {Iles},
  \citenamefont {Ouladdiaf},\ and\ \citenamefont
  {Hansen}}]{chatterji2010magnetoelastic}%
  \BibitemOpen
  \bibfield  {author} {\bibinfo {author} {\bibfnamefont {T.}~\bibnamefont
  {Chatterji}}, \bibinfo {author} {\bibfnamefont {G.~N.}\ \bibnamefont {Iles}},
  \bibinfo {author} {\bibfnamefont {B.}~\bibnamefont {Ouladdiaf}},\ and\
  \bibinfo {author} {\bibfnamefont {T.~C.}\ \bibnamefont {Hansen}},\ }\bibfield
   {title} {\bibinfo {title} {Magnetoelastic effect in mf2 (m= mn, fe, ni)
  investigated by neutron powder diffraction},\ }\href@noop {} {\bibfield
  {journal} {\bibinfo  {journal} {Journal of Physics: Condensed Matter}\
  }\textbf {\bibinfo {volume} {22}},\ \bibinfo {pages} {316001} (\bibinfo
  {year} {2010}{\natexlab{b}})}\BibitemShut {NoStop}%
\bibitem [{\citenamefont {Shen}\ \emph {et~al.}(1999)\citenamefont {Shen},
  \citenamefont {Chac{\'o}n}, \citenamefont {Rosov}, \citenamefont {Elder},
  \citenamefont {Allman},\ and\ \citenamefont {Bartlett}}]{shen1999structure}%
  \BibitemOpen
  \bibfield  {author} {\bibinfo {author} {\bibfnamefont {C.}~\bibnamefont
  {Shen}}, \bibinfo {author} {\bibfnamefont {L.~C.}\ \bibnamefont
  {Chac{\'o}n}}, \bibinfo {author} {\bibfnamefont {N.}~\bibnamefont {Rosov}},
  \bibinfo {author} {\bibfnamefont {S.~H.}\ \bibnamefont {Elder}}, \bibinfo
  {author} {\bibfnamefont {J.~C.}\ \bibnamefont {Allman}},\ and\ \bibinfo
  {author} {\bibfnamefont {N.}~\bibnamefont {Bartlett}},\ }\bibfield  {title}
  {\bibinfo {title} {The structure of r-nif3, and synthesis, and magnetism of
  new r3 miiniivf6 (m= fe, co, cu, zn), and miiniiif4 (m= co, cu)},\
  }\href@noop {} {\bibfield  {journal} {\bibinfo  {journal} {Comptes Rendus de
  l'Acad{\'e}mie des Sciences-Series IIC-Chemistry}\ }\textbf {\bibinfo
  {volume} {2}},\ \bibinfo {pages} {557} (\bibinfo {year} {1999})}\BibitemShut
  {NoStop}%
\bibitem [{\citenamefont {Fischer}\ \emph {et~al.}(1974)\citenamefont
  {Fischer}, \citenamefont {H{\"a}lg}, \citenamefont {Schwarzenbach},\ and\
  \citenamefont {Gamsj{\"a}ger}}]{fischer1974magnetic}%
  \BibitemOpen
  \bibfield  {author} {\bibinfo {author} {\bibfnamefont {P.}~\bibnamefont
  {Fischer}}, \bibinfo {author} {\bibfnamefont {W.}~\bibnamefont {H{\"a}lg}},
  \bibinfo {author} {\bibfnamefont {D.}~\bibnamefont {Schwarzenbach}},\ and\
  \bibinfo {author} {\bibfnamefont {H.}~\bibnamefont {Gamsj{\"a}ger}},\
  }\bibfield  {title} {\bibinfo {title} {Magnetic and crystal structure of
  copper (ii) fluoride},\ }\href@noop {} {\bibfield  {journal} {\bibinfo
  {journal} {Journal of Physics and Chemistry of Solids}\ }\textbf {\bibinfo
  {volume} {35}},\ \bibinfo {pages} {1683} (\bibinfo {year}
  {1974})}\BibitemShut {NoStop}%
\bibitem [{\citenamefont {Ebert}\ and\ \citenamefont
  {Woitinek}(1933)}]{ebert1933kristallstrukturen}%
  \BibitemOpen
  \bibfield  {author} {\bibinfo {author} {\bibfnamefont {F.}~\bibnamefont
  {Ebert}}\ and\ \bibinfo {author} {\bibfnamefont {H.}~\bibnamefont
  {Woitinek}},\ }\bibfield  {title} {\bibinfo {title} {Kristallstrukturen von
  fluoriden. ii. hgf, hgf2, cuf und cuf2},\ }\href@noop {} {\bibfield
  {journal} {\bibinfo  {journal} {Zeitschrift f{\"u}r anorganische und
  allgemeine Chemie}\ }\textbf {\bibinfo {volume} {210}},\ \bibinfo {pages}
  {269} (\bibinfo {year} {1933})}\BibitemShut {NoStop}%
\bibitem [{\citenamefont {Baur}\ and\ \citenamefont
  {Khan}(1971)}]{baur1971rutile}%
  \BibitemOpen
  \bibfield  {author} {\bibinfo {author} {\bibfnamefont {W.~H.}\ \bibnamefont
  {Baur}}\ and\ \bibinfo {author} {\bibfnamefont {A.~A.}\ \bibnamefont
  {Khan}},\ }\bibfield  {title} {\bibinfo {title} {Rutile-type compounds. iv.
  sio2, geo2 and a comparison with other rutile-type structures},\ }\href@noop
  {} {\bibfield  {journal} {\bibinfo  {journal} {Acta Crystallographica Section
  B: Structural Crystallography and Crystal Chemistry}\ }\textbf {\bibinfo
  {volume} {27}},\ \bibinfo {pages} {2133} (\bibinfo {year}
  {1971})}\BibitemShut {NoStop}%
\bibitem [{\citenamefont {Baur}(1958)}]{uber1958rutile}%
  \BibitemOpen
  \bibfield  {author} {\bibinfo {author} {\bibfnamefont {W.~H.}\ \bibnamefont
  {Baur}},\ }\bibfield  {title} {\bibinfo {title} {Uber die verfeinerung der
  kristallstrukturbestimmung einiger vertreter des rutiltyps. ii. die
  difluoride von mn, fe, co, ni und zn},\ }\href
  {https://doi.org/https://doi.org/10.1107/S0365110X58001353} {\bibfield
  {journal} {\bibinfo  {journal} {Acta Crystallographica}\ }\textbf {\bibinfo
  {volume} {11}},\ \bibinfo {pages} {488} (\bibinfo {year} {1958})}\BibitemShut
  {NoStop}%
\bibitem [{pro(1957)}]{proce1957}%
  \BibitemOpen
  \bibfield  {title} {\bibinfo {title} {Proceedings of the chemical society.},\
  }\href@noop {} {\bibfield  {journal} {\bibinfo  {journal} {Proc. Chem. Soc.}\
  ,\ \bibinfo {pages} {185}} (\bibinfo {year} {1957})}\BibitemShut {NoStop}%
\bibitem [{\citenamefont {Burns}\ and\ \citenamefont
  {Hawthorne}(1991)}]{burns1991rietveld}%
  \BibitemOpen
  \bibfield  {author} {\bibinfo {author} {\bibfnamefont {P.~C.}\ \bibnamefont
  {Burns}}\ and\ \bibinfo {author} {\bibfnamefont {F.~C.}\ \bibnamefont
  {Hawthorne}},\ }\bibfield  {title} {\bibinfo {title} {Rietveld refinement of
  the crystal structure of cuf2},\ }\href@noop {} {\bibfield  {journal}
  {\bibinfo  {journal} {Powder Diffraction}\ }\textbf {\bibinfo {volume} {6}},\
  \bibinfo {pages} {156} (\bibinfo {year} {1991})}\BibitemShut {NoStop}%
\bibitem [{\citenamefont {Siegel}(1956)}]{Siegel:a01775}%
  \BibitemOpen
  \bibfield  {author} {\bibinfo {author} {\bibfnamefont {S.}~\bibnamefont
  {Siegel}},\ }\bibfield  {title} {\bibinfo {title} {{The structure of TiF${\sb
  3}$}},\ }\href@noop {} {\bibfield  {journal} {\bibinfo  {journal} {Acta
  Crystallographica}\ }\textbf {\bibinfo {volume} {9}},\ \bibinfo {pages} {684}
  (\bibinfo {year} {1956})}\BibitemShut {NoStop}%
\bibitem [{\citenamefont {Jack}\ and\ \citenamefont
  {Gutmann}(1951)}]{jack1951crystal}%
  \BibitemOpen
  \bibfield  {author} {\bibinfo {author} {\bibfnamefont {K.}~\bibnamefont
  {Jack}}\ and\ \bibinfo {author} {\bibfnamefont {V.}~\bibnamefont {Gutmann}},\
  }\bibfield  {title} {\bibinfo {title} {The crystal structure of vanadium
  trifluoride vf3},\ }\href@noop {} {\bibfield  {journal} {\bibinfo  {journal}
  {Acta Crystallographica}\ }\textbf {\bibinfo {volume} {4}},\ \bibinfo {pages}
  {246} (\bibinfo {year} {1951})}\BibitemShut {NoStop}%
\bibitem [{\citenamefont {Knox}(1960)}]{knox1960structures}%
  \BibitemOpen
  \bibfield  {author} {\bibinfo {author} {\bibfnamefont {K.}~\bibnamefont
  {Knox}},\ }\bibfield  {title} {\bibinfo {title} {Structures of chromium (iii)
  fluoride},\ }\href@noop {} {\bibfield  {journal} {\bibinfo  {journal} {Acta
  Crystallographica}\ }\textbf {\bibinfo {volume} {13}},\ \bibinfo {pages}
  {507} (\bibinfo {year} {1960})}\BibitemShut {NoStop}%
\bibitem [{\citenamefont {Hepworth}\ \emph {et~al.}(1957)\citenamefont
  {Hepworth}, \citenamefont {Jack}, \citenamefont {Peacock},\ and\
  \citenamefont {Westland}}]{hepworth1957crystal}%
  \BibitemOpen
  \bibfield  {author} {\bibinfo {author} {\bibfnamefont {M.}~\bibnamefont
  {Hepworth}}, \bibinfo {author} {\bibfnamefont {K.}~\bibnamefont {Jack}},
  \bibinfo {author} {\bibfnamefont {R.}~\bibnamefont {Peacock}},\ and\ \bibinfo
  {author} {\bibfnamefont {G.}~\bibnamefont {Westland}},\ }\bibfield  {title}
  {\bibinfo {title} {The crystal structures of the trifluorides of iron,
  cobalt, ruthenium, rhodium, palladium and iridium},\ }\href@noop {}
  {\bibfield  {journal} {\bibinfo  {journal} {Acta Crystallographica}\ }\textbf
  {\bibinfo {volume} {10}},\ \bibinfo {pages} {63} (\bibinfo {year}
  {1957})}\BibitemShut {NoStop}%
\bibitem [{\citenamefont {Leblanc}\ \emph {et~al.}(1985)\citenamefont
  {Leblanc}, \citenamefont {Pannetier}, \citenamefont {Ferey},\ and\
  \citenamefont {De~Pape}}]{leblanc1985single}%
  \BibitemOpen
  \bibfield  {author} {\bibinfo {author} {\bibfnamefont {M.}~\bibnamefont
  {Leblanc}}, \bibinfo {author} {\bibfnamefont {J.}~\bibnamefont {Pannetier}},
  \bibinfo {author} {\bibfnamefont {G.}~\bibnamefont {Ferey}},\ and\ \bibinfo
  {author} {\bibfnamefont {R.}~\bibnamefont {De~Pape}},\ }\bibfield  {title}
  {\bibinfo {title} {Single crystal refinement of the structure of rhombohedral
  fef3},\ }\href@noop {} {\bibfield  {journal} {\bibinfo  {journal} {Revue de
  chimie min{\'e}rale}\ }\textbf {\bibinfo {volume} {22}},\ \bibinfo {pages}
  {107} (\bibinfo {year} {1985})}\BibitemShut {NoStop}%
\bibitem [{\citenamefont {Hepworth}\ and\ \citenamefont
  {Jack}(1957)}]{Hepworth:a01996}%
  \BibitemOpen
  \bibfield  {author} {\bibinfo {author} {\bibfnamefont {M.~A.}\ \bibnamefont
  {Hepworth}}\ and\ \bibinfo {author} {\bibfnamefont {K.~H.}\ \bibnamefont
  {Jack}},\ }\bibfield  {title} {\bibinfo {title} {{The crystal structure of
  manganese trifluoride, MnF${\sb 3}$}},\ }\href@noop {} {\bibfield  {journal}
  {\bibinfo  {journal} {Acta Crystallographica}\ }\textbf {\bibinfo {volume}
  {10}},\ \bibinfo {pages} {345} (\bibinfo {year} {1957})}\BibitemShut
  {NoStop}%
\bibitem [{\citenamefont {Bialowons}\ \emph {et~al.}(1995)\citenamefont
  {Bialowons}, \citenamefont {M{\"u}ller},\ and\ \citenamefont
  {M{\"u}ller}}]{bialowons1995titantetrafluorid}%
  \BibitemOpen
  \bibfield  {author} {\bibinfo {author} {\bibfnamefont {H.}~\bibnamefont
  {Bialowons}}, \bibinfo {author} {\bibfnamefont {M.}~\bibnamefont
  {M{\"u}ller}},\ and\ \bibinfo {author} {\bibfnamefont {B.}~\bibnamefont
  {M{\"u}ller}},\ }\bibfield  {title} {\bibinfo {title}
  {Titantetrafluorid--eine {\"u}berraschend einfache kolumnarstruktur},\
  }\href@noop {} {\bibfield  {journal} {\bibinfo  {journal} {Zeitschrift
  f{\"u}r anorganische und allgemeine Chemie}\ }\textbf {\bibinfo {volume}
  {621}},\ \bibinfo {pages} {1227} (\bibinfo {year} {1995})}\BibitemShut
  {NoStop}%
\bibitem [{\citenamefont {Becker}\ and\ \citenamefont
  {M{\"u}ller}(1990)}]{becker1990vanadiumtetrafluorid}%
  \BibitemOpen
  \bibfield  {author} {\bibinfo {author} {\bibfnamefont {S.}~\bibnamefont
  {Becker}}\ and\ \bibinfo {author} {\bibfnamefont {B.~G.}\ \bibnamefont
  {M{\"u}ller}},\ }\bibfield  {title} {\bibinfo {title}
  {Vanadiumtetrafluorid},\ }\href@noop {} {\bibfield  {journal} {\bibinfo
  {journal} {Angewandte Chemie}\ }\textbf {\bibinfo {volume} {102}},\ \bibinfo
  {pages} {426} (\bibinfo {year} {1990})}\BibitemShut {NoStop}%
\bibitem [{\citenamefont {Kr{\"a}mer}\ and\ \citenamefont
  {M{\"u}ller}(1995)}]{kramer1995struktur}%
  \BibitemOpen
  \bibfield  {author} {\bibinfo {author} {\bibfnamefont {O.}~\bibnamefont
  {Kr{\"a}mer}}\ and\ \bibinfo {author} {\bibfnamefont {B.}~\bibnamefont
  {M{\"u}ller}},\ }\bibfield  {title} {\bibinfo {title} {Zur struktur des
  chromtetrafluorids},\ }\href@noop {} {\bibfield  {journal} {\bibinfo
  {journal} {Zeitschrift f{\"u}r anorganische und allgemeine Chemie}\ }\textbf
  {\bibinfo {volume} {621}},\ \bibinfo {pages} {1969} (\bibinfo {year}
  {1995})}\BibitemShut {NoStop}%
\bibitem [{\citenamefont {M{\"u}ller}\ and\ \citenamefont
  {Serafin}(1987)}]{muller1987kristallstruktur}%
  \BibitemOpen
  \bibfield  {author} {\bibinfo {author} {\bibfnamefont {B.~G.}\ \bibnamefont
  {M{\"u}ller}}\ and\ \bibinfo {author} {\bibfnamefont {M.}~\bibnamefont
  {Serafin}},\ }\bibfield  {title} {\bibinfo {title} {Die kristallstruktur von
  mangantetrafluorid the crystal/structure of manganese tetrafluoride},\
  }\href@noop {} {\bibfield  {journal} {\bibinfo  {journal} {Zeitschrift
  f{\"u}r Naturforschung B}\ }\textbf {\bibinfo {volume} {42}},\ \bibinfo
  {pages} {1102} (\bibinfo {year} {1987})}\BibitemShut {NoStop}%
\bibitem [{\citenamefont {Hsu}\ \emph {et~al.}(2009)\citenamefont {Hsu},
  \citenamefont {Umemoto}, \citenamefont {Cococcioni},\ and\ \citenamefont
  {Wentzcovitch}}]{hsu2009first}%
  \BibitemOpen
  \bibfield  {author} {\bibinfo {author} {\bibfnamefont {H.}~\bibnamefont
  {Hsu}}, \bibinfo {author} {\bibfnamefont {K.}~\bibnamefont {Umemoto}},
  \bibinfo {author} {\bibfnamefont {M.}~\bibnamefont {Cococcioni}},\ and\
  \bibinfo {author} {\bibfnamefont {R.}~\bibnamefont {Wentzcovitch}},\
  }\bibfield  {title} {\bibinfo {title} {First-principles study for low-spin
  lacoo 3 with a structurally consistent hubbard u},\ }\href@noop {} {\bibfield
   {journal} {\bibinfo  {journal} {Physical Review B}\ }\textbf {\bibinfo
  {volume} {79}},\ \bibinfo {pages} {125124} (\bibinfo {year}
  {2009})}\BibitemShut {NoStop}%
\bibitem [{\citenamefont {Lide}(2004)}]{lide2004crc}%
  \BibitemOpen
  \bibfield  {author} {\bibinfo {author} {\bibfnamefont {D.~R.}\ \bibnamefont
  {Lide}},\ }\href@noop {} {\emph {\bibinfo {title} {CRC handbook of chemistry
  and physics}}},\ Vol.~\bibinfo {volume} {85}\ (\bibinfo  {publisher} {CRC
  press},\ \bibinfo {year} {2004})\BibitemShut {NoStop}%
\bibitem [{\citenamefont {Kraus}\ \emph {et~al.}(2020)\citenamefont {Kraus},
  \citenamefont {Ivlev}, \citenamefont {Bandemehr}, \citenamefont {Sachs},
  \citenamefont {Pietzonka}, \citenamefont {Conrad}, \citenamefont {Serafin},\
  and\ \citenamefont {M{\"u}ller}}]{kraus2020synthesis}%
  \BibitemOpen
  \bibfield  {author} {\bibinfo {author} {\bibfnamefont {F.}~\bibnamefont
  {Kraus}}, \bibinfo {author} {\bibfnamefont {S.~I.}\ \bibnamefont {Ivlev}},
  \bibinfo {author} {\bibfnamefont {J.}~\bibnamefont {Bandemehr}}, \bibinfo
  {author} {\bibfnamefont {M.}~\bibnamefont {Sachs}}, \bibinfo {author}
  {\bibfnamefont {C.}~\bibnamefont {Pietzonka}}, \bibinfo {author}
  {\bibfnamefont {M.}~\bibnamefont {Conrad}}, \bibinfo {author} {\bibfnamefont
  {M.}~\bibnamefont {Serafin}},\ and\ \bibinfo {author} {\bibfnamefont {B.~G.}\
  \bibnamefont {M{\"u}ller}},\ }\bibfield  {title} {\bibinfo {title} {Synthesis
  and characterization of manganese tetrafluoride $\beta$-mnf4},\ }\href@noop
  {} {\bibfield  {journal} {\bibinfo  {journal} {Zeitschrift f{\"u}r
  anorganische und allgemeine Chemie}\ }\textbf {\bibinfo {volume} {646}},\
  \bibinfo {pages} {1481} (\bibinfo {year} {2020})}\BibitemShut {NoStop}%
\bibitem [{\citenamefont {Nikitin}\ and\ \citenamefont
  {Alikhanyan}(2019)}]{nikitin2019thermochemistry}%
  \BibitemOpen
  \bibfield  {author} {\bibinfo {author} {\bibfnamefont {M.}~\bibnamefont
  {Nikitin}}\ and\ \bibinfo {author} {\bibfnamefont {A.}~\bibnamefont
  {Alikhanyan}},\ }\bibfield  {title} {\bibinfo {title} {Thermochemistry of
  nickel trifluoride},\ }\href@noop {} {\bibfield  {journal} {\bibinfo
  {journal} {Russian Journal of Inorganic Chemistry}\ }\textbf {\bibinfo
  {volume} {64}},\ \bibinfo {pages} {641} (\bibinfo {year} {2019})}\BibitemShut
  {NoStop}%
\bibitem [{\citenamefont {Hummel}(2011)}]{hummel2011electronic}%
  \BibitemOpen
  \bibfield  {author} {\bibinfo {author} {\bibfnamefont {R.~E.}\ \bibnamefont
  {Hummel}},\ }\href@noop {} {\emph {\bibinfo {title} {Electronic Properties of
  Materials}}}\ (\bibinfo  {publisher} {Springer},\ \bibinfo {address} {New
  York, NY},\ \bibinfo {year} {2011})\BibitemShut {NoStop}%
\bibitem [{\citenamefont {Devey}\ \emph {et~al.}(2009)\citenamefont {Devey},
  \citenamefont {Grau-Crespo},\ and\ \citenamefont
  {De~Leeuw}}]{devey2009electronic}%
  \BibitemOpen
  \bibfield  {author} {\bibinfo {author} {\bibfnamefont {A.}~\bibnamefont
  {Devey}}, \bibinfo {author} {\bibfnamefont {R.}~\bibnamefont {Grau-Crespo}},\
  and\ \bibinfo {author} {\bibfnamefont {N.}~\bibnamefont {De~Leeuw}},\
  }\bibfield  {title} {\bibinfo {title} {Electronic and magnetic structure of
  fe 3 s 4: Gga+ u investigation},\ }\href@noop {} {\bibfield  {journal}
  {\bibinfo  {journal} {Physical Review B}\ }\textbf {\bibinfo {volume} {79}},\
  \bibinfo {pages} {195126} (\bibinfo {year} {2009})}\BibitemShut {NoStop}%
\bibitem [{\citenamefont {Rodr{\'\i}guez}\ \emph {et~al.}(2020)\citenamefont
  {Rodr{\'\i}guez}, \citenamefont {Zandalazini}, \citenamefont {Navarro},
  \citenamefont {Vadiraj},\ and\ \citenamefont
  {Albanesi}}]{rodriguez2020first}%
  \BibitemOpen
  \bibfield  {author} {\bibinfo {author} {\bibfnamefont {S.}~\bibnamefont
  {Rodr{\'\i}guez}}, \bibinfo {author} {\bibfnamefont {C.}~\bibnamefont
  {Zandalazini}}, \bibinfo {author} {\bibfnamefont {J.}~\bibnamefont
  {Navarro}}, \bibinfo {author} {\bibfnamefont {K.}~\bibnamefont {Vadiraj}},\
  and\ \bibinfo {author} {\bibfnamefont {E.}~\bibnamefont {Albanesi}},\
  }\bibfield  {title} {\bibinfo {title} {First principles calculations and
  experimental study of the optical properties of ni-doped zns},\ }\href@noop
  {} {\bibfield  {journal} {\bibinfo  {journal} {Materials research express}\
  }\textbf {\bibinfo {volume} {7}},\ \bibinfo {pages} {016303} (\bibinfo {year}
  {2020})}\BibitemShut {NoStop}%
\bibitem [{\citenamefont {Seo}\ \emph {et~al.}(2015)\citenamefont {Seo},
  \citenamefont {Urban}, \citenamefont {Ceder} \emph
  {et~al.}}]{seo2015calibrating}%
  \BibitemOpen
  \bibfield  {author} {\bibinfo {author} {\bibfnamefont {D.-H.}\ \bibnamefont
  {Seo}}, \bibinfo {author} {\bibfnamefont {A.}~\bibnamefont {Urban}}, \bibinfo
  {author} {\bibfnamefont {G.}~\bibnamefont {Ceder}}, \emph {et~al.},\
  }\bibfield  {title} {\bibinfo {title} {Calibrating transition-metal energy
  levels and oxygen bands in first-principles calculations: Accurate prediction
  of redox potentials and charge transfer in lithium transition-metal oxides},\
  }\href@noop {} {\bibfield  {journal} {\bibinfo  {journal} {Physical Review
  B}\ }\textbf {\bibinfo {volume} {92}},\ \bibinfo {pages} {115118} (\bibinfo
  {year} {2015})}\BibitemShut {NoStop}%
\bibitem [{\citenamefont {Dimov}\ \emph {et~al.}(2013)\citenamefont {Dimov},
  \citenamefont {Nishimura}, \citenamefont {Chihara}, \citenamefont {Kitajou},
  \citenamefont {Gocheva},\ and\ \citenamefont {Okada}}]{dimov2013transition}%
  \BibitemOpen
  \bibfield  {author} {\bibinfo {author} {\bibfnamefont {N.}~\bibnamefont
  {Dimov}}, \bibinfo {author} {\bibfnamefont {A.}~\bibnamefont {Nishimura}},
  \bibinfo {author} {\bibfnamefont {K.}~\bibnamefont {Chihara}}, \bibinfo
  {author} {\bibfnamefont {A.}~\bibnamefont {Kitajou}}, \bibinfo {author}
  {\bibfnamefont {I.~D.}\ \bibnamefont {Gocheva}},\ and\ \bibinfo {author}
  {\bibfnamefont {S.}~\bibnamefont {Okada}},\ }\bibfield  {title} {\bibinfo
  {title} {Transition metal namf3 compounds as model systems for studying the
  feasibility of ternary li-mf and na-mf single phases as cathodes for
  lithium--ion and sodium--ion batteries},\ }\href@noop {} {\bibfield
  {journal} {\bibinfo  {journal} {Electrochimica Acta}\ }\textbf {\bibinfo
  {volume} {110}},\ \bibinfo {pages} {214} (\bibinfo {year}
  {2013})}\BibitemShut {NoStop}%
\bibitem [{\citenamefont {Mahajan}\ \emph {et~al.}(2021)\citenamefont
  {Mahajan}, \citenamefont {Timrov}, \citenamefont {Marzari},\ and\
  \citenamefont {Kashyap}}]{mahajan2021importance}%
  \BibitemOpen
  \bibfield  {author} {\bibinfo {author} {\bibfnamefont {R.}~\bibnamefont
  {Mahajan}}, \bibinfo {author} {\bibfnamefont {I.}~\bibnamefont {Timrov}},
  \bibinfo {author} {\bibfnamefont {N.}~\bibnamefont {Marzari}},\ and\ \bibinfo
  {author} {\bibfnamefont {A.}~\bibnamefont {Kashyap}},\ }\bibfield  {title}
  {\bibinfo {title} {Importance of intersite hubbard interactions in $\beta$-
  mno 2: A first-principles dft+ u+ v study},\ }\href@noop {} {\bibfield
  {journal} {\bibinfo  {journal} {Physical Review Materials}\ }\textbf
  {\bibinfo {volume} {5}},\ \bibinfo {pages} {104402} (\bibinfo {year}
  {2021})}\BibitemShut {NoStop}%
\bibitem [{\citenamefont {Timrov}\ \emph {et~al.}(2022)\citenamefont {Timrov},
  \citenamefont {Aquilante}, \citenamefont {Cococcioni},\ and\ \citenamefont
  {Marzari}}]{timrov2022accurate}%
  \BibitemOpen
  \bibfield  {author} {\bibinfo {author} {\bibfnamefont {I.}~\bibnamefont
  {Timrov}}, \bibinfo {author} {\bibfnamefont {F.}~\bibnamefont {Aquilante}},
  \bibinfo {author} {\bibfnamefont {M.}~\bibnamefont {Cococcioni}},\ and\
  \bibinfo {author} {\bibfnamefont {N.}~\bibnamefont {Marzari}},\ }\bibfield
  {title} {\bibinfo {title} {Accurate electronic properties and intercalation
  voltages of olivine-type li-ion cathode materials from extended hubbard
  functionals},\ }\href@noop {} {\bibfield  {journal} {\bibinfo  {journal} {PRX
  Energy}\ }\textbf {\bibinfo {volume} {1}},\ \bibinfo {pages} {033003}
  (\bibinfo {year} {2022})}\BibitemShut {NoStop}%
\bibitem [{\citenamefont {Ricca}\ \emph {et~al.}(2020)\citenamefont {Ricca},
  \citenamefont {Timrov}, \citenamefont {Cococcioni}, \citenamefont {Marzari},\
  and\ \citenamefont {Aschauer}}]{ricca2020phyrevl}%
  \BibitemOpen
  \bibfield  {author} {\bibinfo {author} {\bibfnamefont {C.}~\bibnamefont
  {Ricca}}, \bibinfo {author} {\bibfnamefont {I.}~\bibnamefont {Timrov}},
  \bibinfo {author} {\bibfnamefont {M.}~\bibnamefont {Cococcioni}}, \bibinfo
  {author} {\bibfnamefont {N.}~\bibnamefont {Marzari}},\ and\ \bibinfo {author}
  {\bibfnamefont {U.}~\bibnamefont {Aschauer}},\ }\bibfield  {title} {\bibinfo
  {title} {Self-consistent $\mathrm{DFT}+u+v$ study of oxygen vacancies in
  ${\mathrm{srtio}}_{3}$},\ }\href@noop {} {\bibfield  {journal} {\bibinfo
  {journal} {Phys. Rev. Res.}\ }\textbf {\bibinfo {volume} {2}},\ \bibinfo
  {pages} {023313} (\bibinfo {year} {2020})}\BibitemShut {NoStop}%
\bibitem [{\citenamefont {Tancogne-Dejean}\ and\ \citenamefont
  {Rubio}(2020)}]{tancogne2020parameter}%
  \BibitemOpen
  \bibfield  {author} {\bibinfo {author} {\bibfnamefont {N.}~\bibnamefont
  {Tancogne-Dejean}}\ and\ \bibinfo {author} {\bibfnamefont {A.}~\bibnamefont
  {Rubio}},\ }\bibfield  {title} {\bibinfo {title} {Parameter-free hybridlike
  functional based on an extended hubbard model: Dft+ u+ v},\ }\href@noop {}
  {\bibfield  {journal} {\bibinfo  {journal} {Physical Review B}\ }\textbf
  {\bibinfo {volume} {102}},\ \bibinfo {pages} {155117} (\bibinfo {year}
  {2020})}\BibitemShut {NoStop}%
\bibitem [{\citenamefont {Tiago}\ \emph {et~al.}(2004)\citenamefont {Tiago},
  \citenamefont {Ismail-Beigi},\ and\ \citenamefont {Louie}}]{tiago2004effect}%
  \BibitemOpen
  \bibfield  {author} {\bibinfo {author} {\bibfnamefont {M.~L.}\ \bibnamefont
  {Tiago}}, \bibinfo {author} {\bibfnamefont {S.}~\bibnamefont
  {Ismail-Beigi}},\ and\ \bibinfo {author} {\bibfnamefont {S.~G.}\ \bibnamefont
  {Louie}},\ }\bibfield  {title} {\bibinfo {title} {Effect of semicore orbitals
  on the electronic band gaps of si, ge, and gaas within the gw
  approximation},\ }\href@noop {} {\bibfield  {journal} {\bibinfo  {journal}
  {Physical Review B}\ }\textbf {\bibinfo {volume} {69}},\ \bibinfo {pages}
  {125212} (\bibinfo {year} {2004})}\BibitemShut {NoStop}%
\bibitem [{\citenamefont {Oshikiri}\ and\ \citenamefont
  {Aryasetiawan}(1999)}]{oshikiri1999band}%
  \BibitemOpen
  \bibfield  {author} {\bibinfo {author} {\bibfnamefont {M.}~\bibnamefont
  {Oshikiri}}\ and\ \bibinfo {author} {\bibfnamefont {F.}~\bibnamefont
  {Aryasetiawan}},\ }\bibfield  {title} {\bibinfo {title} {Band gaps and
  quasiparticle energy calculations on zno, zns, and znse in the zinc-blende
  structure by the gw approximation},\ }\href@noop {} {\bibfield  {journal}
  {\bibinfo  {journal} {Physical Review B}\ }\textbf {\bibinfo {volume} {60}},\
  \bibinfo {pages} {10754} (\bibinfo {year} {1999})}\BibitemShut {NoStop}%
\bibitem [{\citenamefont {Aryasetiawan}\ and\ \citenamefont
  {Gunnarsson}(1995)}]{aryasetiawan1995electronic}%
  \BibitemOpen
  \bibfield  {author} {\bibinfo {author} {\bibfnamefont {F.}~\bibnamefont
  {Aryasetiawan}}\ and\ \bibinfo {author} {\bibfnamefont {O.}~\bibnamefont
  {Gunnarsson}},\ }\bibfield  {title} {\bibinfo {title} {Electronic structure
  of nio in the gw approximation},\ }\href@noop {} {\bibfield  {journal}
  {\bibinfo  {journal} {Physical review letters}\ }\textbf {\bibinfo {volume}
  {74}},\ \bibinfo {pages} {3221} (\bibinfo {year} {1995})}\BibitemShut
  {NoStop}%
\end{thebibliography}%

\end{document}